\documentclass[usenatbib]{mnras2}
\usepackage{newtxtext,newtxmath}
\usepackage[T1]{fontenc}
\usepackage{ae,aecompl}
\usepackage{pdfpages}
\usepackage{graphicx}
\usepackage{aas_macros}
\usepackage{amsmath}
\usepackage{color}

\usepackage{hyperref}
\usepackage{changes}
\usepackage{todonotes}
\newcommand{\datastatement}[1]{\begin{small}\section*{Data Availability Statement}\end{small}{\noindent #1}\vspace{5pt}}

\usepackage{newtxtext,newtxmath}

\usepackage[T1]{fontenc}

\DeclareRobustCommand{\VAN}[3]{#2}
\let\VANthebibliography\thebibliography
\def\thebibliography{\DeclareRobustCommand{\VAN}[3]{##3}\VANthebibliography}

\usepackage{graphicx}	
\usepackage{amsmath}	

\newcommand{\Alf}{{Alfv\'en}}

\title[RDIs and the variability of AGB and RCB stars]{On the optical properties of Resonant Drag Instabilities: Variability of Asymptotic Giant Branch and R Coronae Borealis stars}

\author[Ulrich P. Steinwandel et al.]{Ulrich P. Steinwandel$^{1}$\thanks{E-mail: usteinwandel@flatironinstitute.org (UPS)} 
Alexander A. Kaurov$^{2,3}$, 
Philip F. Hopkins$^{4,5}$, 
and Jonathan Squire$^{6}$ 
\\
  $^{1}$Center for Computational Astrophysics, Flatiron Institute, 162 5th Avenue, New York, NY 10010\\
  $^{2}$Blue Marble Space Institute of Science, Seattle, WA 98104, USA\\
  $^{3}$Institute for Advanced Study, 1 Einstein Drive, Princeton, NJ 08540, USA \\
  $^{4}$TAPIR, Mailcode 350-17, California Institute of Technology, Pasadena, CA 91125, USA \\
  $^{5}$Walter Burke Institute for Theoretical Physics, Pasadena, CA 91125, USA \\
  $^{6}$Physics Department, University of Otago, Dunedin 9010, New Zealand \\
  \vspace{-0.5cm}
}

\date{Accepted XXX. Received YYY; in original form ZZZ}

\pubyear{2021}

\begin{document}
\label{firstpage}
\pagerange{\pageref{firstpage}--\pageref{lastpage}}
\maketitle

\begin{abstract}
In dusty cool-star outflow or ejection events around AGB or RCB-like stars, dust is accelerated by radiation from the star and coupled to the gas via collisional drag forces. But it has recently been shown that such dust-gas mixtures are unstable to a super-class of instabilities called the resonant drag instabilities (RDIs), which promote dust clustering. We therefore consider idealized simulations of the RDIs operating on a spectrum of dust grain sizes subject to radiative acceleration (allowing for different grain optical properties), coupled to the gas with a realistic drag law, including or excluding the effects of magnetic fields and charged grains, and calculate for the first time how the RDIs could contribute to observed variability. We show that the RDIs naturally produce significant variations ($\sim 10-20\%$ $1\sigma$-level) in the extinction, corresponding to $\sim 0.1-1\,$mag level in the stellar types above, on timescales of order months to a year. The fluctuations are surprisingly robust to finite source-size effects as they are dominated by large-scale modes, which also means their spatial structure could be resolved in some nearby systems. We also quantify how this produces variations in the line-of-sight grain size-distribution. All of these variations are similar to those observed, suggesting that the RDIs may play a key role driving observed variability in dust extinction within dusty outflow/ejection events around cool stars. We further propose that the measured variations in grain sizes could directly be used to identify the presence of the RDIs in close by systems with observations.
\end{abstract}





\section{Introduction}

Dust is a fundamental ingredient within many systems in astrophysics and contributes heavily to star and planet formation, astro-chemistry, extinction and attenuation of interstellar light, interstellar medium (ISM) heating and cooling processes, dynamics of outflows around active galactic nuclei, and more \citep[for a review, see][]{Draine2003}. 

One important regime where dust plays a major role is in ``dusty outflows'' from cool stars, such as asymptotic giant branch (AGB) or R Coronae Borealis (RCB) stars. While the winds of hotter more massive giant stars are driven by gas line opacities \citep[e.g.][]{Kudritzki2000, Puls2008, Owocki2010, Owocki2014, Smith2014, Puls2015}, in RCB- and AGB-star systems some initial processes (e.g.\ pulsational instabilities) allow the envelope temperature to drop below the condensation temperature and form dust grains, which are then strongly accelerated by absorbing the stellar radiation, potentially driving the outflow by collisionally dragging gas along \citep{Reimers1975,Willson1979, Hill1979, Wood1979,Engels1983, Baud1983,1986NInfo..61...48F}. 

The broad outlines of these qualitative scenarios have been supported by a wide range of observations of both time-resolved \citep{Whitelock1991, Whitelock2003, Glass2009, McDonald2016} and spatially-resolved \citep{Tej2003, Wittkowski2007, Zhao-Geisler2012, Karovicova2013, Stewart2016, Khouri2016, Ohnaka2017A&A, Ohnaka2017Natur} structures in the dust \citep[for a review, see][]{2018A&ARv..26....1H}.

Recently, however, \citet{squire.hopkins:RDI} showed that dust grains being accelerated through gas in this manner are subject to an enormous super-class of instabilities, which they termed the Resonant Drag Instabilities (RDIs). The ``resonance'' here refers to the fact that the growth rates are maximized wherever the natural frequencies of dust and gas modes are equal, so different resonances arise for all possible ``mode pairs.'' The analytic linear theory of these instabilities was studied further in \citet{hopkins:2018.mhd.rdi,squire:rdi.ppd, hopkins:2017.acoustic.RDI}, who noted that cool-star outflows should be linearly unstable to RDIs on essentially all spatial scales, with linear growth rates much shorter than the bulk outflow timescales. In fact, certain special cases of the ``acoustic quasi-sound and quasi-drift'' RDIs were first identified in the cool star outflow literature by \citet{morris:1993.cool.wind.dust.drag.instability.slow.saturated.mode,mastrodemos:dust.gas.coupling.cool.winds.spherical.symmetry,deguchi:1997.dust.envelope.pne.spherical.drag.instability.quasi.resonant}, but some of the simplifying assumptions imposed (e.g.\ the dust always following the local equilibrium drift velocity) prevented identification of the key resonance structures that allow rapid RDI growth. 
More recently, \citet{moseley:2018.acoustic.rdi.sims,seligman:2018.mhd.rdi.sims,hopkins:2019.mhd.rdi.periodic.box.sims} presented first idealized numerical studies of RDIs, and argued that in the non-linear state these could drive strong dust clumping which, they speculated, might explain some observed variability in the dusty structures in cool-star envelopes and outflows. However those studies did not allow for a spectrum of grain sizes, crucial for predicting extinction properties, and did not investigate any observable diagnostics or parameters most relevant for these physical systems. 

In this paper, we therefore present idealized simulations of the RDIs, with a spectrum of grain sizes and parameters chosen to be order-of-magnitude relevant for cool-star outflows, and for the first time calculate the behavior of temporal and spatial variations in the dust extinction along different sightlines to the star. We stress that these are not ``full physics'' simulations of dusty outflows: there has been extensive work on multi-physics simulations and calculations of wind dynamics with many other instabilities that can drive variability \citep[see e.g.][]{macgregor:grains.cool.star.eventually.decouple,hartquist:bfield.dust.coupling.cool.star.winds,
sandin:agb.wind.sims,soker:agb.star.magnetic.cycle.instabilities,soker:2002.arcs.around.agb.stars.from.instability,simis:2001.shells.around.dust.agb.winds,woitke:2d.agb.wind.simulations,woitke:2d.rad.pressure.dust.agb.wind.models}. As discussed in \citet{hopkins:2017.acoustic.RDI}, some of these numerical studies likely saw some of the RDIs, though they were not explicitly identified as such. Rather our goal is to specifically study idealized simulations of the RDIs which allow us to isolate their role, so that we can ask whether or not these could plausibly explain dust variability in certain physical systems.


\section{Numerical Methods and Simulations} \label{sec:simulation_method}

In this section we briefly discuss the numerical methods that are needed to model the RDIs in certain regions of interest. For this cause we partially use the simulations that are carried out by \cite{hopkins:2018.mhd.rdi} and run new simulations with different dust-to-gas ratios but include the same physics to investigate a larger region of the parameter space.
The simulations are carried out in a dimensionless unit system. This makes it possible to apply the simulations to several astrophysical regions of interest where dust is believed to have a major impact on the underlying physics.

\subsection{System of equations}

We present the system of equations that is taken into account within our simulations. In general there are three parts that we need to consider to treat the system properly. The first is the equation of motion (EOM) of the dust. The second is the EOM of the gas (Euler`s equation). The third is the coupling of the dust with the gas (and vice versa). For the dust grains the EOM can be directly integrated for each individual dust grain by solving: 
\begin{align}
  \frac{d\mathbf{v}_\mathrm{d}}{dt} = - \frac{\mathbf{w}_\mathrm{s}}{t_\mathrm{s}} + \mathbf{a}, 
  \label{eq:EOMdust}
\end{align}
in co-moving coordinates with the constant external acceleration $\mathbf{a}$ on the dust grain, the drift velocity $\mathbf{w}_\mathrm{d} = \mathbf{v}_\mathrm{d} - \mathbf{u}_\mathrm{g}$, which is given as the relative movement between dust $(\mathbf{v}_\mathrm{d})$ and gas particles $\mathbf{u}_\mathrm{g}$ and the stopping time $t_\mathrm{s}$. In our simulations the stopping time is given by the assumption of Epstein-drag which leads to the following expression in both sub and supersonic regimes following \citep{Draine1979}:
\begin{align}
  t_\mathrm{s} = \sqrt{\frac{\pi \gamma}{8}}\frac{\bar{\rho}_\mathrm{d}\epsilon_\mathrm{d}}{\rho_\mathrm{g} c_\mathrm{s}} \left(1 + \frac{9\pi\gamma}{128}\frac{\mathbf{w}_\mathrm{d}^{2}}{c_\mathrm{s}^2} \right)^{-1/2},  
\end{align}  
with the gas density $\rho_\mathrm{g}$, the adiabatic index $\gamma$ of the gas given by $\partial \log{P}_\mathrm{g}/\partial \log{\rho_\mathrm{g}}$ with the pressure $P_\mathrm{g}$ of the fluid (we adopt an isothermal $\gamma=1$), the grain density $\bar{\rho}_\mathrm{d}$, the dust grain radius $\epsilon_\mathrm{d}$ and the sound speed $c_\mathrm{s} = \sqrt{\partial P_\mathrm{g}/ \partial \rho_\mathrm{g}}$.
The gas component is modelled via Euler`s equation, including a term that models the coupling (drag force) of the dust on the gas particles:
\begin{align}
  \rho_\mathrm{g} \left(\frac{\partial}{\partial t} + \mathbf{u}_\mathrm{g} \cdot \nabla  \right)\mathbf{u}_\mathrm{g} = -\nabla P_\mathrm{g} + \int d^3 v_\mathrm{d}f_\mathrm{d}(\mathbf{v}_\mathrm{d})\frac{\mathbf{w}_\mathrm{s}}{t_\mathrm{s}}.
\label{eq:eomdustgas}
\end{align} 
The last term on the right-hand side of the equation \ref{eq:eomdustgas} is the coupling term for the drag force that the gas particles feel in the presence of the dust. The function $f_\mathrm{d}(\mathbf{v}_\mathrm{d})$ is the given phase-space density of the underlying distribution of the dust particles, where the grain density is given by evaluating the expression $\bar{\rho}_\mathrm{d} = \int d^3 v_\mathrm{d}f_\mathrm{d}(\mathbf{v}_\mathrm{d})$. In recent studies the focus was on the growth rate of the linear regime \citep{hopkins:2017.acoustic.RDI, squire:rdi.ppd} or the non-linear regime \citep{moseley:2018.acoustic.rdi.sims, seligman:2018.mhd.rdi.sims, hopkins2021:mhd.rdi.stratified.box.sims}, where it was convenient to consider just a single grain size. However, in our studies we apply a grain size distribution of the form:
\begin{align}
  \frac{dN}{d\epsilon_\mathrm{d}} \propto \epsilon_\mathrm{d}^{-3.5},
  \label{eq:grainsizedist}
\end{align}
following the grain size distribution from the interstellar dust model of \cite{Mathis1977}, with a minimum grain size a factor $=100$ smaller than the maximum \citep[see e.g.][]{Weingartner2001}. Including a grain size spectrum that takes the form which is presented in equation \ref{eq:grainsizedist} gives us the opportunity to study the influence of different grain sizes on the geometrical optical depth.

Some simulations are carried out including the effects of magnetic fields. Thus we solve the equations of magnetohydrodynamics instead of hydrodynamics and replace equation \ref{eq:EOMdust} by:
\begin{align}
  \frac{d\mathbf{v}_\mathrm{d}}{dt} = - \frac{\mathbf{w}_\mathrm{s}}{t_\mathrm{s}} + \frac{\mathbf{w}_\mathrm{s} \times \hat{\mathbf{B}}}{t_\mathrm{L}} + \mathbf{a}, 
  \label{eq:EOMdust_mhd}
\end{align}
where t$_\mathrm{L}$ is the Larmor-time given as 
\begin{align}
    t_\mathrm{L} = \frac{m_\mathrm{d} c}{|q_\mathrm{d} \mathbf{B}|} = \frac{4 \pi \bar{\rho}_\mathrm{d} \epsilon_\mathrm{d}^{3} c}{3 e |Z_\mathrm{d} \textbf{B}|},
\end{align}
with $m_{d} \equiv (4\pi/3)\,\bar{\rho}_{\mathrm{d}}\,\epsilon_{\mathrm{d}}^{3}$ for a mean internal grain density $\bar{\rho}_\mathrm{d}$ is the grain mass and Z$_{d}$ as the grain charge. This Larmor-time corresponds to a Larmor-frequency $\omega_\mathrm{L}$ at which charged particles gyrate under the influence of the magnetic field. In the MHD case, equation \ref{eq:eomdustgas} is replaced with 
\begin{align}
  \rho_\mathrm{g} \left(\frac{\partial}{\partial t} + \mathbf{u}_\mathrm{g} \cdot \nabla  \right)\mathbf{u}_\mathrm{g} = -\nabla P_\mathrm{g} + \frac{\mathbf{B} \times (\nabla \times \mathbf{B})}{4 \pi} + \int d^3 v_\mathrm{d}f_\mathrm{d}(\mathbf{v}_\mathrm{d})\frac{\mathbf{w}_\mathrm{s}}{t_\mathrm{s}}.
\label{eq:eomdustgas_mhd}
\end{align}
We assume collisional grain charging following \citet{draine:1987.grain.charging} to calculate $Z_\mathrm{d}$ (see \citealt{seligman:2018.mhd.rdi.sims}). 

We are interested here in applications to dust-driven winds from cool stars, so the relevant acceleration comes from radiation pressure, i.e.\ $\hat{\bf a} \sim (L_{\ast}/(4\pi\,r^{2}\,c))\,(Q\,\pi\,\epsilon_{\rm d}^{2}/m_{\rm d})\,\hat{\bf r}$, where $L_{\ast}$ is the stellar luminosity, ${\bf r}$ is the direction of the radiation pressure from the star, and $Q=Q(\epsilon_\mathrm{d},\,\lambda,\,...)$ is the usual dimensionless absorption efficiency (which is a function of the grain composition and incident spectrum). We will refer to the ``geometric'' absorption limit, $Q=1$, as appropriate when $\epsilon_\mathrm{d} \gg \lambda$. However, in many situations the driving radiation may have $\lambda \gg \epsilon_{\rm d}$, in which case $Q\sim \epsilon_\mathrm{d}/\lambda$ \citep{draine:ism.book}, and we can model this case by taking $Q=Q_{0}\,(\epsilon_{\rm d}/\epsilon_{\rm d}^{\rm max})$.

\subsection{Numerical Methods}
\begin{figure*}
\makebox[0.97\textwidth]{
\includegraphics[width=0.51\textwidth]{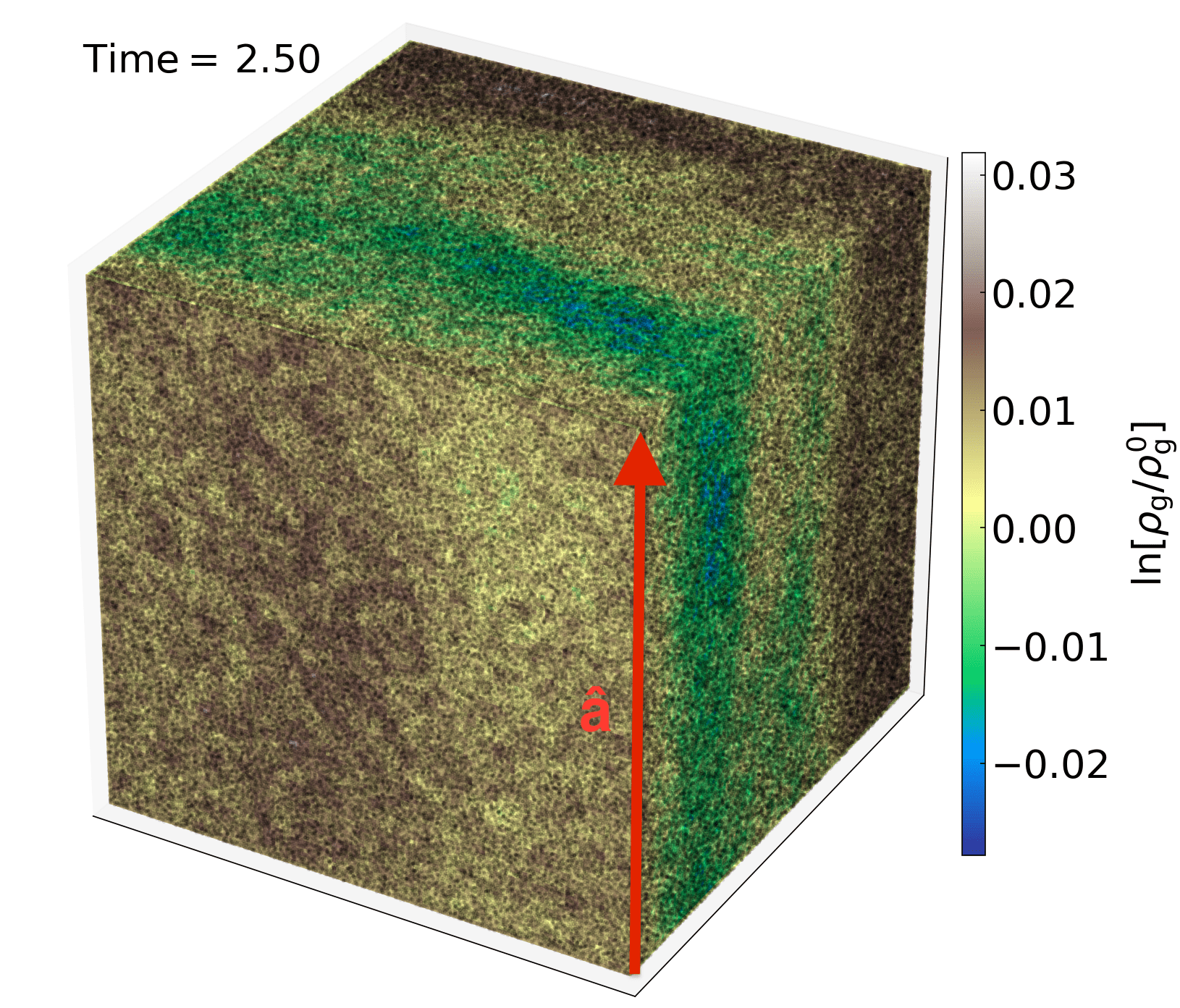}
\includegraphics[width=0.51\textwidth]{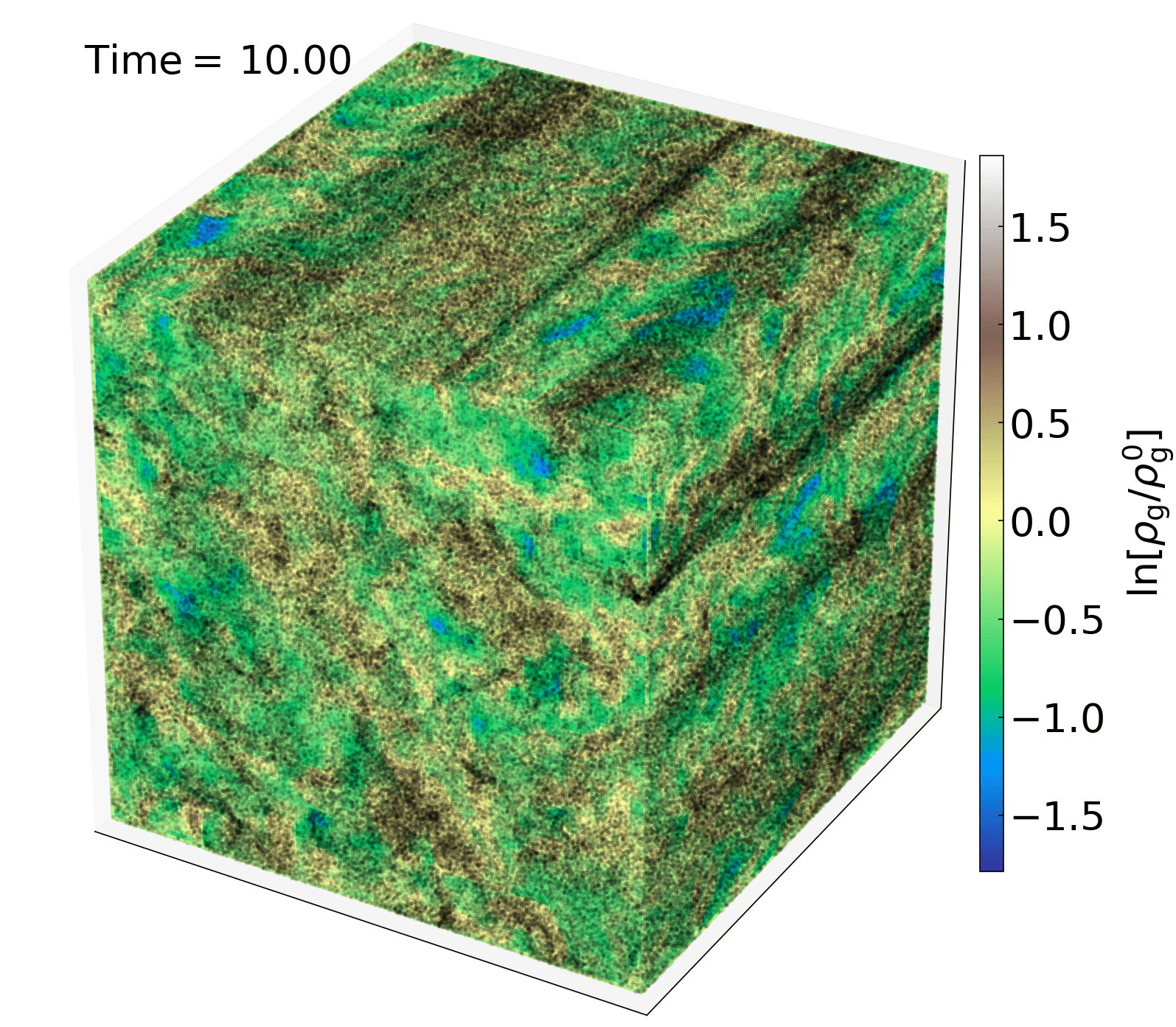}
}\\[0cm]
\caption{Evolution of the RDIs in simulation {\bf HD-Q} (in which the acceleration is independent of grain size; see \S~\ref{sec:params}) at an early time where modes are still roughly linear ({\em left}) and well into non-linear saturated state ({\rm right}). Time is in code units $L_{\rm box}/c_{s}$. Colors show gas density $\rho_{\rm g}$ relative to the mean, in our idealized initially-homogeneous periodic box, and black points show dust grains. Initially, small grains form sheet-like morphologies owing to their sub-sonic drift, but this collapses to more complex structures and saturates in elongated dust filaments aligned with the direction of the radiative acceleration of grains (\S~\ref{sec:morphology}). 
}
\label{fig:RDI-const}
\end{figure*}
\begin{figure*}
\makebox[0.97\textwidth]{
\includegraphics[width=0.51\textwidth]{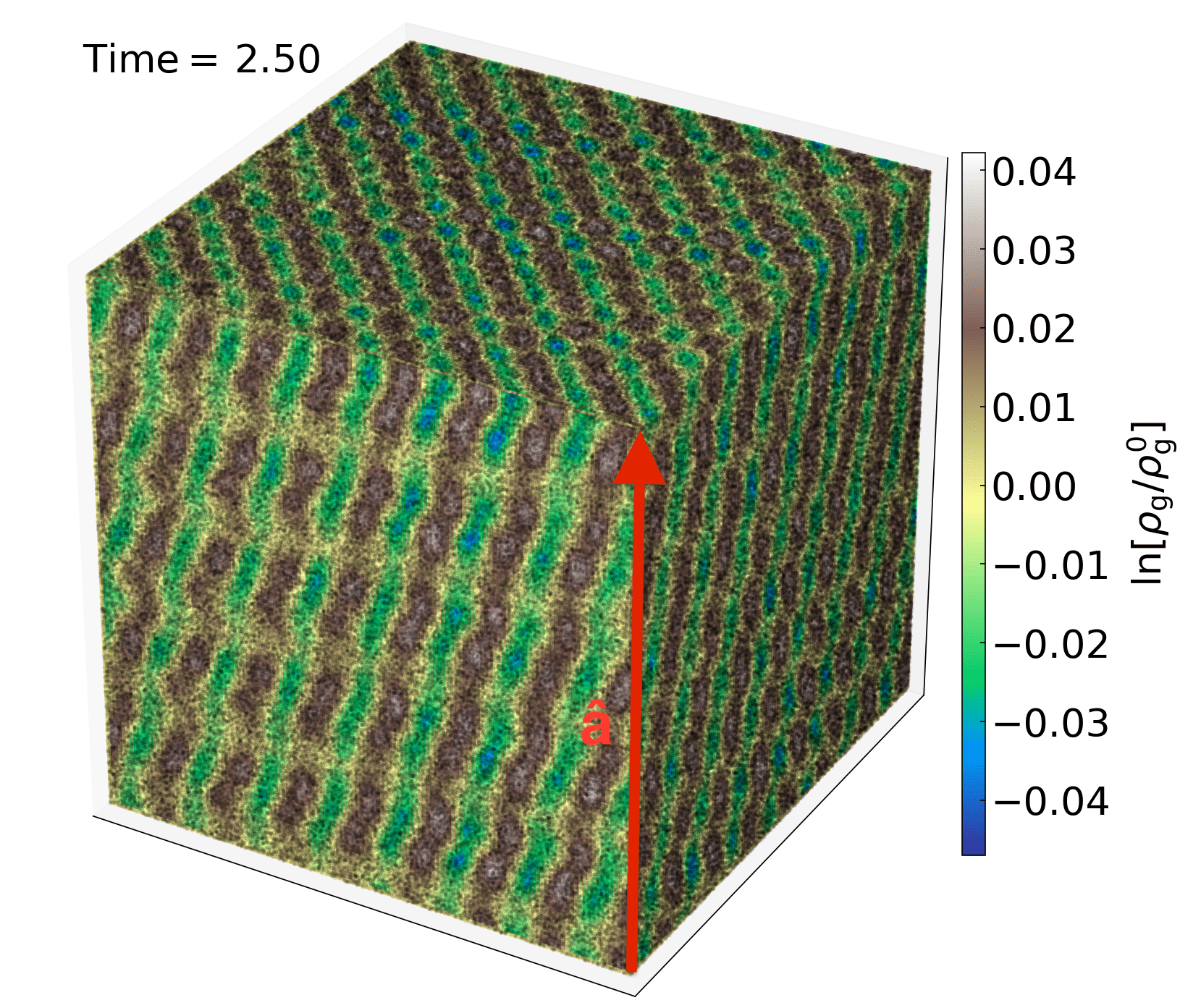}
\includegraphics[width=0.51\textwidth]{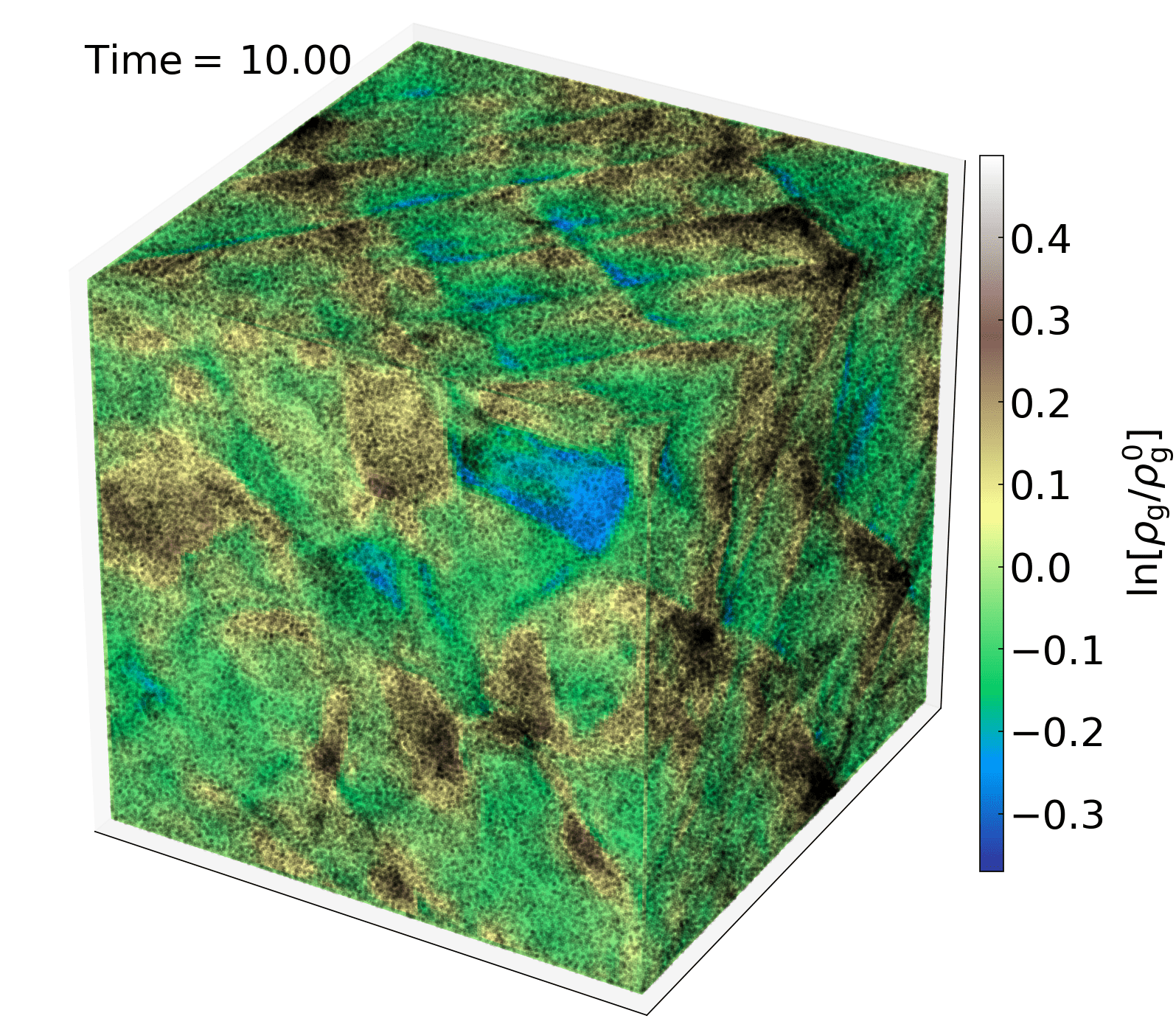}
}\\[0cm]
\caption{Evolution of simulation {\bf HD} (with variable grain size dependent acceleration, see \S~\ref{sec:params}), as Fig.~\ref{fig:RDI-const}. Here the grains have different accelerations but this produces the same initial super-sonic equilibrium drift speed, so the RDI resonant angles align into coherent structures at early times, but this breaks into less coherent ``patches'' instead of filaments in the saturated state. 
}
\label{fig:RDI-variable}
\end{figure*}
\begin{figure*}
        \makebox[0.97\textwidth]{
\includegraphics[width=0.51\textwidth]{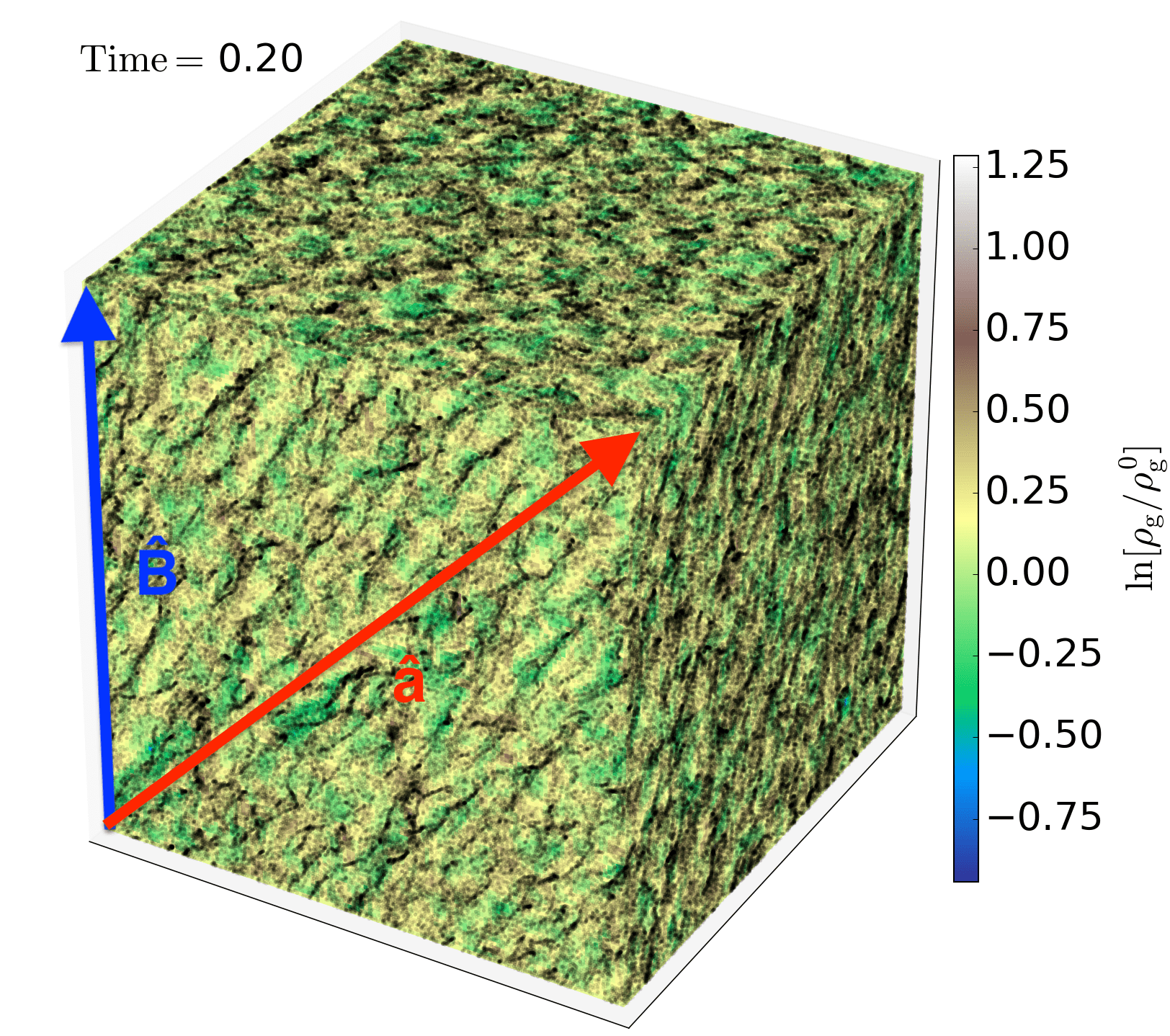}
\includegraphics[width=0.51\textwidth]{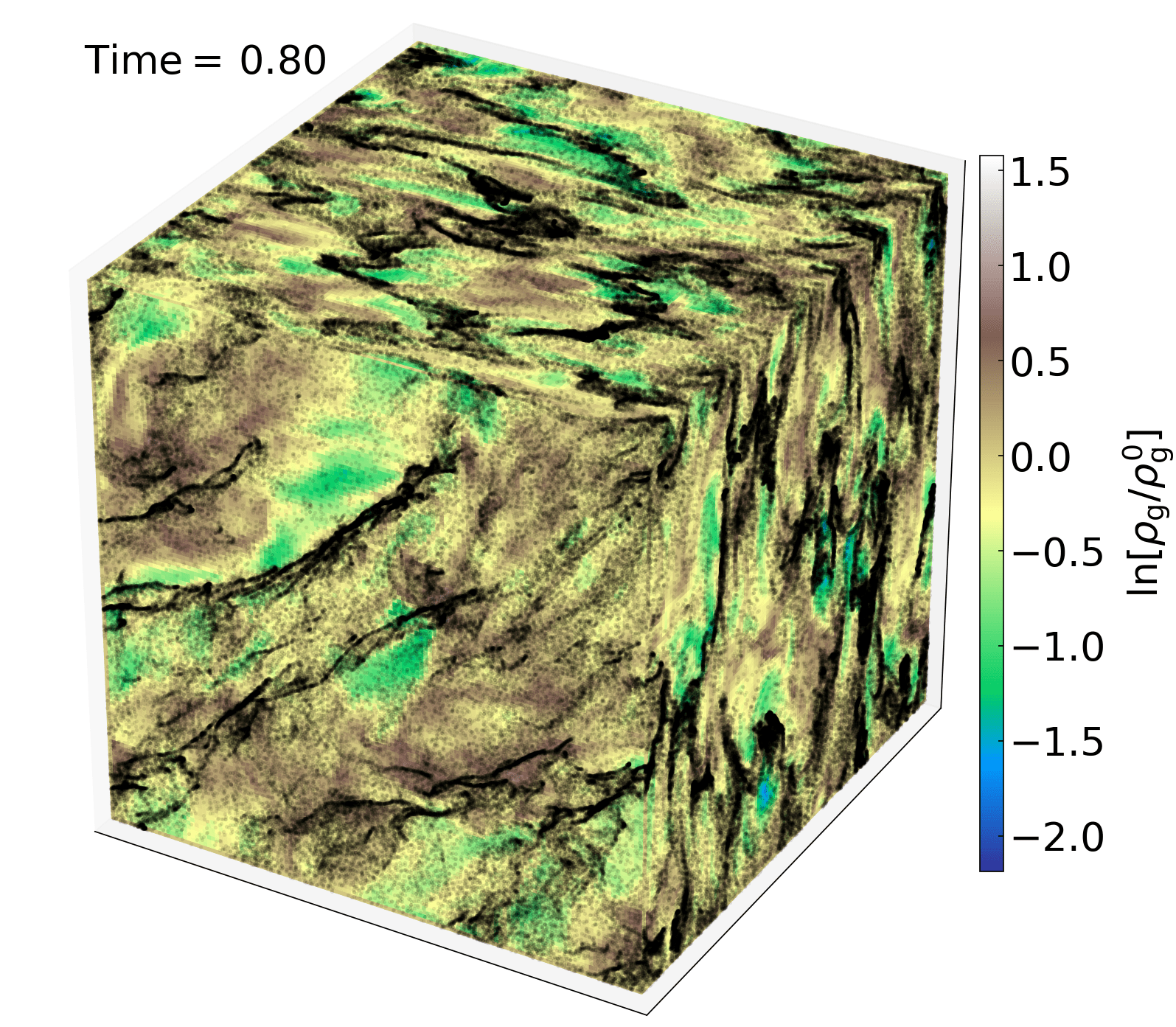}
}\\[0cm]
        \caption{Evolution of simulation {\bf MHD-Q} same as Fig.~\ref{fig:RDI-const}. This simulation enables magnetic fields (initial ${\bf B}$ points in the vertical direction) with charged grains, in ideal MHD, which produces new RDIs that clump the dust rapidly on small scales. The non-linear state is more corrugated-sheet-like in the direction of the acceleration. 
        }
        \label{fig:RDI-MHD-const}
\end{figure*}
\begin{figure*}
        \makebox[0.97\textwidth]{
\includegraphics[width=0.51\textwidth]{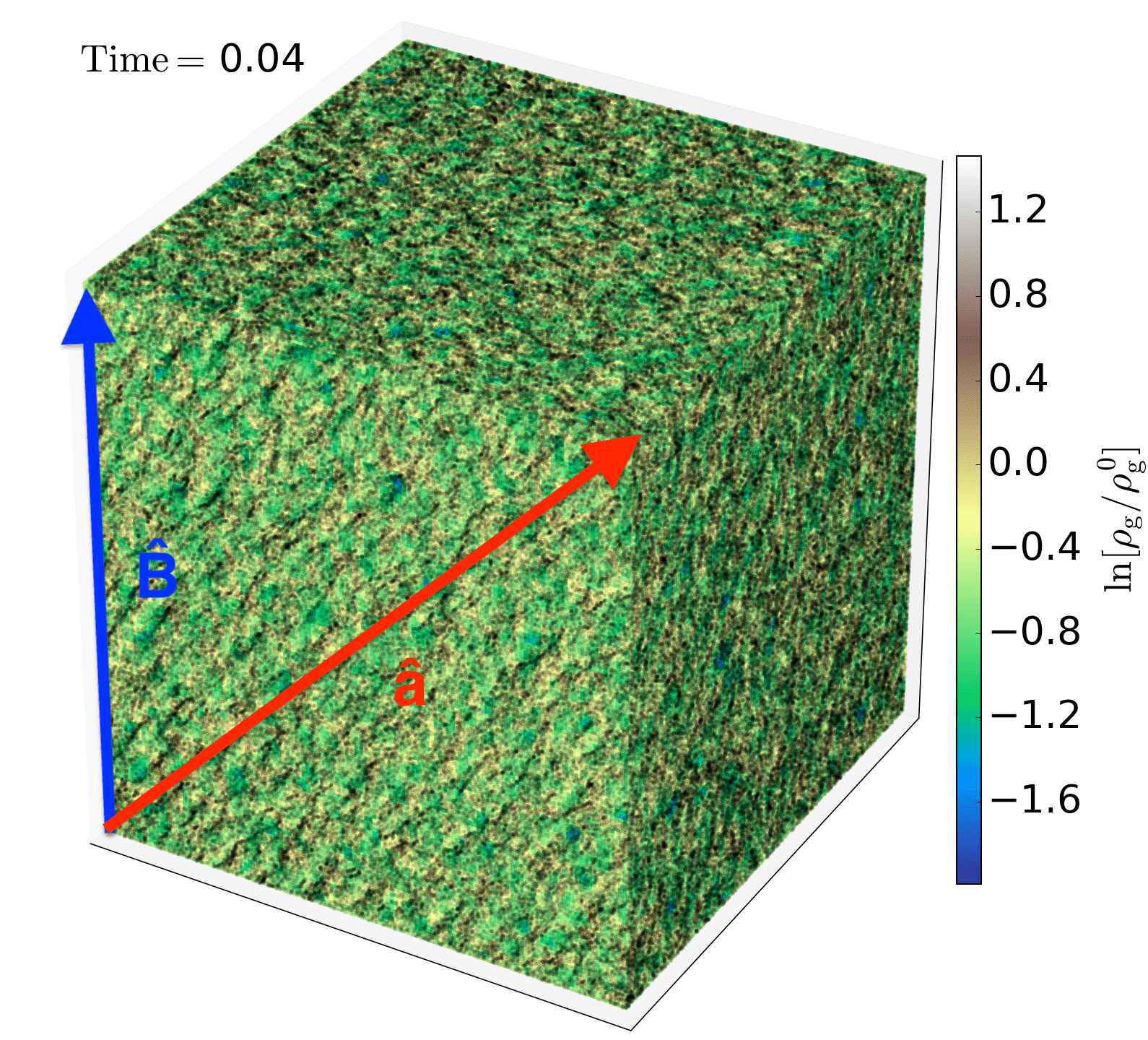}
\includegraphics[width=0.51\textwidth]{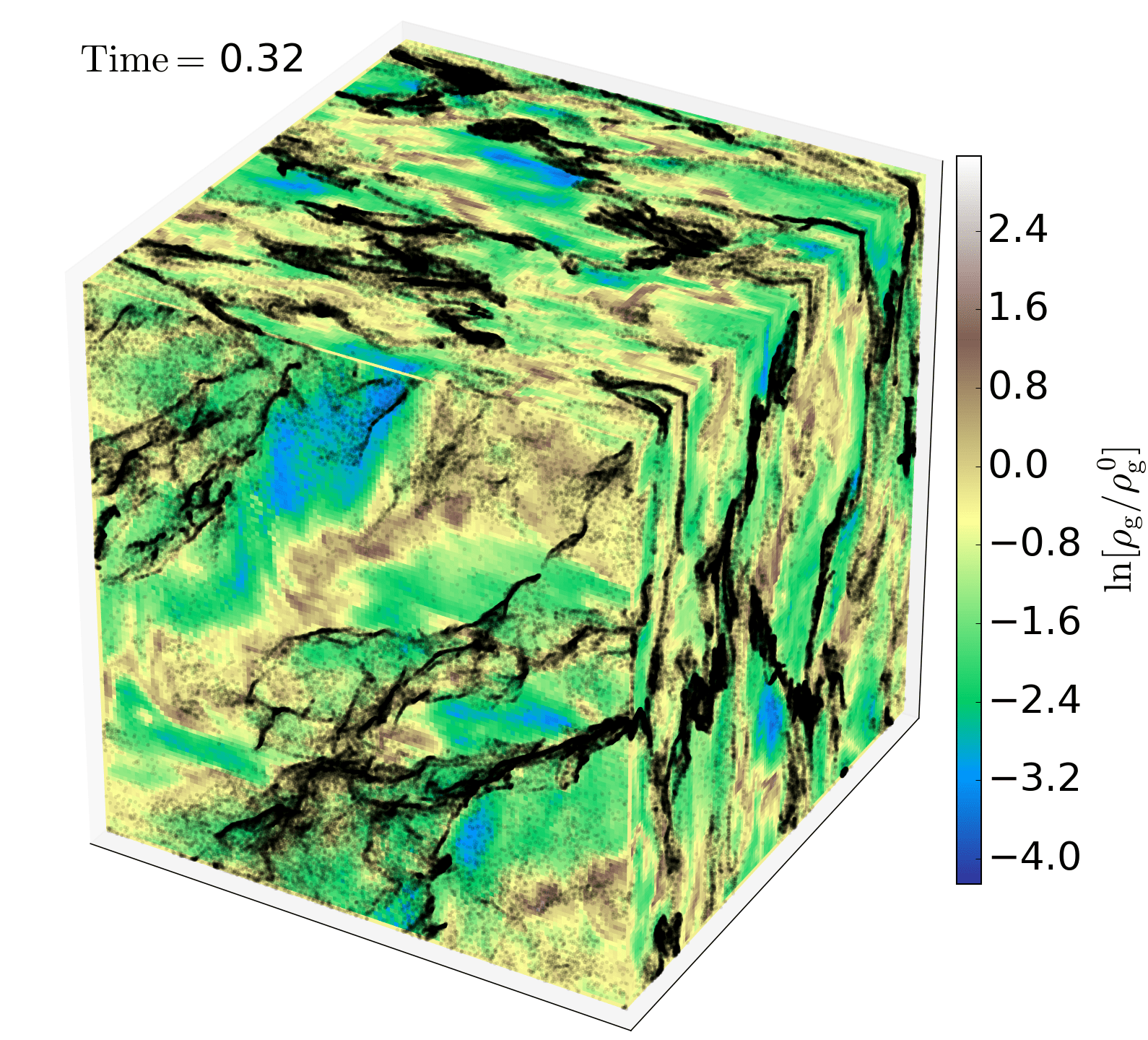}
}\\[0cm]
        \caption{Evolution of simulation {\bf MHD} same as Fig.~\ref{fig:RDI-MHD-const}. Similar to simulation {\bf MHD} this enables magnetic fields (again initial ${\bf B}$ points in the vertical direction) with charged grains, in ideal MHD. The structural differences between {\bf MHD} and {\bf MHD-Q} are less clear as in {\bf HD-Q} and {\bf HD} gauging the importance of the presence of magnetic fields on the build-up of the RDIs. Similar to {\bf MHD-Q} we observe more sheet like structures when the RDIs are in saturation. 
        }
        \label{fig:RDI-MHD}
\end{figure*}
\begin{figure*}
        \centering
        \includegraphics[scale=0.435]{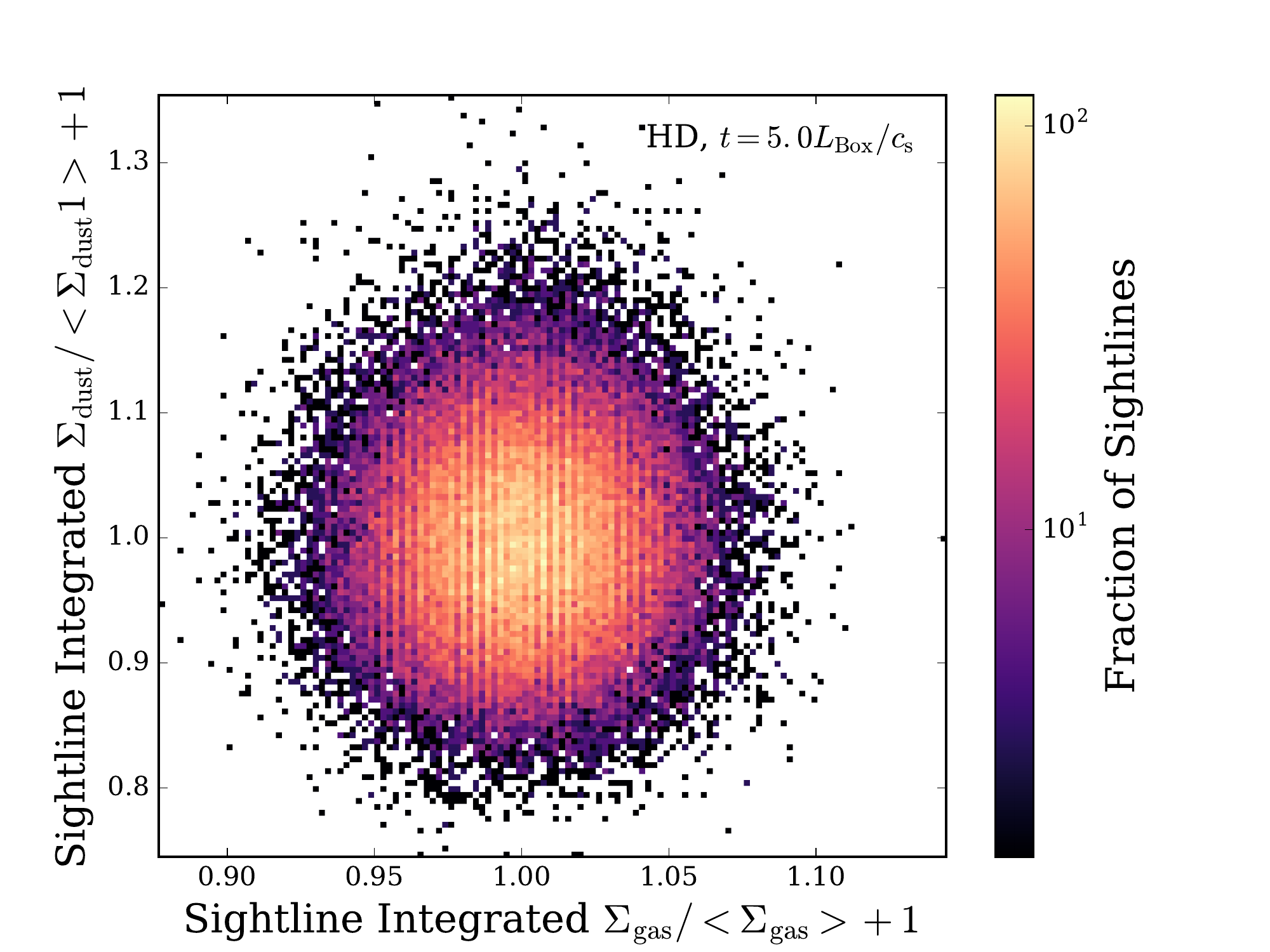}
        \includegraphics[scale=0.435]{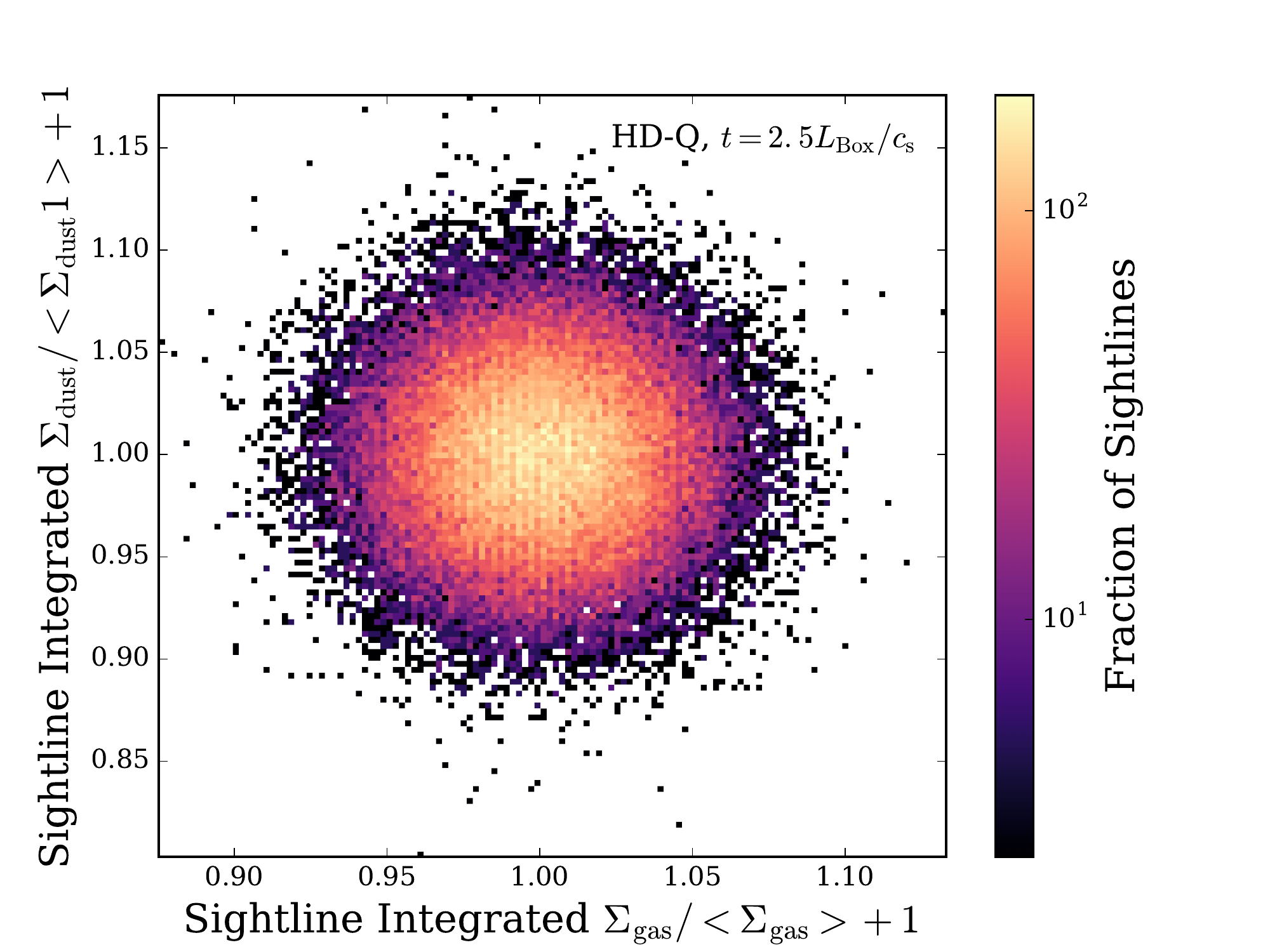}
        \includegraphics[scale=0.435]{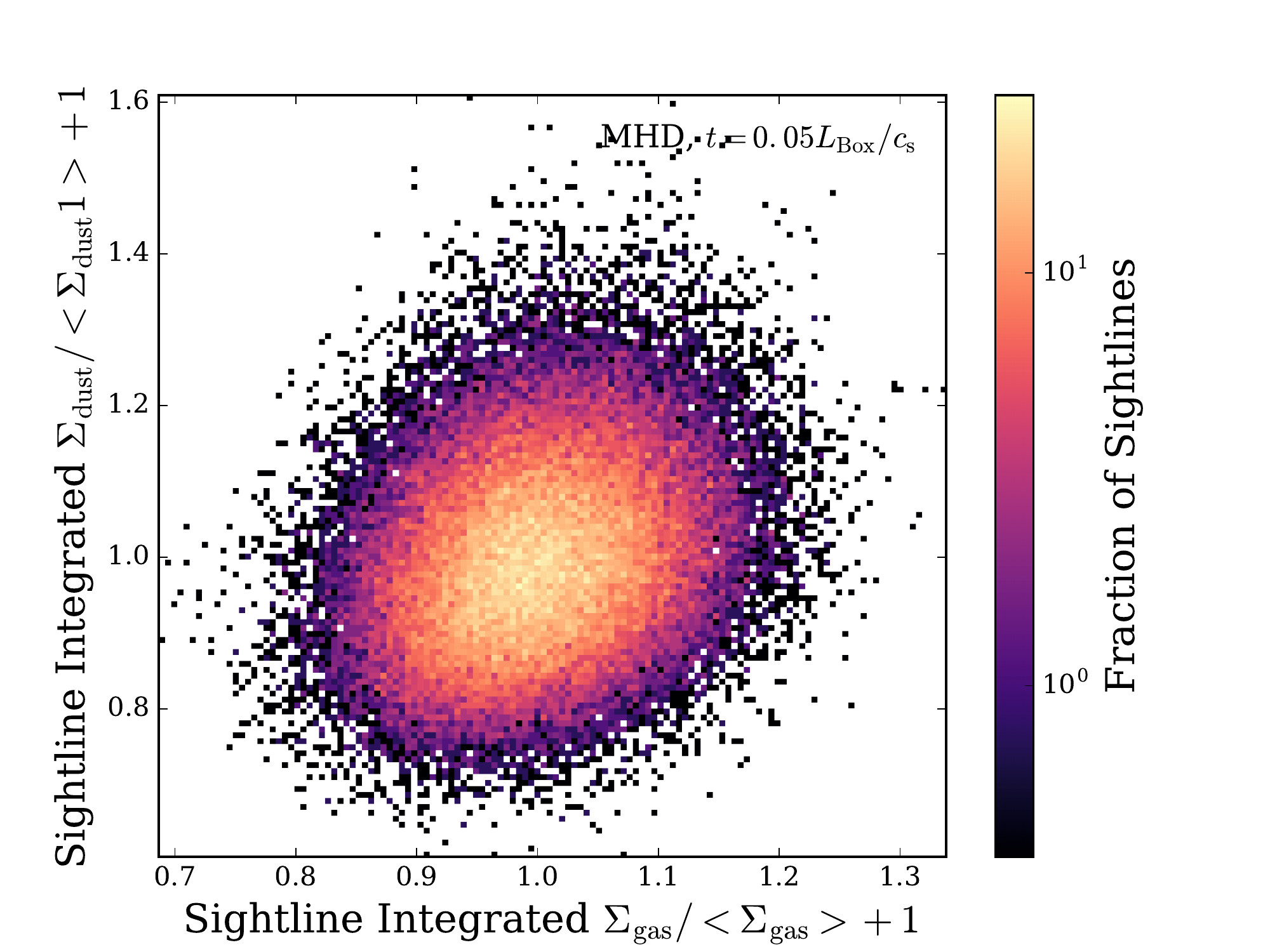}
        \includegraphics[scale=0.435]{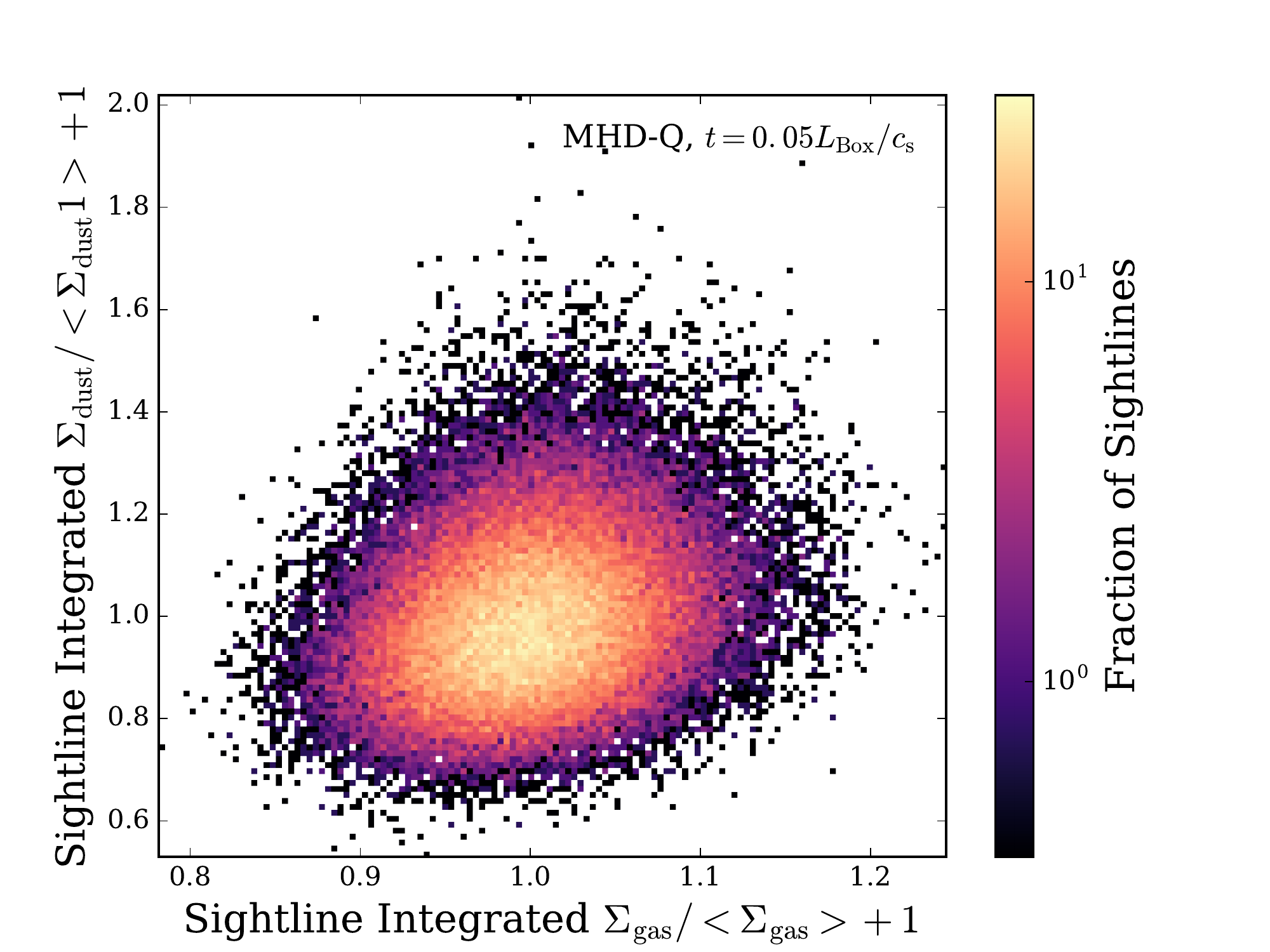}
        \caption{Bivariate distribution of dust column density $\Sigma_{\rm d}$ and gas column density $\Sigma_{\rm g}$ (\S~\ref{sec:gasden}), integrated along sightlines in the acceleration/outflow/radial direction $\hat{\bf a}$, in the \textit{initial growth phase} of the instabilities in all simulations {\bf HD, HD-Q, MHD} and {\bf MHD-Q}. Although there are large local variations in the dust-to-gas ratio, the sightline-integrated dust and gas surface densities are well-correlated and trace one another, initially without significant fluctuations. We note that at very early times there is only a very weak time-evolution present but there is a difference between the {\bf HD-Q, HD} and {\bf MHD-Q, MHD} models where one can already see weak scattering of the dust component the slightly higher column density, already indicating the quicker growth of the RDIs at an early evolutionary stage of the instabilities.
        }
        \label{fig:bias_color_early}
\end{figure*}
\begin{figure*}
        \centering
        \includegraphics[scale=0.435]{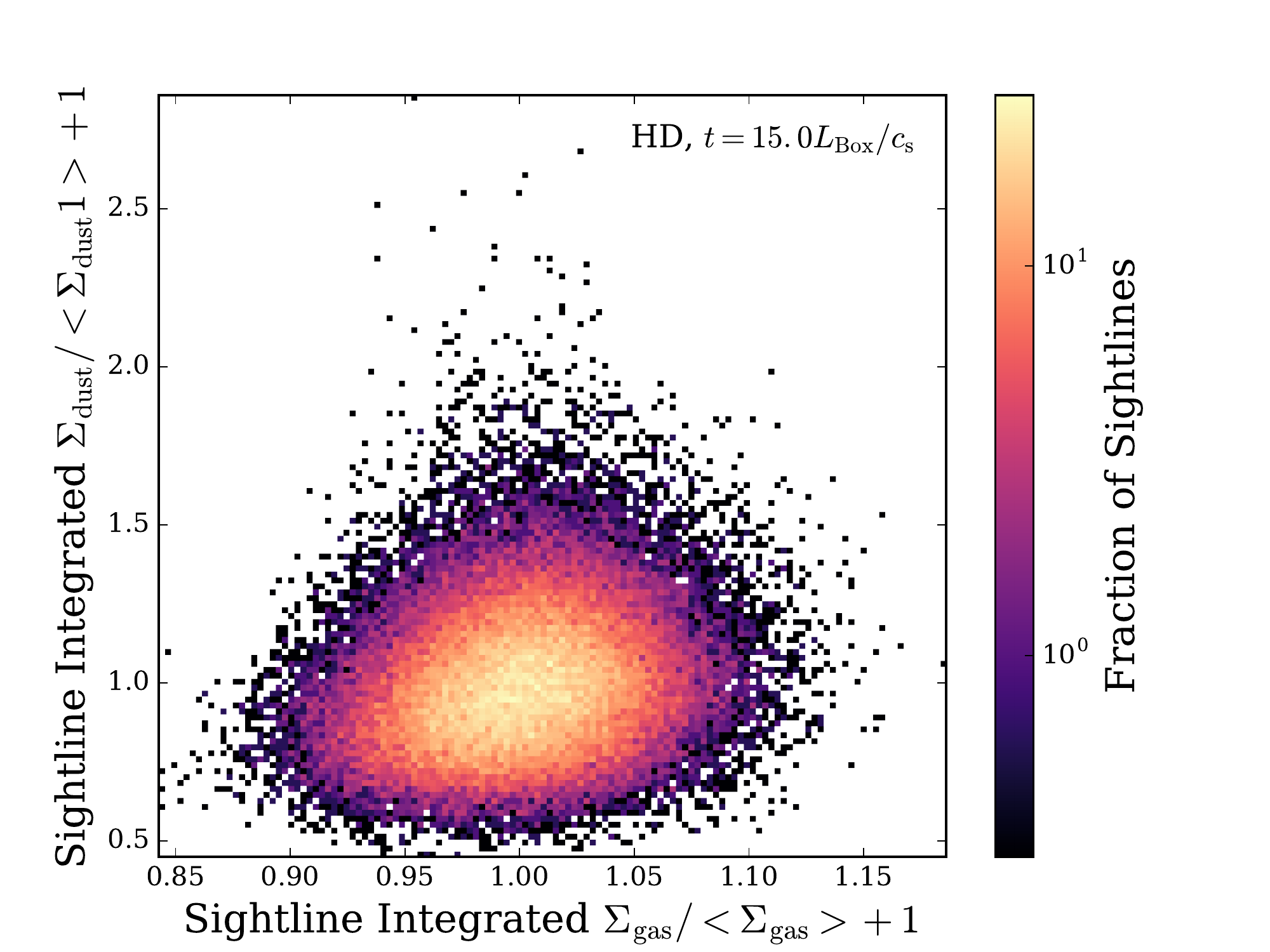}
        \includegraphics[scale=0.435]{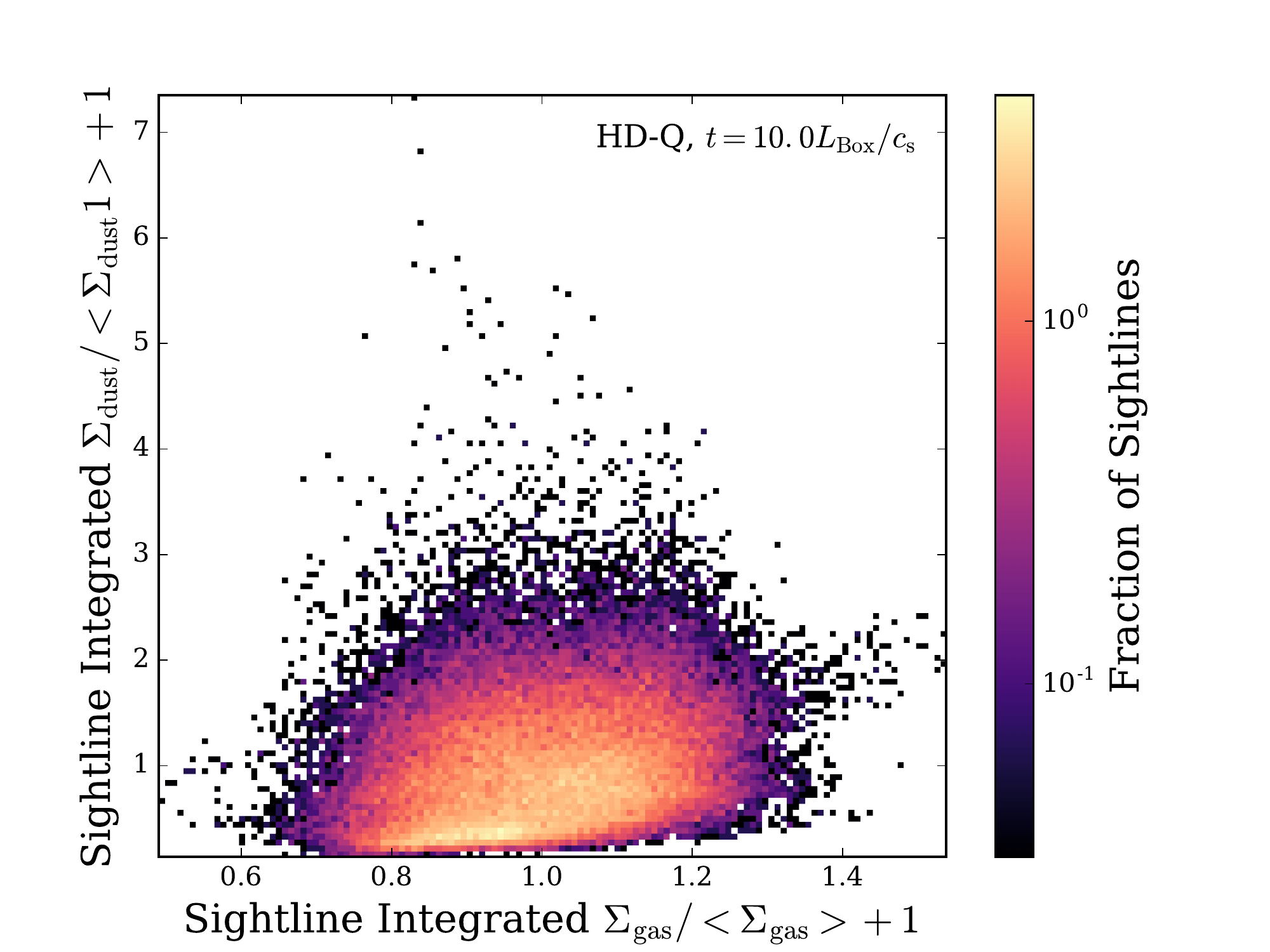}
        \includegraphics[scale=0.435]{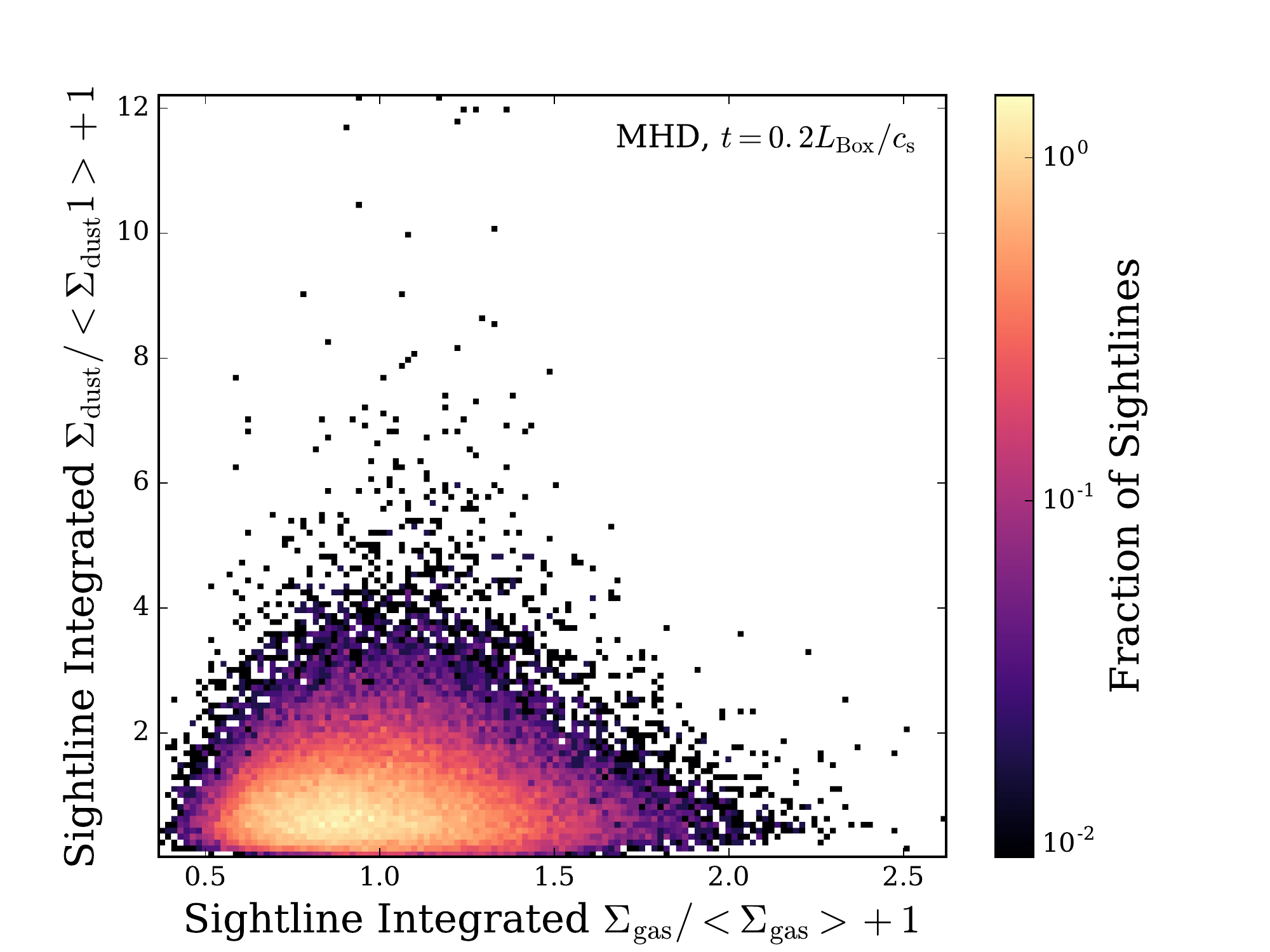}
        \includegraphics[scale=0.435]{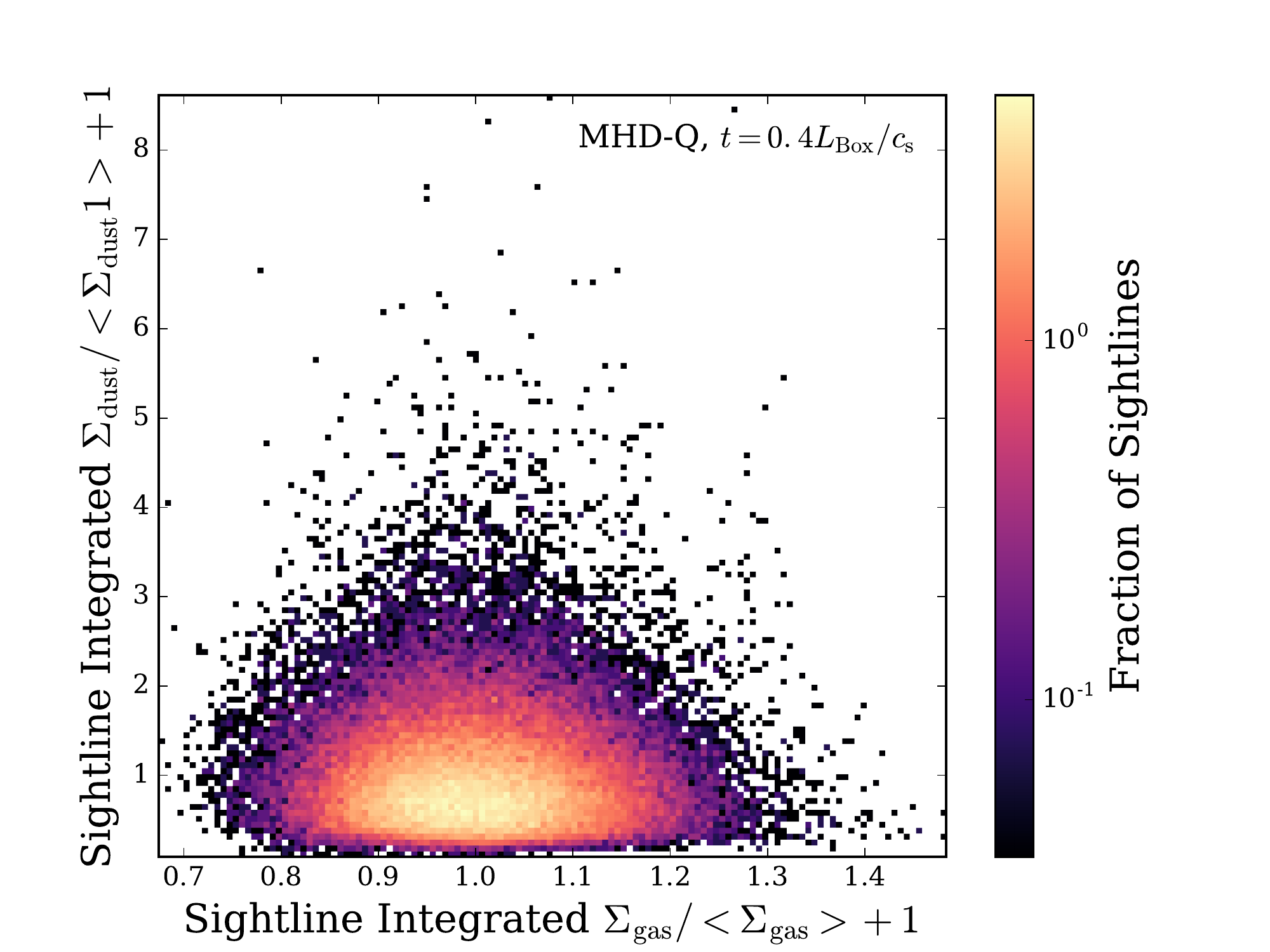}
        \caption{Bivariate distribution of dust column density $\Sigma_{\rm d}$ and gas column density $\Sigma_{\rm g}$ (\S~\ref{sec:gasden}), integrated along sightlines in the acceleration/outflow/radial direction $\hat{\bf a}$, \textit{in the saturated state} for all simulations {\bf HD, HD-Q, MHD} and {\bf MHD-Q}. Although there are large local variations in the dust-to-gas ratio, the sightline-integrated dust and gas surface densities are well-correlated and trace one another, with significant fluctuations. In the saturated state of the RDIs there is only a weak time evolution present.
        }
        \label{fig:bias_color_late}
\end{figure*}
\begin{figure}
        \includegraphics[scale=0.43]{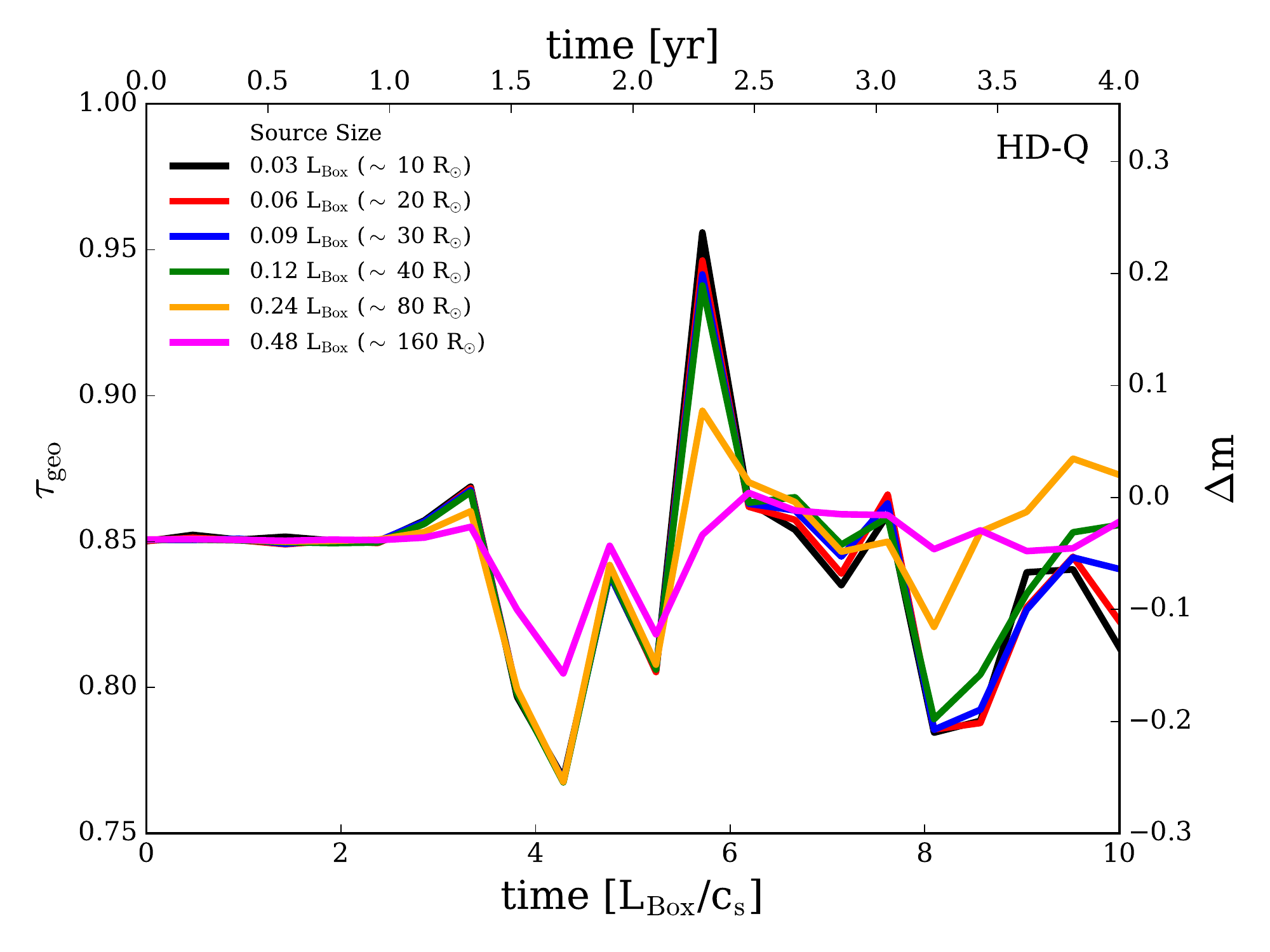}
        \includegraphics[scale=0.43]{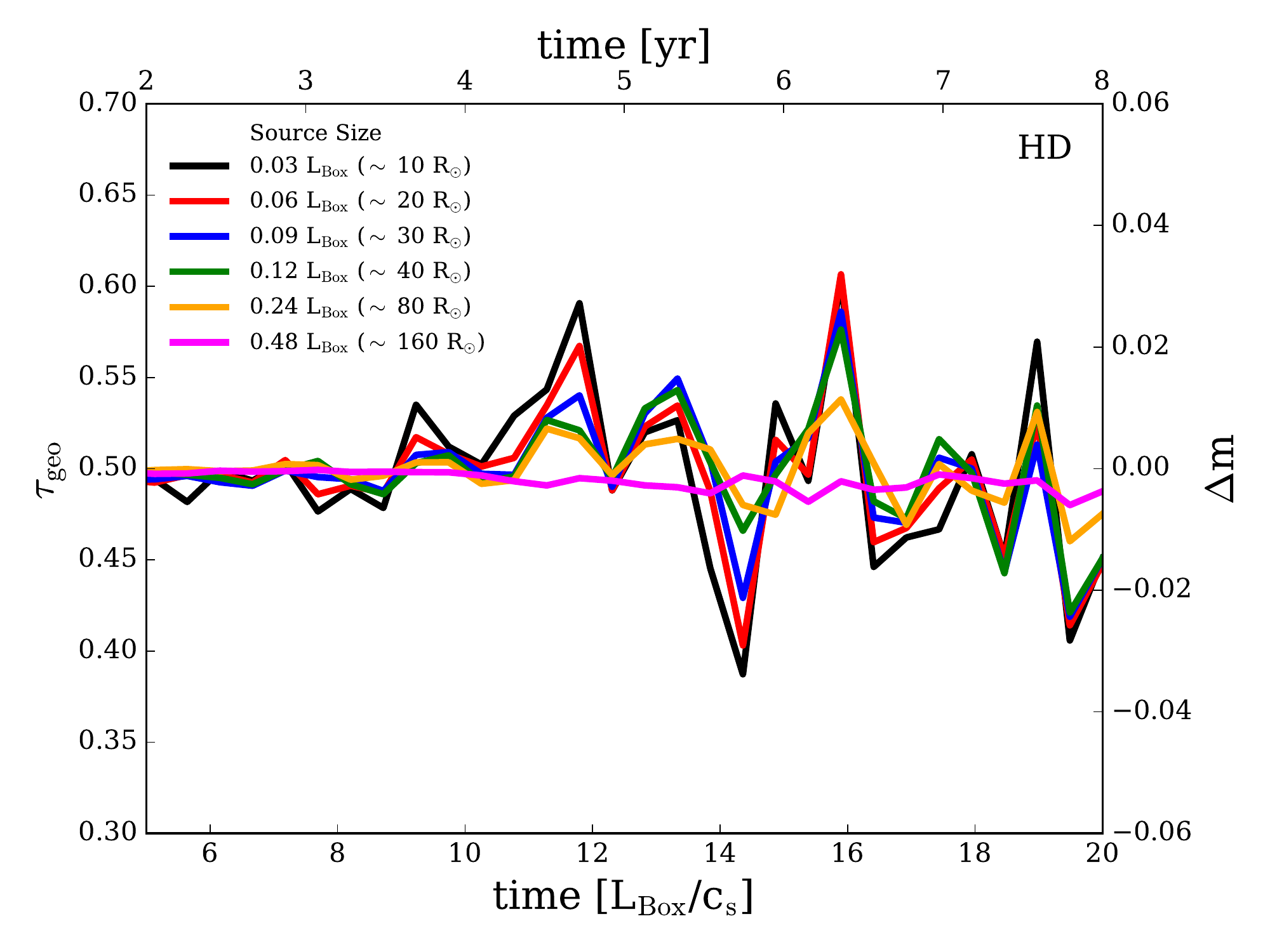}
    	\caption{Effective geometric optical depth $\tilde{\tau}_{\rm geo}$ (see \S~\ref{sec:extinction}) for a finite-size source disc of radius $R_{\ast}$, integrated along a fixed sightline through the box center along the outflow direction $\hat{\bf a}$, as a function of time (ignoring early times before saturation), for simulations {\bf HD-Q} and {\bf HD}. The alternative $y$-axis shows the equivalent change in magnitudes $\delta m = -(2.5/\ln{10})\,\delta \tilde{\tau}_{\rm geo}$; the alternative $x$-axis and legend labels show an approximate equivalent physical unit scale by taking $L_{\rm box} \sim 300\,R_{\odot}$ and $c_{s} \sim 4\,{\rm km\,s^{-1}}$. The mean geometric optical depth is $\sim 0.5$ for the model {\bf HD} and $ \sim 0.85$ for the model {\bf HD-Q}, with variations at the $\sim0.1$\,mag level, similar to observed AGB stars.
    	}
	\label{fig:tau_hd}
\end{figure}
\begin{figure}
    	\includegraphics[scale=0.43]{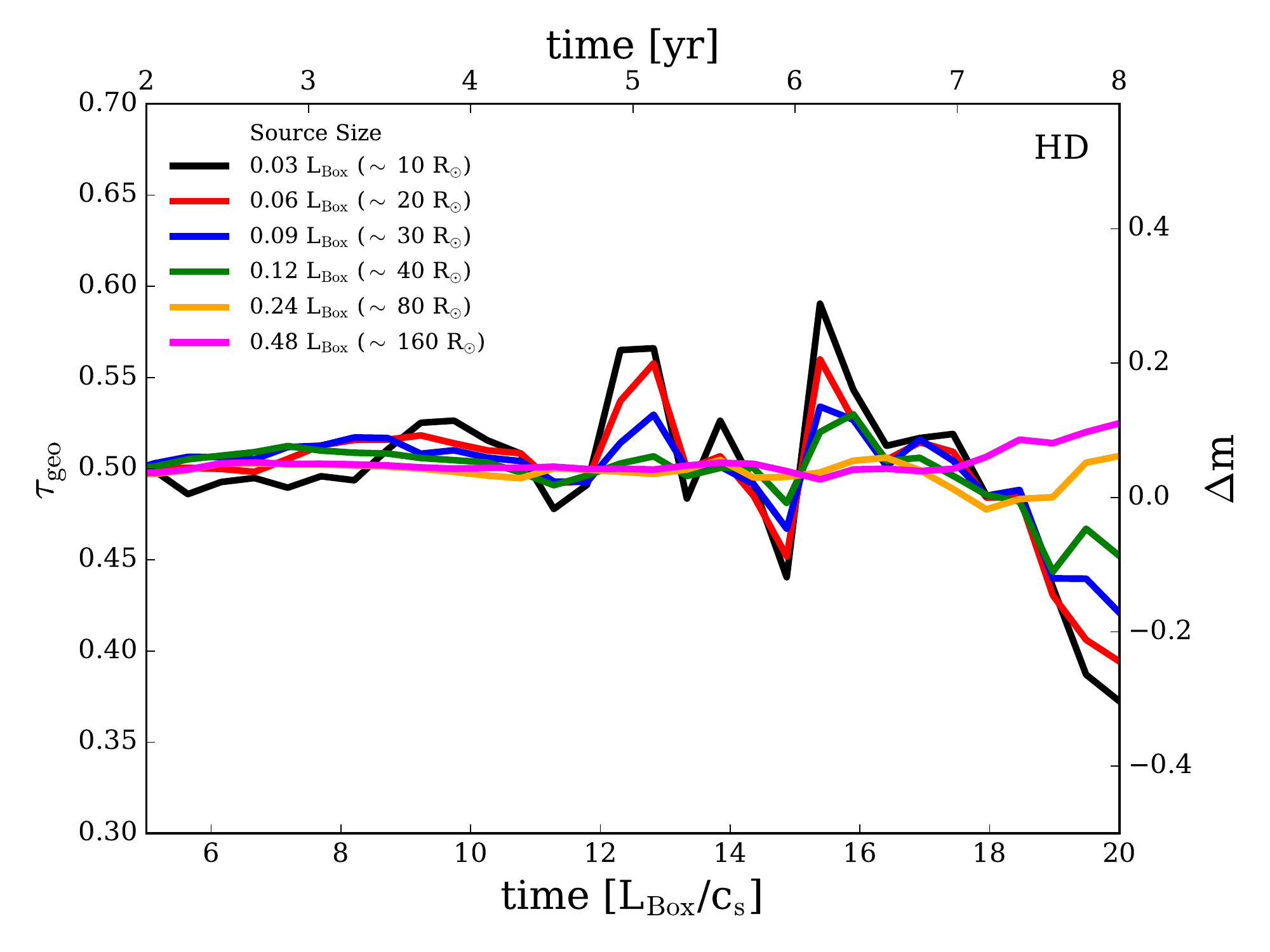}
    	\caption{Effective geometric optical depth $\tilde{\tau}_{\rm geo}$ (see \S~\ref{sec:extinction}) for a finite-size source disc of radius $R_{\ast}$, integrated along a fixed sightline through the box center perpendicular to the outflow direction $\hat{\bf a}$, as a function of time (ignoring early times before saturation). We show one example for this for the simulation {\bf HD}. The alternative $y$-axis shows the equivalent change in magnitudes $\delta m = -(2.5/\ln{10})\,\delta \tilde{\tau}_{\rm geo}$; the alternative $x$-axis and legend labels show an approximate equivalent physical unit scale by taking $L_{\rm box} \sim 300\,R_{\odot}$ and $c_{s} \sim 4\,{\rm km\,s^{-1}}$. The mean geometric optical depth is $\sim 0.5$, with variations at the $\sim0.1$\,mag level, similar to observed AGB stars. We note that the third spatial direction is redundant for the simulations {\bf HD} and {\bf HD-Q} since it is statistically identical with the first perpendicular sightline. Hence, we include this as a proof to gauge that we get similar optical depth fluctuations for the perpendicular sightlines. 
    	}
	\label{fig:tau_perp}
\end{figure}
\begin{figure}
        \includegraphics[scale=0.43]{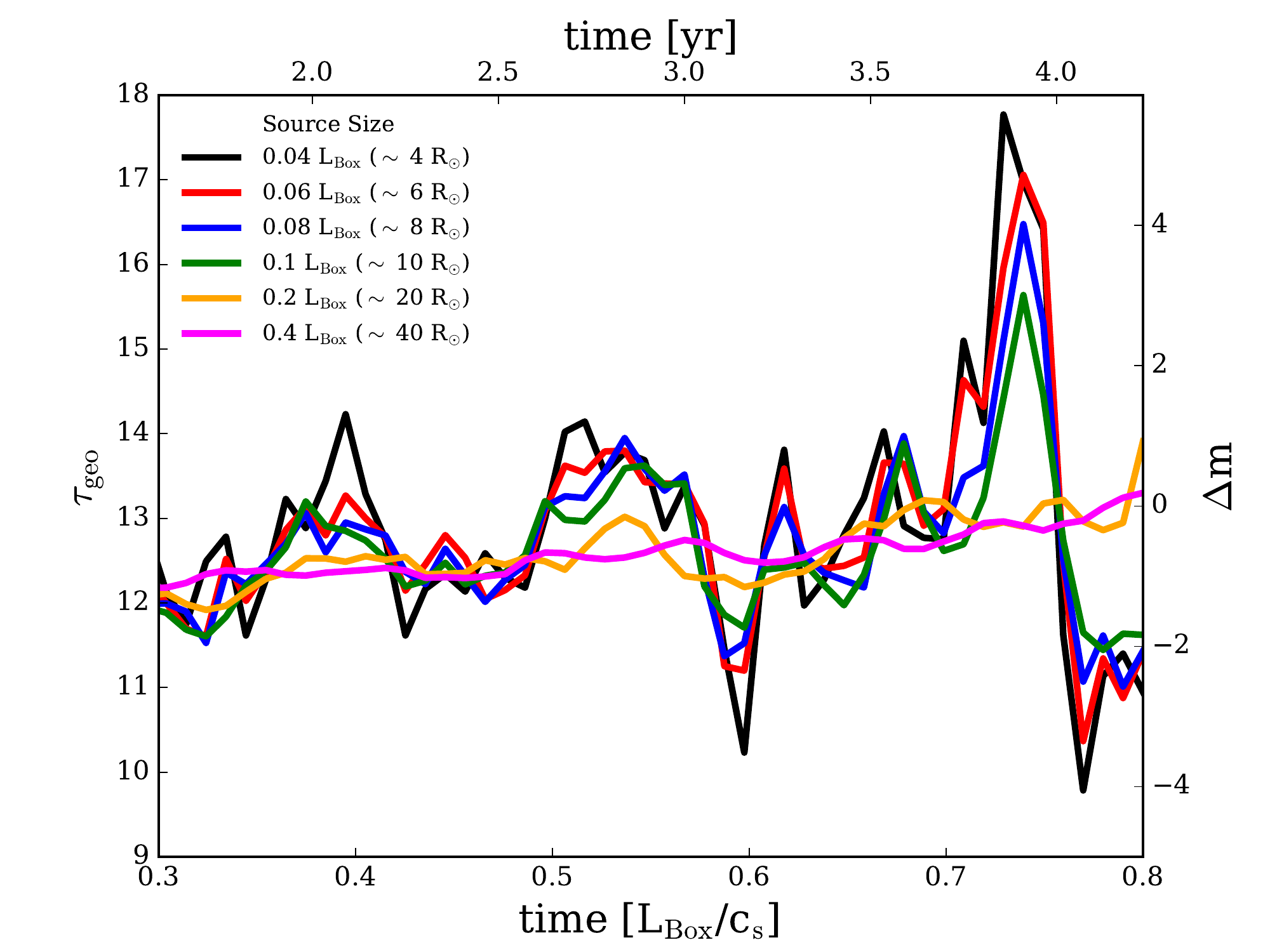}
    	\includegraphics[scale=0.43]{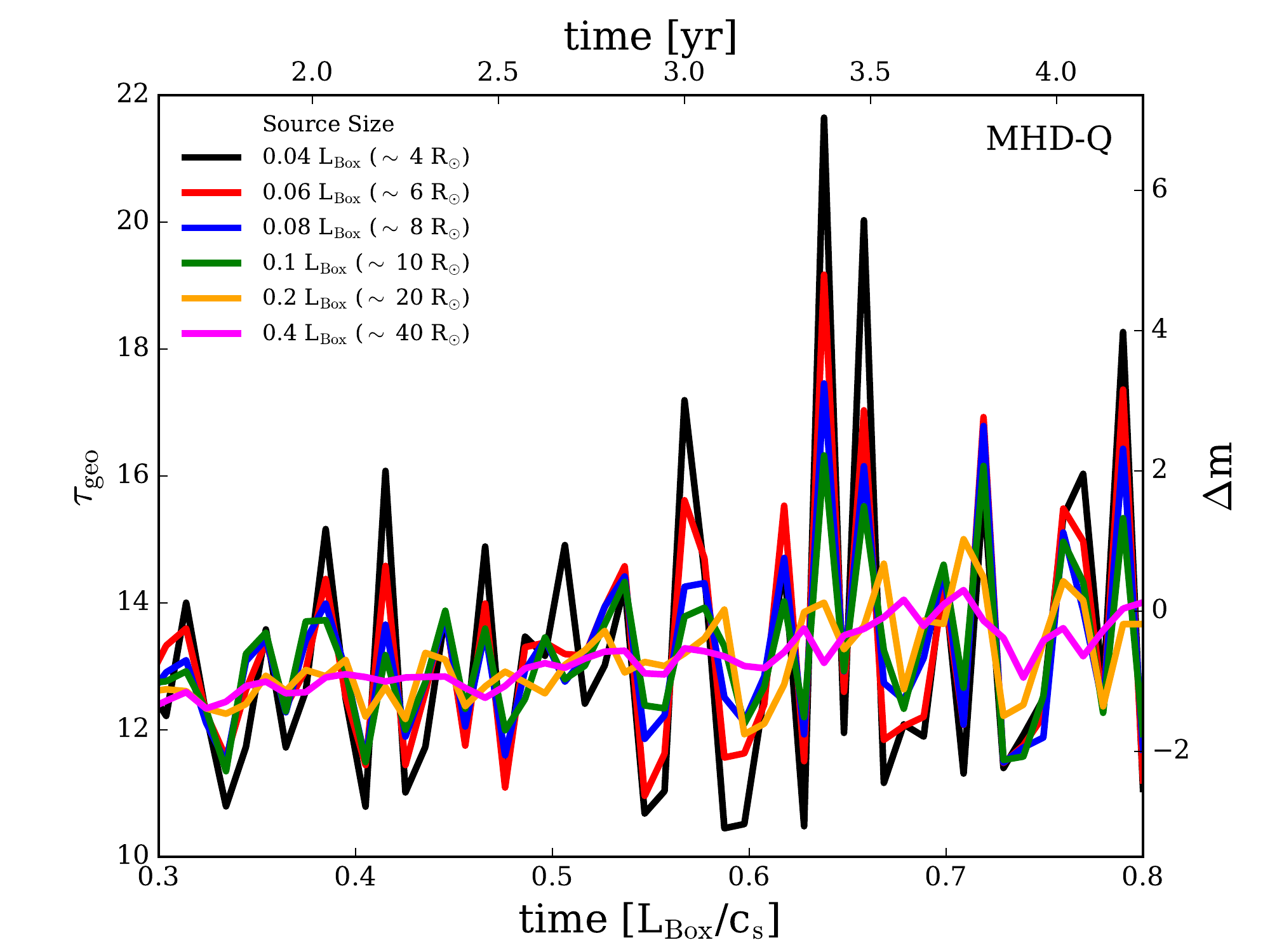}
    	\includegraphics[scale=0.43]{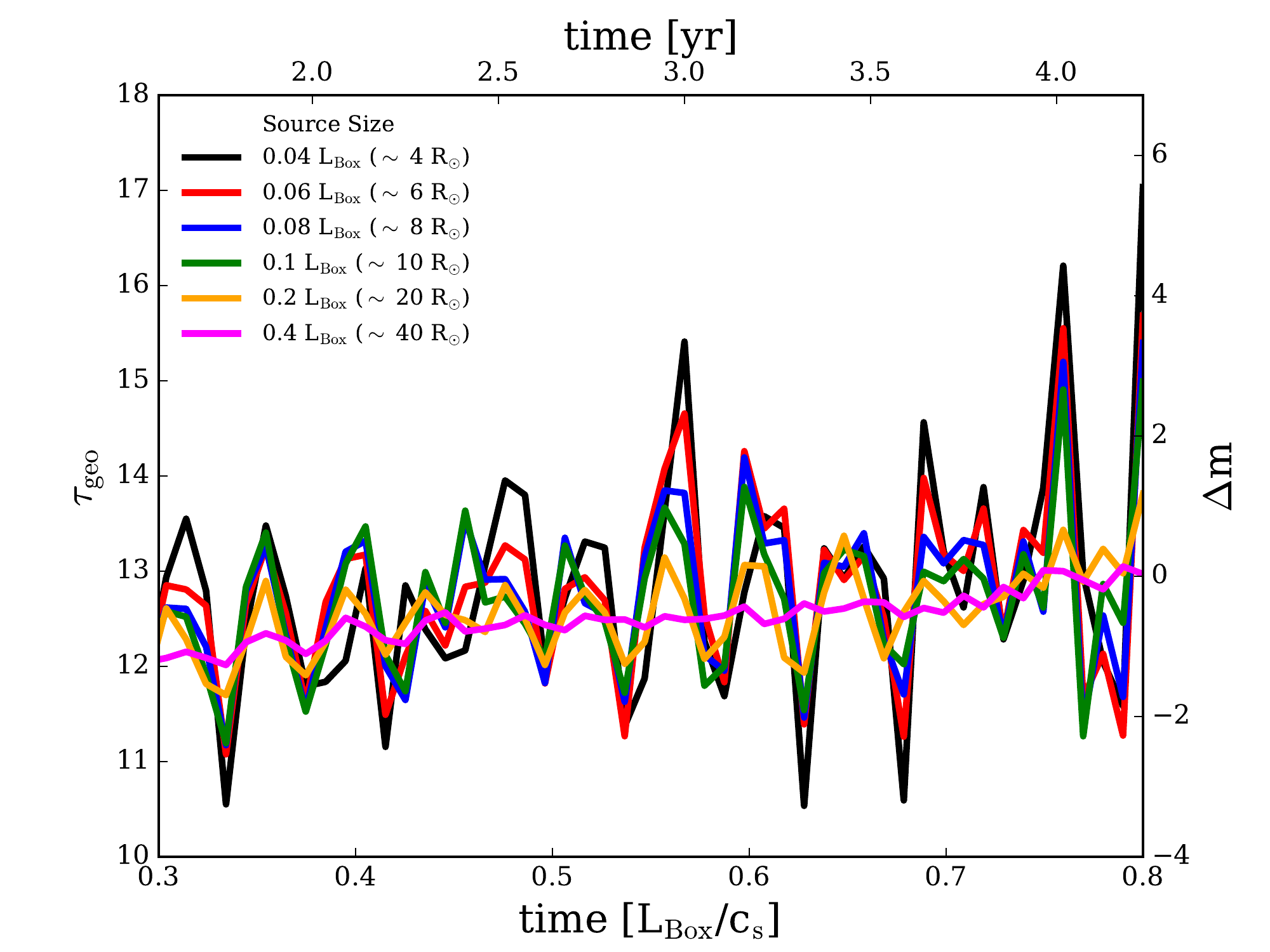}
    	\caption{Time variation in $\tilde{\tau}_{\rm geo}$ as Fig.~\ref{fig:tau_hd}, for our {{\bf MHD-Q}} simulation. We compare all three sightlines in the $\hat{\bf a}$ direction ({\em top}) and both perpendicular directions ({\em center, bottom}), since in the MHD-case we have much more available modes (e.g. magnetosonic or \Alf) We again show some approximate physical-equivalent units assuming $L_{\rm box} \sim 100\,R_{\odot}$ and $c_{s} \sim 5\,{\rm km\,s^{-1}}$. The mean volume-averaged $\tau_{\rm geo} \sim 10$ is higher in this run, more comparable to RCB stars in a dusty outburst, and the RDIs produce $\sim 1\,$mag fluctuations on $\sim$\,months timescales.
    	}
    	\label{fig:tau_mhd}
\end{figure}
\begin{figure}
        \includegraphics[scale=0.43]{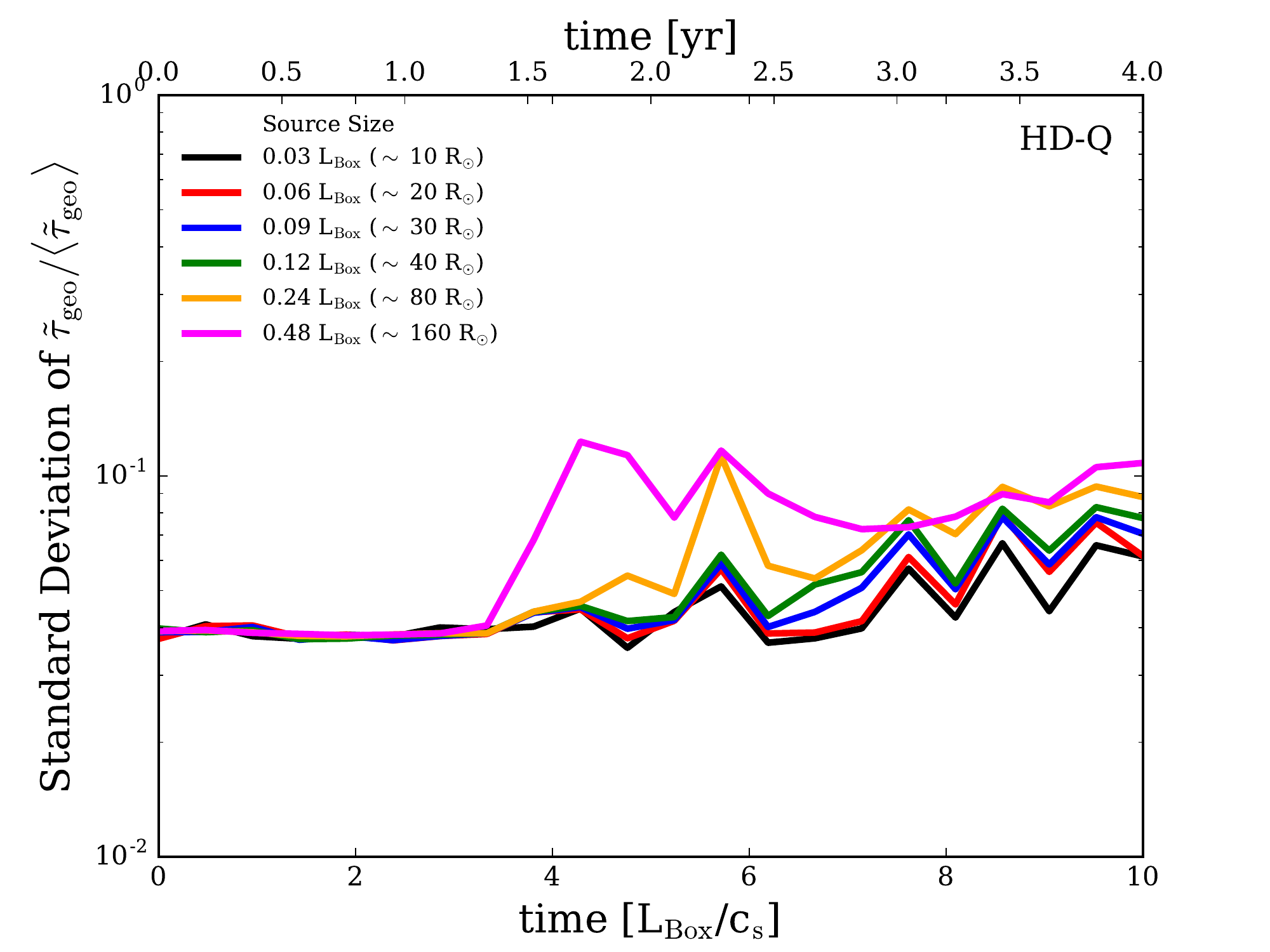}
    	\includegraphics[scale=0.43]{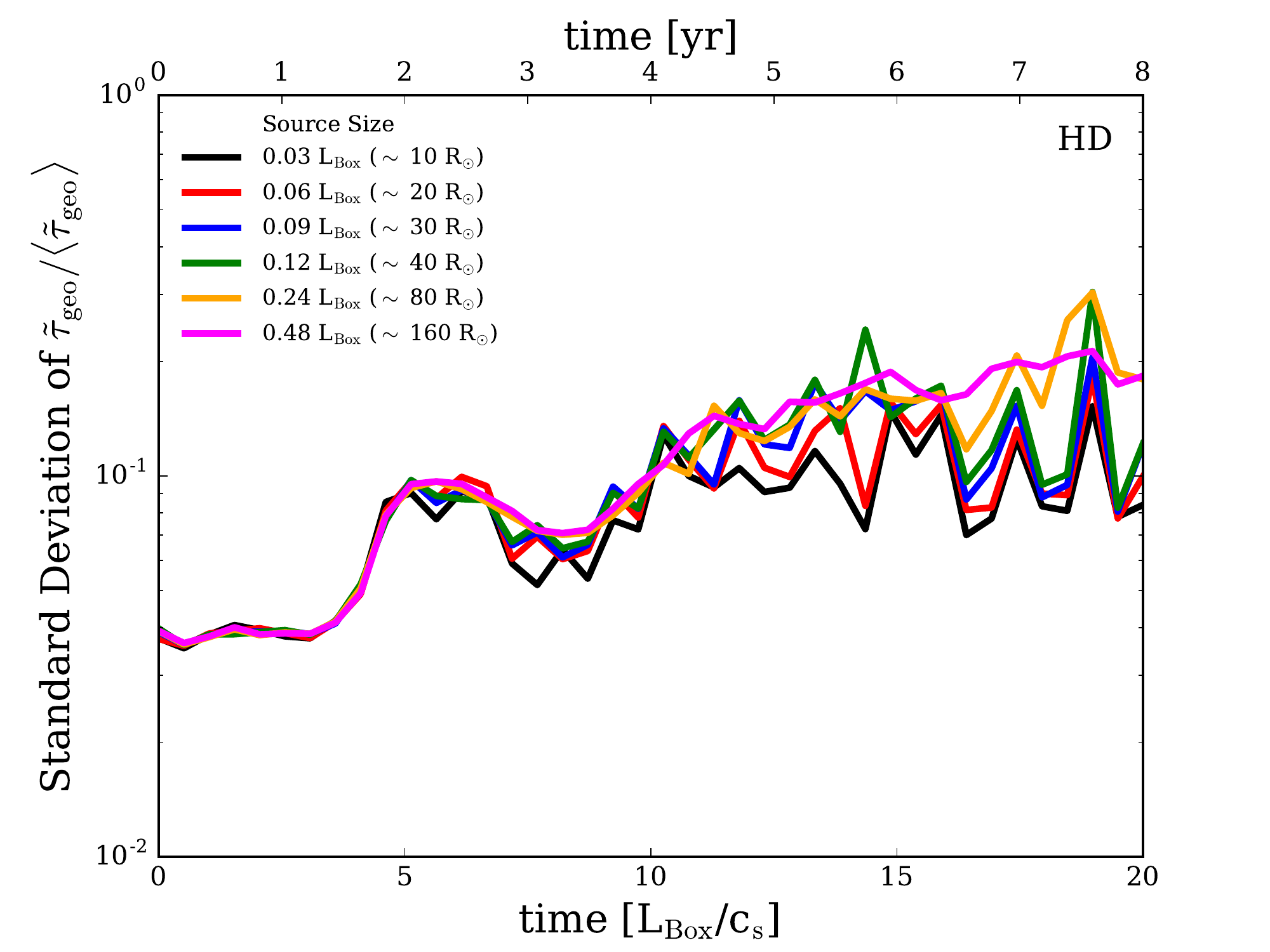}
        \includegraphics[scale=0.43]{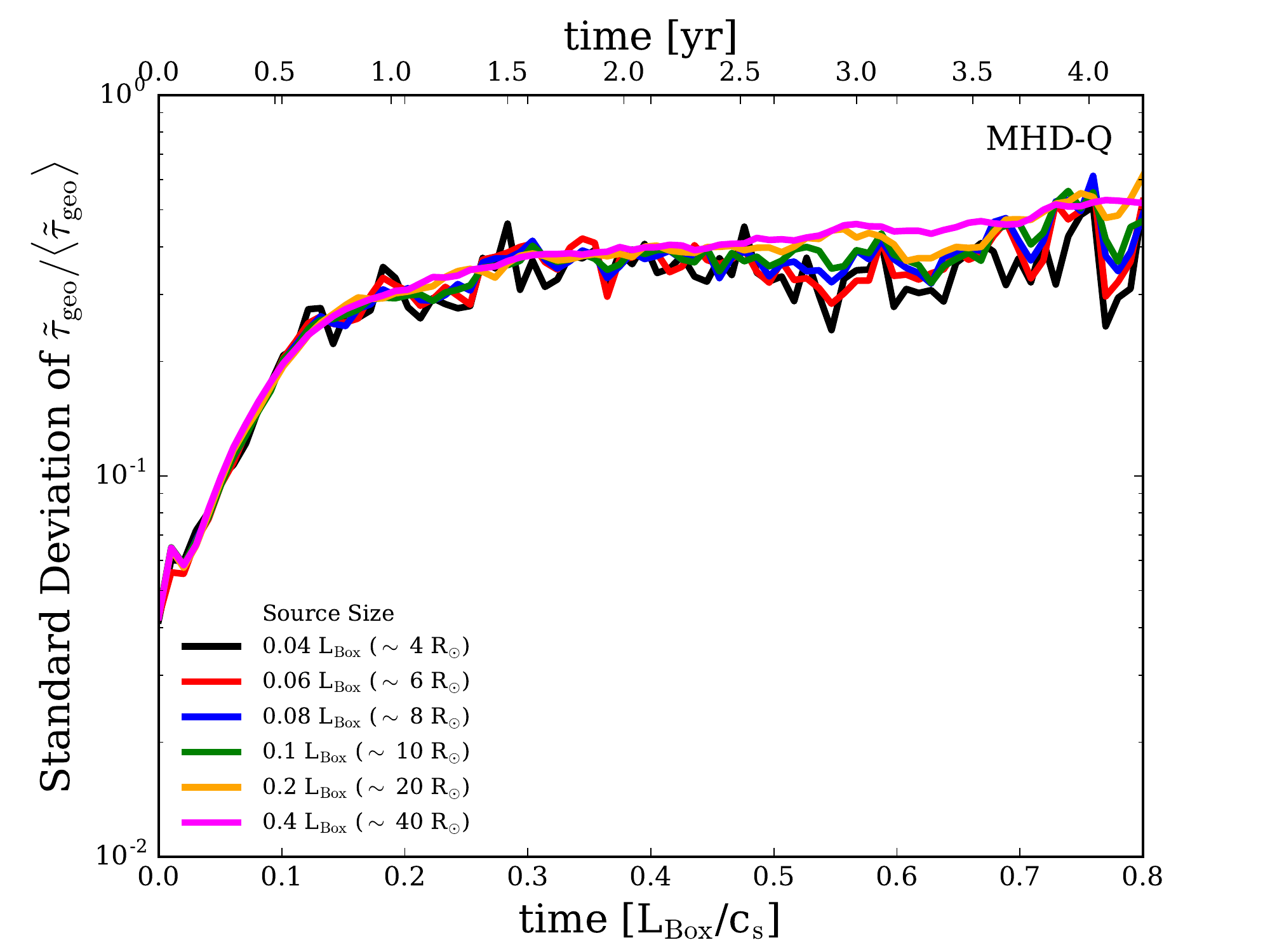}
    	\caption{Standard deviation of $\tilde{\tau}_{\rm geo} / \langle \tilde{\tau}_{\rm geo} \rangle$, i.e.\ $1\sigma$ fractional fluctuations in $\tilde{\tau}_{\rm geo}$, averaged over independent sightlines at each time along $\hat{\bf a}$, versus time. All simulations reach saturation on a couple box sound-crossing times, with saturated $\tilde{\tau}_{\rm geo} / \langle \tilde{\tau}_{\rm geo} \rangle \sim 0.1$. The fluctuations are dominated by large structures, so do not depend strongly on source size for sources with $R_{\ast} \lesssim 0.2\,L_{\rm box}$ (i.e.\ envelopes a factor of a couple or more larger than the star).
    	}
    	\label{fig:std_hd}
\end{figure}
\begin{figure*}
    \centering
    \includegraphics[scale=0.435]{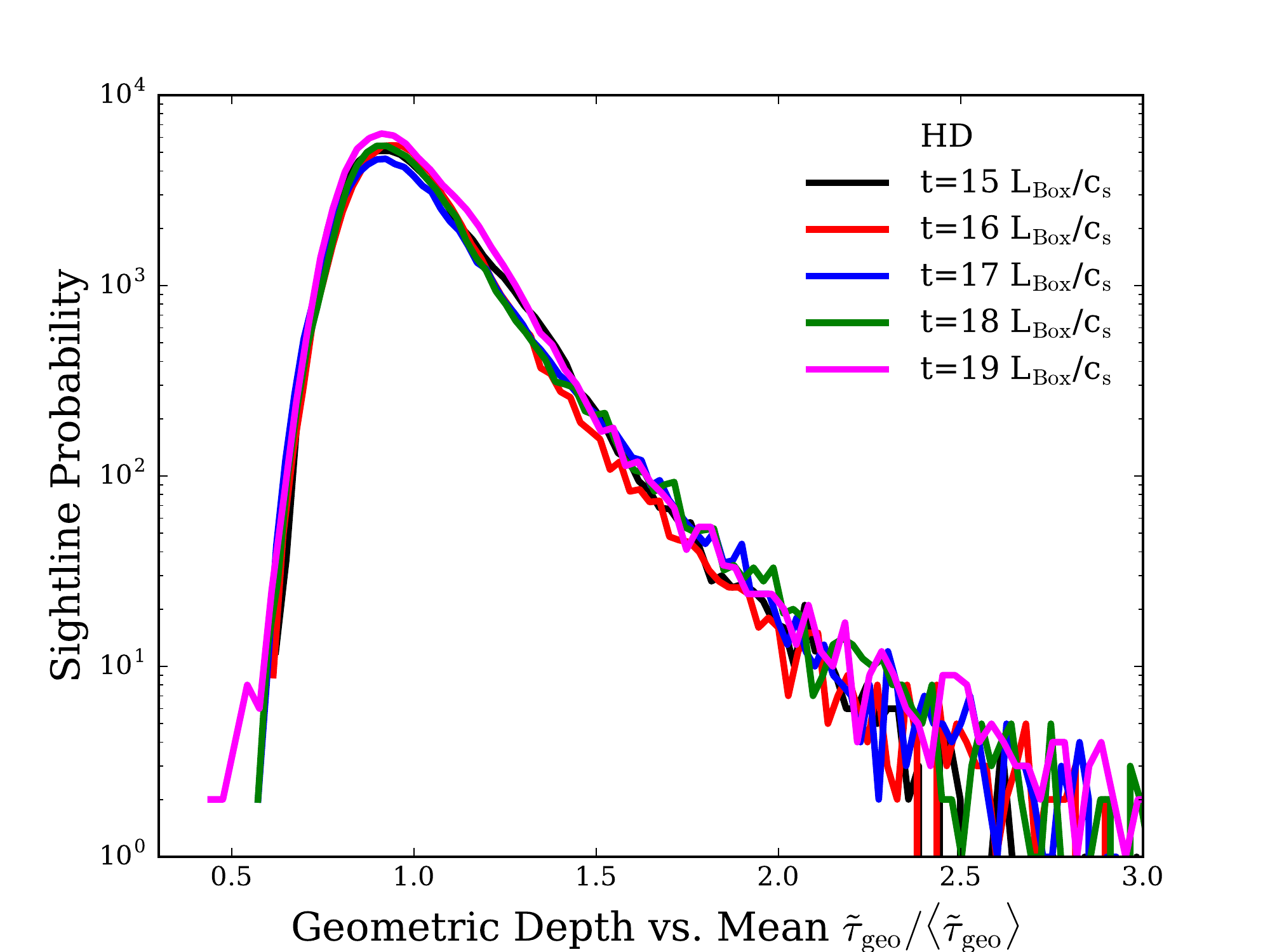}
    \includegraphics[scale=0.435]{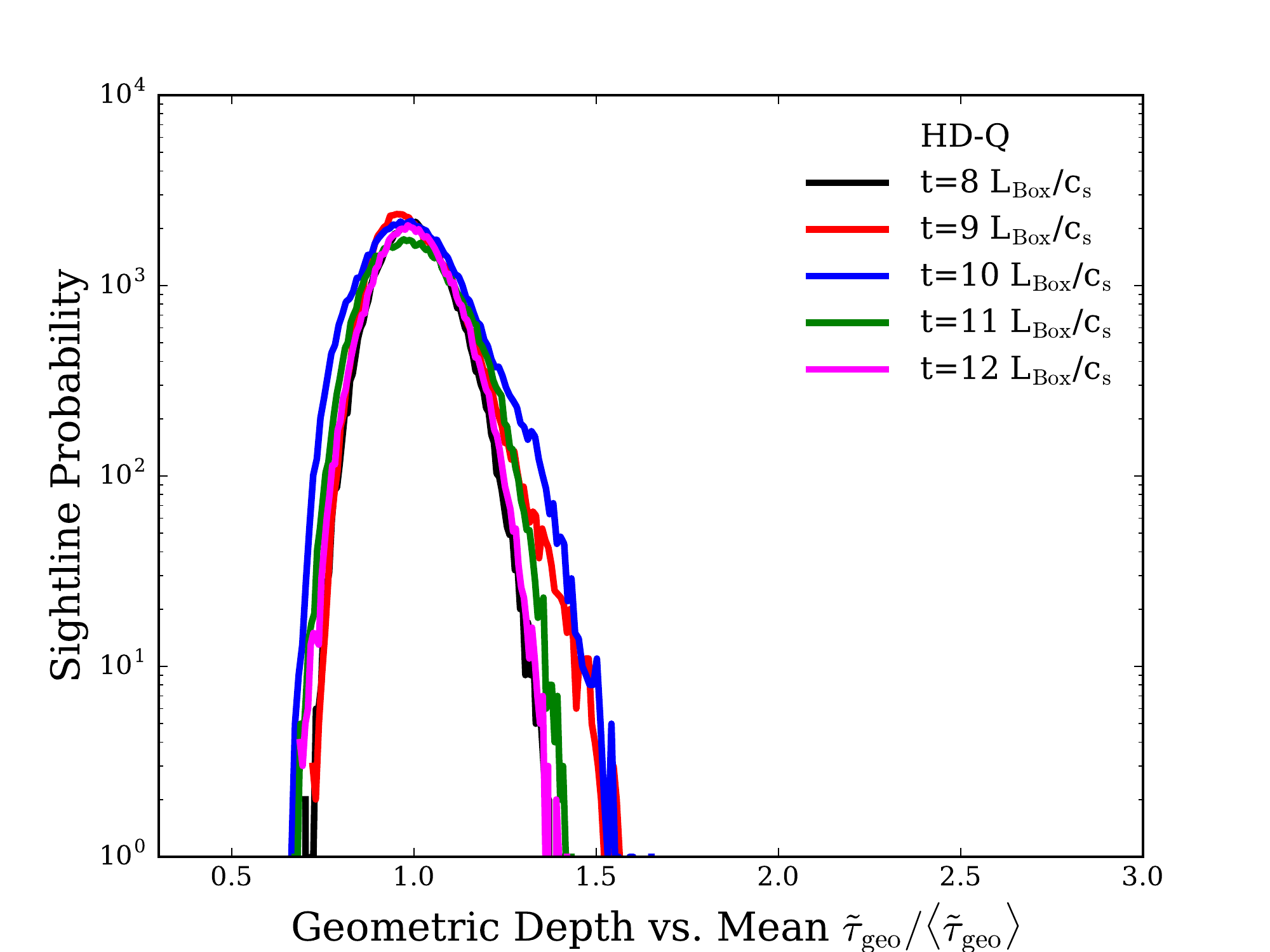}
    \includegraphics[scale=0.435]{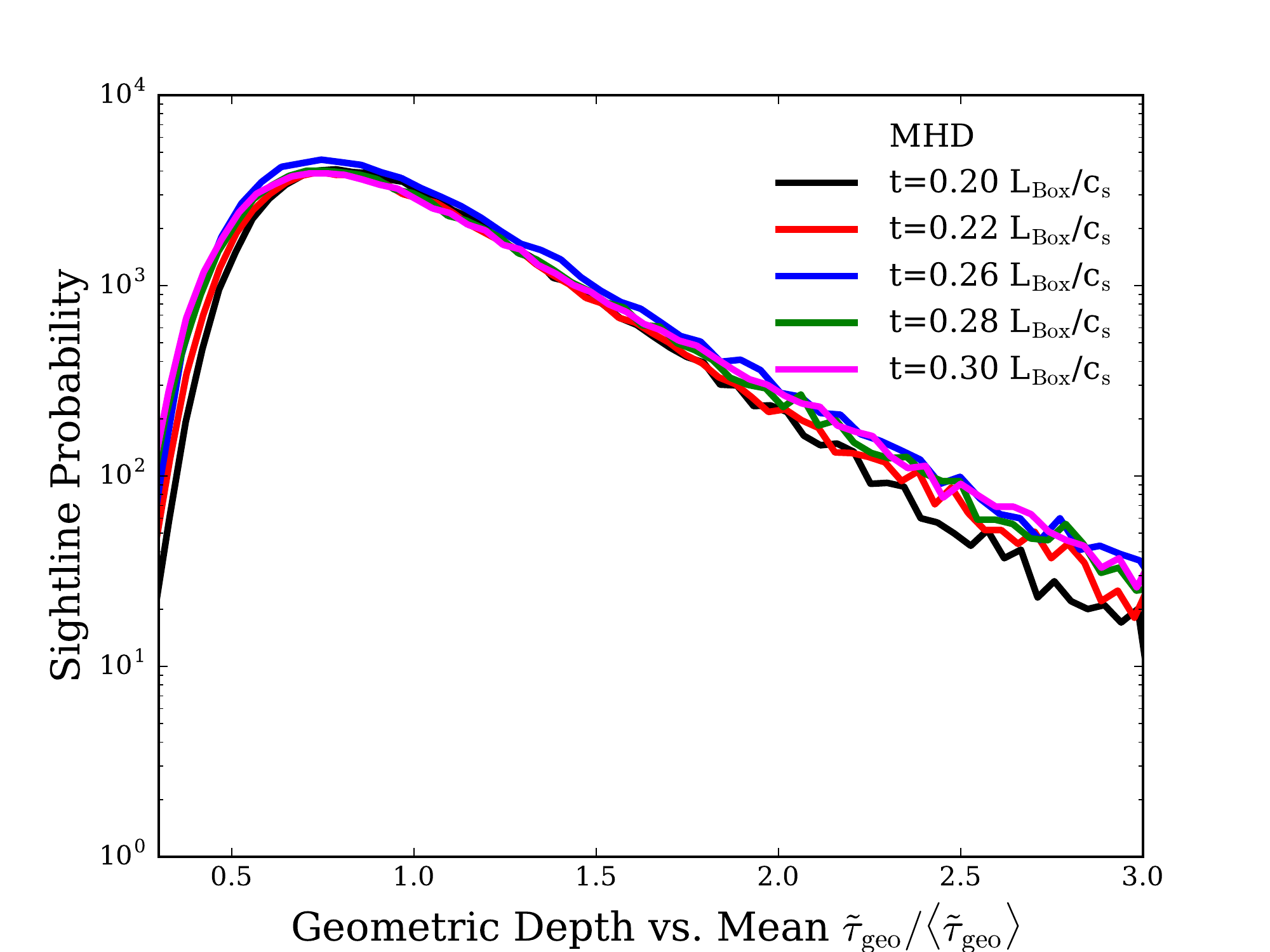}
    \includegraphics[scale=0.435]{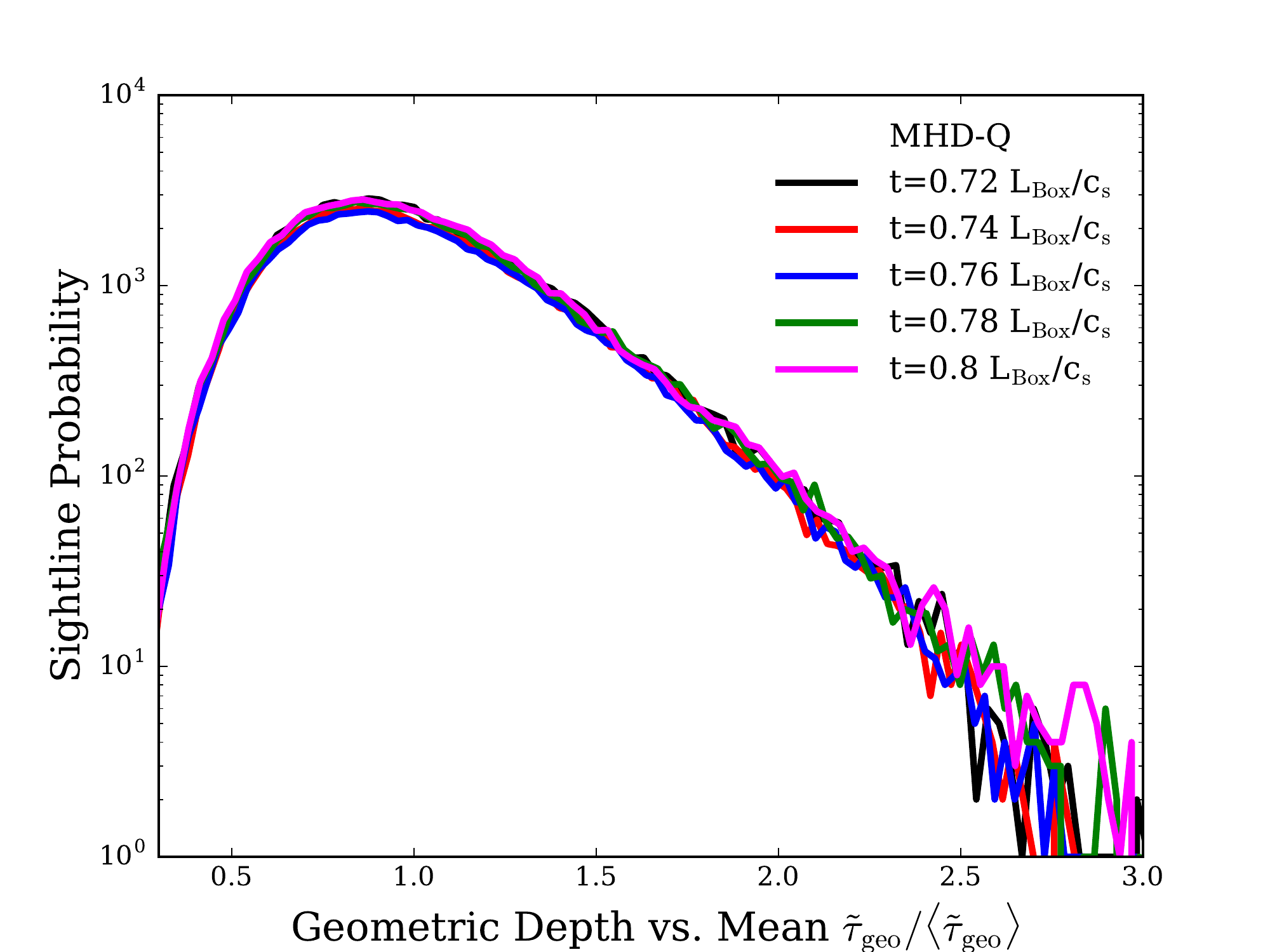}
    \caption{Mass weighted probability distribution function (PDF) of $\tilde{\tau}_{\rm geo} / \langle \tilde{\tau}_{\rm geo} \rangle$ for several different times during saturation for all simulations {\bf HD, HD-Q, MHD} and {\bf MHD-Q}, along individual narrow sightlines. The PDFs are non-Gaussian and some rare sightlines can have much higher extinction.
    }
    \label{fig:pdf_hd}
\end{figure*}
\begin{figure*}
\includegraphics[scale=0.435]{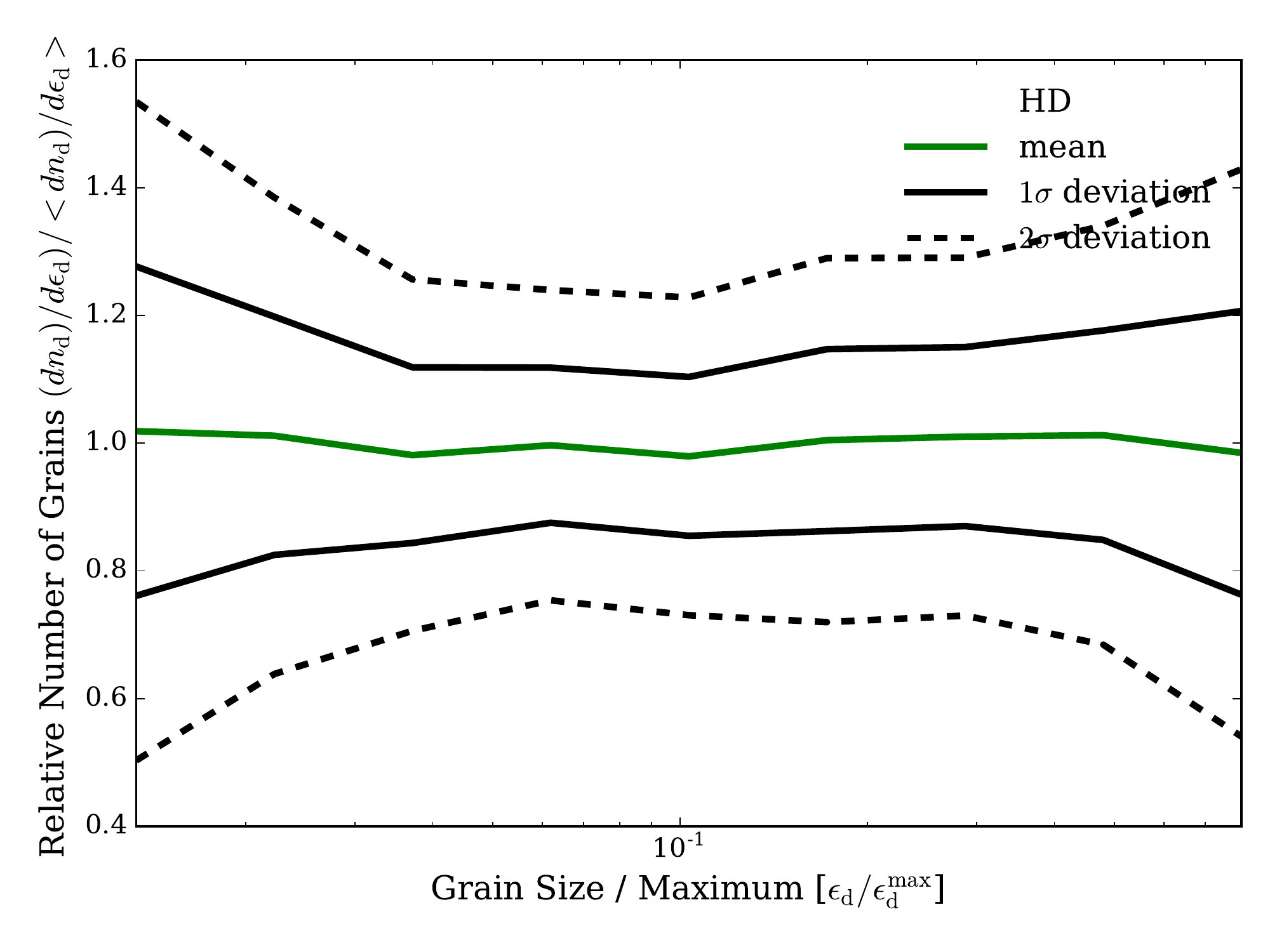}
\includegraphics[scale=0.435]{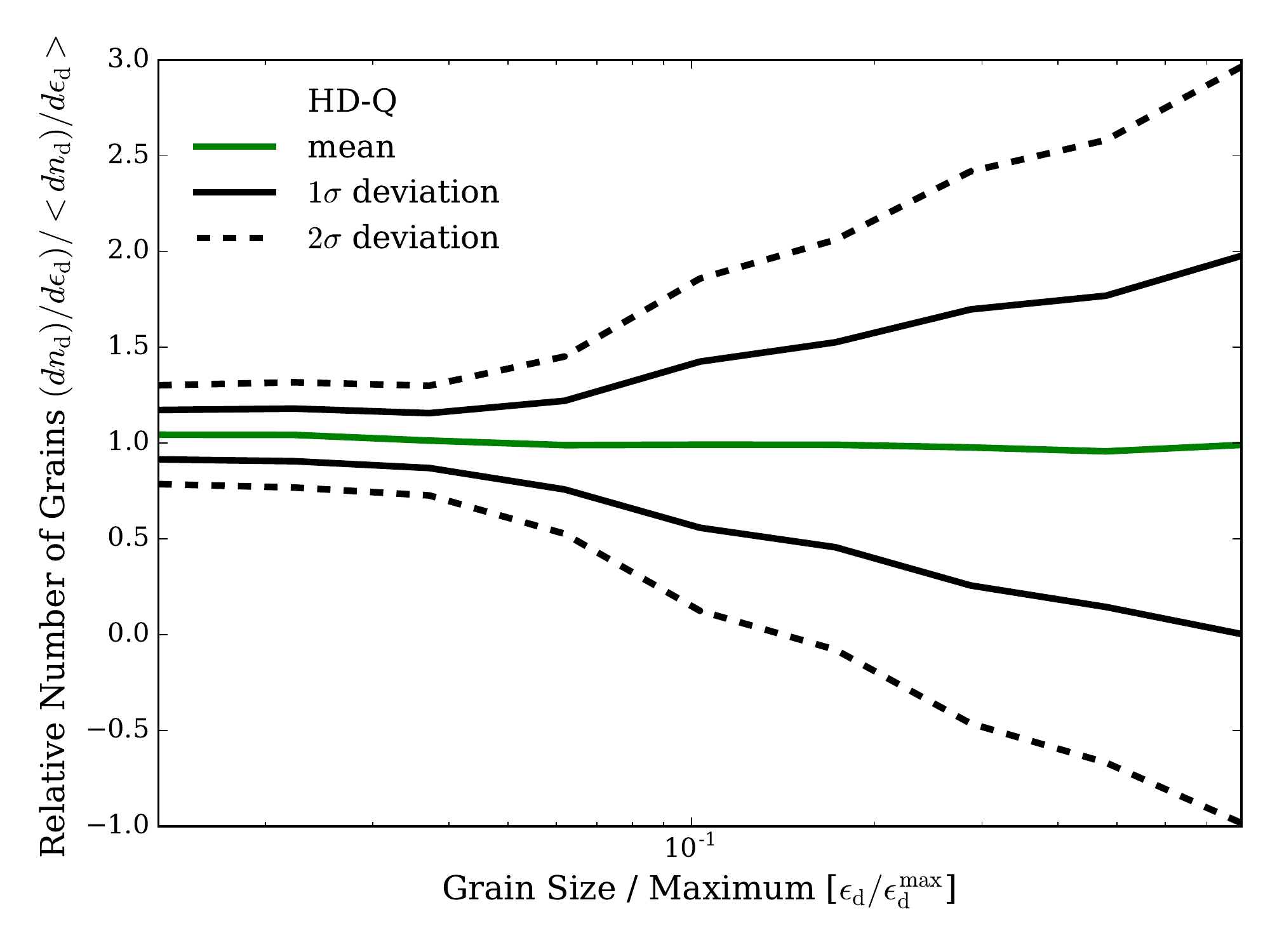}
\includegraphics[scale=0.435]{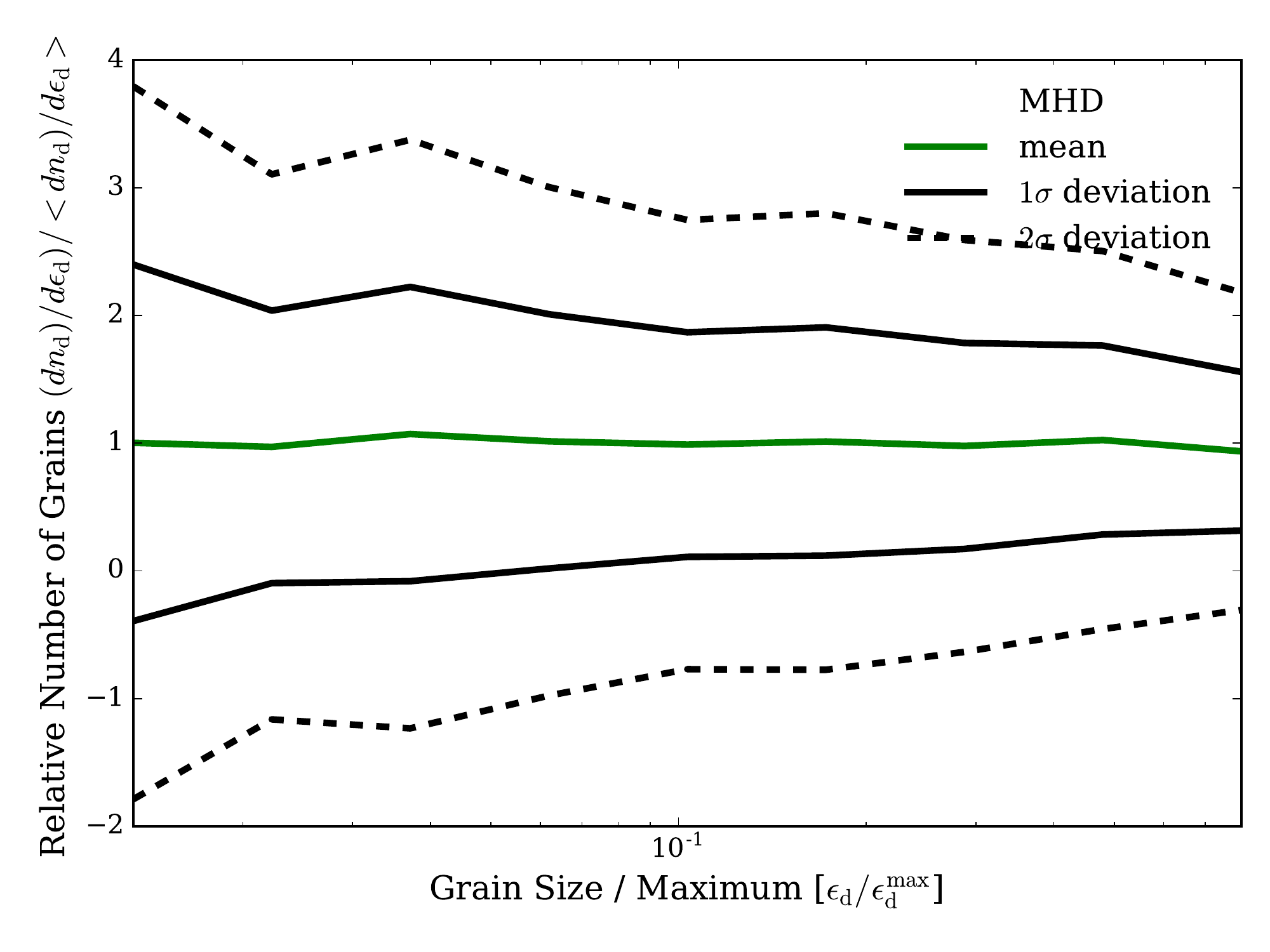}
\includegraphics[scale=0.435]{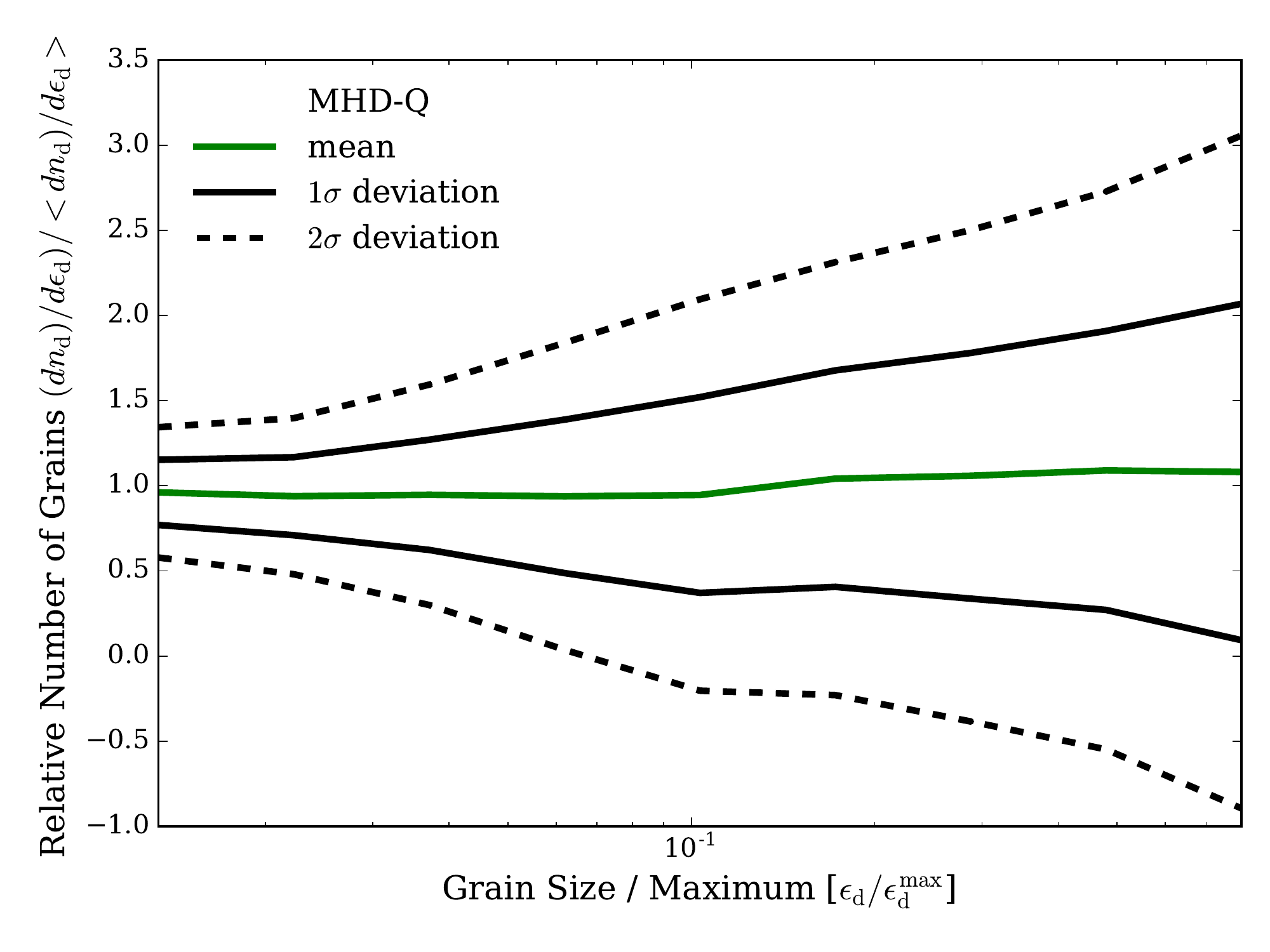}
        \caption{Variations in the grain size distribution (GSD; see \S~\ref{sec:hist}). We plot differential number of grains per unit size $\epsilon_{\rm d}$, $d n_{\rm d}/d \epsilon_{\rm d}$, relative to the box-average distribution $\langle d n_{\rm d}/d \epsilon_{\rm d} \rangle$ (MRN-like $\langle d n_{\rm d}/d \epsilon_{\rm d} \rangle \propto \epsilon_{\rm d}^{1/2}$) normalized to have the same grain mass, along individual sightlines. The mean ({\em dashed}), $\pm1\,\sigma$ ({\em shaded}), and $\pm2\,\sigma$ ({\em dotted}) values over all sightlines (in times after saturation) are shown, for each simulation (labeled). The GSD, and therefore extinction curve shapes, fluctuates along with extinction, with up to factor $\sim 2$ variation in the ratio of smallest-to-largest grain number. When the acceleration is grain size independent ({\bf HD-Q}, {\bf MHD-Q}), fluctuations are largest in the large grains, while when the acceleration is grain size dependent, the opposite occurs. 
        }
        \label{fig:grainsize.dist}
\end{figure*}
The simulations are carried out with the multi-method code \textsc{Gizmo} \citep{Hopkins2015}\footnote{A public version of the code, including all methods used in this paper, is available at \href{http://www.tapir.caltech.edu/~phopkins/Site/GIZMO.html}{http://www.tapir.caltech.edu/phopkins/Site/GIZMO.html}}. The equations of (magneto) hydrodynamics are solved with the second-order Lagrangian Godunov finite volume method, the meshless finite volume (MFV; \citealt{Gaburov2011, Hopkins2015}). The dust particles in the simulations are modeled using the super particle approach \citep[e.g.][]{Carballido2008, Johansen2009, Bai2010, Bai2010DiskI, Bai2010DiskII, Pan2011}, where each super-particle represents a group of grains with some size and charge obeying the single-grain equations above. Extensive tests of the methods we use for both MHD and dust dynamics in \textsc{Gizmo} are presented in \citet{hopkins:mhd.gizmo,hopkins:cg.mhd.gizmo,hopkins:gizmo.public.release,deng:2019.mri.turb.sims.gizmo.methods,hopkins.2016:dust.gas.molecular.cloud.dynamics.sims,lee:dynamics.charged.dust.gmcs,moseley:2018.acoustic.rdi.sims,seligman:2018.mhd.rdi.sims,hopkins:2019.mhd.rdi.periodic.box.sims}. 
We initialize a homogeneous, period 3D box (side-length $L_{\rm box}$), obeying the homogeneous equilibrium dust drift solution:
\begin{align}
 \label{eqn:driftvel}   {\bf w}_{s}^{0} = \frac{|\hat{\bf a}|\, t_{s}^{0}}{1+\mu}\left[\frac{\hat{\textbf{a}} - t_{s}^{0}/t_{L}^{0} (\hat{\textbf{a}} \times \hat{\textbf{B}}_{0}) +(t_{s}^{0}/t_{L}^{0})^2 (\hat{\textbf{a}} \cdot \hat{\textbf{B}}_{0})}{1+ (t_{s}^{0}/t_{L}^{0})^2}\right].
\end{align}
then allow it to evolve until the statistics of all properties studied here reach saturation.

\subsection{Parameter Choices \&\ Units}
\label{sec:params}

Ignoring magnetic fields for now, if we work in code units of the isothermal sound speed $c_{s}$, mean initial gas density $\langle \rho_{\rm g}\rangle$, and box size $L_{\rm box}$, then our simulations are scale-free and the dynamics are entirely specified by three dimensionless parameters: (1) the volume-averaged dust-to-gas mass ratio $\mu \equiv \langle \rho_{\rm d} \rangle/\langle \rho_{\rm g} \rangle$; (2) the ``size parameter'' of the largest grains, $\alpha \equiv \bar{\rho}_{\rm d}\,\epsilon_{\rm d}^{\rm max} / \langle \rho_{\rm g} \rangle\,L_{\rm box}$, and (3) the ``acceleration parameter'' $\tilde{a} \equiv |\hat{\bf a}|\,L_{\rm box}/c_{s}^{2}$ (the direction of $\hat{\bf a}$ defines our axes). To motivate some initial parameter choices for our study, and allow us to make some order-of-magnitude translation to physical units, we consider a system where the box represents a region $L_{\rm box} \sim r$ at a distance $r$ in a cool-star outflow with $\langle \rho_{\rm g} \rangle \sim \dot{M}_{\ast}/(4\pi\,r^{2}\,v_{\rm wind})$, and luminosity $L_{\ast}$, with typical maximum grain sizes $\epsilon_{\rm d}^{\rm max} \sim 1\,{\rm \mu m}$ and $\bar{\rho}_{\rm d} \sim 2\,{\rm g\,cm^{-3}}$. From e.g.\ \citet{2018A&ARv..26....1H}, some typical AGB parameters might be stellar radius $R_{\ast} \sim 10-250\,R_{\odot}$, effective temperature $T_{\rm eff} \sim 1000-3000\,$K, luminosity $L_{\ast} \sim 10^{3}-10^{4}\,L_{\odot}$, wind velocity $v_{\rm wind} \sim 5-30\,{\rm km\,s^{-1}}$, mass-loss rate $\dot{M}_{\ast} \sim 10^{-8}-10^{-6}\,{\rm M_{\odot}\,yr^{-1}}$, dust-to-gas mass ratio $\mu \sim 10^{-4} - 10^{-2}$, geometric optical depth (defined below) $\tau_{\rm geo} \sim 0.5-1.2$, while for RCB stars, typical parameters might be $R_{\ast} \sim 100\,R_{\odot}$, $T_{\rm eff} \sim 4000\,$K, $L_{\ast} \sim 10^{4}\,L_{\odot}$, $v_{\rm wind} \sim 100\,{\rm km\,s^{-1}}$, $\dot{M}_{\ast} \sim 10^{-6}\,{\rm M_{\odot}\,yr^{-1}}$, $\tau_{\rm geo} \sim 10$. Inserting these, we obtain:
\begin{align}
    \alpha \sim \left(\frac{\epsilon_{\rm d}^{\rm max}}{\rm \mu m}\right) \left(\frac{L_{\rm Box}}{500 R_{\odot}}\right) \left(\frac{v_{\rm wind}}{10 \rm km s^{-1}}\right) \left(\frac{\dot{M}}{10^{-8} M_{\odot} \rm yr^{-1}}\right),
\end{align}
and
\begin{align}
    \tilde{a} \sim 3400 \left(\frac{L_{\ast}}{10^3 \rm L_{\odot}}\right) \left(\frac{Q_{0}}{0.2}\right) \left(\frac{\epsilon_{\rm d}^{\rm max}}{\rm \mu m}\right) \left(\frac{L_{\rm Box}}{500 R_{\odot}}\right) \left(\frac{T_{\rm gas}}{1000 \rm K}\right). 
\end{align}
We also note that magnetic fields in cool star-star atmospheres are generally believed to be weak \citep[e.g.][]{Sabin2015}. However, there seems to be some evidence that there is some magnetic activity on the surface of RCB stars that might even contribute to the stellar wind due to heating via magnetosonic waves \citep[e.g.][]{Rao2008}. In full detail we would need to model this self-consistently accounting for non-ideal MHD effects which depend sensitively on the ionization fraction, which requires physics well beyond those included in our idealized studies. We therefore simply consider both ``pure hydro'' and ``ideal MHD'' cases below, to span the range of interest.

With this in mind consider three simulations from the parameter survey in \citet{hopkins:2019.mhd.rdi.periodic.box.sims} that could plausibly represent cool-star outflows. In each we take $\mu=0.01$, but we have also considered simulations with $\mu=0.001$, which behave qualitatively similarly (the linear growth is slightly slower, but non-linear behavior similar, with somewhat stronger fractional dust clumping, as shown in \citealt{hopkins:2019.mhd.rdi.periodic.box.sims}), and have somewhat lower optical depths than typical cool-star envelopes, so do not study those in detail.

\begin{itemize}

\item{\bf HD-Q}: This is a pure-hydrodynamic ($|{\bf B}|\rightarrow0$) simulation, with $(\alpha,\,\tilde{a})=(0.088,\,1400)$. From the above, these choices are broadly plausible for a typical weaker AGB outflow ($\dot{M} \sim 10^{-8}\,{\rm M_{\odot}\,yr^{-1}}$) at radii $\sim 500\,R_{\odot}$ from the stellar photosphere. We assume in this run that $Q \propto (\epsilon_\mathrm{d}/\lambda)$ i.e.\ that the system is in the ``small grain limit'' (grain sizes much smaller than the wavelengths driving the outflow), so the radiative acceleration on grains is {\em independent} of grain size $\epsilon_\mathrm{d}$. It is important to note that this directly implies that the grain velocity is \textit{dependent} on grain size $\epsilon_\mathrm{d}$.

\item{\bf HD} This is again pure-hydrodynamic, but with $(\alpha,\,\tilde{a})=(0.15,\,20)$, and we assume $Q=\,$constant, i.e.\ the ``large grain'' or ``geometric absorption'' limit, likely more relevant for the large grains ($\sim {\rm \mu m}$) that dominate the grain mass. Of course, real systems can have grains in both limits and $Q$ can be a complicated function of grain size and other properties, but given the idealized nature of our simulations, this is simply a useful comparison to survey the extremes of different dependencies we might expect in more physical cases. The lower value of $\tilde{a}$ is chosen so that the mean drift velocity of the {\em small} grains, which dominate the opacity, is similar to case {\bf HD-Q}. In this case the acceleration is \textit{dependent} on grain size while the velocity is \textit{independent} of grain size.  

We note that the simulations {\bf HD-Q} and {\bf MHD-Q} are identical with the simulations discussed in \citep{squire2021:hd.periodic.grain.spectrum.sims}.

\item{\bf MHD-Q} This case includes magnetic fields, and adopts $(\alpha,\,\tilde{a})=(0.0067,\,5800)$, assuming $Q\propto (\epsilon_\mathrm{d}/\lambda)$ like {\bf HD-Q}. For the MHD case, we also must initialize ${\bf B}$, which we do by assuming weak-but-not-negligible fields (plasma $\beta \equiv c_{s}^{2}/v_{A}^{2} = 10$) with an initial angle of $45^{\deg}$ between $\hat{\bf B}$ and $\hat{\bf a}$ (the latter is largely irrelevant to the dynamics; see \citealt{hopkins:2019.mhd.rdi.periodic.box.sims}), and the dimensionless grain ``charge parameter'' for the collisional charge law and our fiducial parameters becomes $\tilde{\phi} = 3 Z_\mathrm{d} e /4\pi c \epsilon_\mathrm{d}^2 \rho_\mathrm{g}^{1/2} \approx 40$. More important than magnetic fields, as we show below these simulations have about $\sim 10$ times higher mean geometric optical depth of dust, which places them closer to RCB-like conditions in physical parameter space. 

\item{\bf MHD} This case includes magnetic fields, and adopts $(\alpha,\,\tilde{a})=(0.0067,\,5800)$, assuming $Q=\,$constant, i.e.\ the ``large grain'' or ``geometric absorption'' limit, like {\bf HD}. Apart from this we note that we initialise the magnetic field \textit{exactly} the same way as in the case {\bf MHD-Q} with identical optical depth between these two runs making this run again similar to what we know from the high optical depth limits in RCB-star atmospheres.
\end{itemize}


\section{Results}

\subsection{Morphology of the instabilities}
\label{sec:morphology}

Figs.~\ref{fig:RDI-const}, \ref{fig:RDI-variable}, \ref{fig:RDI-MHD-const} and \ref{fig:RDI-MHD}, visualize the simulations at early (linear) and late (non-linear/saturated) times. The morphologies of the RDIs are distinct in each simulation, though there are some qualitative similarities. At early times, linear mode structures dominate, with the smallest-scale modes (which have the fastest growth times) most visible. As time passes larger modes grow until the box is saturated. In {\bf HD-Q}, the smallest grains drift sub-sonically, so the dominant initial modes are the low-$k$ aligned modes from \citet{hopkins:2017.acoustic.RDI} which builds up sheets (in the plane perpendicular to $\hat{\bf a}$), but the largest grains drift highly super-sonically, which non-linearly drives the elongated filament structures seen for super-sonic drift in single grain simulations in \citet{moseley:2018.acoustic.rdi.sims}. In {\bf HD}, although the acceleration depends on grain size, the initial drift velocity is independent of grain size (because the combination $|\hat{\bf a}|\,t_{s}$ in Eq.~\ref{eqn:driftvel} is independent of $\epsilon_{\rm d}$) with trans-sonic $|{\bf w}_{s}| \sim 1.5\,c_{s}$, so the initial modes have the same acoustic-RDI resonant angle, which appears strongly at early times as the characteristic angle of the sheets; the non-linear evolution being more trans-sonic produces less coherent filaments with a more ``patchy'' morphology. Morphology, growth rate and resulting turbulence are discussed in more detail in \citet{squire2021:hd.periodic.grain.spectrum.sims}. The {\bf MHD} and {\bf MHD-Q} cases are driven by a more complicated mix of modes, since with MHD and charged grains many more RDI families (resonances with gyro motion, \Alf\ or slow or fast magnetosonic waves) are available. The details of these modes, which is similar to the single grain size case ``AGB-XL'' in \citet{hopkins:2018.mhd.rdi}, are discussed therein. The MHD-RDIs allow for even faster mode growth, as more resonances are available, and resonances exist even for grains at arbitrarily sub-sonic drift velocities. Non-linearly, these eventually produce a corrugated sheet morphology, with the plane of the sheets aligned with $\hat{\bf a}$.

Across these behaviors, it is certainly possible that the RDI could source both large-scale features commonly seen in cool-star outflows including arcs or shells (the spherical-geometry version of the ``sheets'' seen here) as in \citet{morris:1993.cool.wind.dust.drag.instability.slow.saturated.mode,1994A&A...288..255W,deguchi:1997.dust.envelope.pne.spherical.drag.instability.quasi.resonant,balick:2002.pne.shape.structure.review}, as well as commonly-seen dust filaments or streamers with ``knots'' of overdensity \citep{odell:1996.pne.cometary.knots.from.instabilities,odell:2002.pne.knots.review,balick:1998.knots.fliers.cometary.tails.pne,matsuura:2009.pne.knots}. And of course turbulence and clumping are evident with similar magnitude to observational claims in e.g.\ \citet{2003ApJ...582L..39F,young:2003.clumpy.wind.models,2007Natur.447.1094Z,2010ApJ...724L.133A,2012A&A...537A..35C}, though this is far from the only mechanism that can produce turbulence in outflows.

\subsection{Gas and Dust Densities} 
\label{sec:gasden}

On micro-scales, the three-dimensional local dust-to-gas ratios can fluctuate by orders of magnitude in the RDIs, and this is studied in \citet{hopkins:2019.mhd.rdi.periodic.box.sims}. In systems like cool-star outflows, that could be extremely important for e.g.\ grain growth, coagulation, shadowing, and other phenomena which would then non-linearly influence the dynamics or even the ability to launch outflows, but we do not study any of these processes here. 

A more observable quantity is the sightline-integrated dust column $\Sigma_{\rm d}$ or gas column $\Sigma_{\rm g}$. We show this for all simulations for an early and a late evolutionary state of the RDIs in each respective simulations base on the bivariate distribution of these in Fig.~\ref{fig:bias_color_early}  and Fig.~ \ref{fig:bias_color_late}. We note that this does not significantly change for the simulations {\bf HD} and {\bf MHD-Q}. Moreover we note that in the saturated growth stage of the RDIs this the distribution is not strongly fluctuating as a function of time. 

We project our simulation assuming direction of $\hat{\bf a}$ is the direction of a (radial) sightline (appropriate for e.g.\ ``down the barrel'' observations of the driving star), and divide our box into a 2D 256x256 grid integrating along each grid point through the box (the results are converged with respect to our grid sampling). We choose a snapshot in the saturated phase; the results are insensitive in this phase to the time chosen. There are still substantial variations in $\Sigma_{\rm d}$ and $\Sigma_{\rm g}$, but most of the local three-dimensional fluctuation in their ratio is factored out: the sightline-integrated correlation between dust and gas surface densities is quite good, in fact, a result of the dust driving the gas into overdensities in the same columns and structures of the outflow. 

We consider sightline-integrated dust column $\Sigma_{\rm d}$ or gas column $\Sigma_{\rm g}$ that are directly related to the observables such as dust extinction. To evaluate these quantities we project our simulation assuming direction of $\hat{\bf a}$ is the direction of a (radial) sightline (appropriate for e.g.\ ``down the barrel'' observations of the driving star), and divide our box into a 2D 256x256 grid integrating along each grid point through the box (the results are converged with respect to our grid sampling). We choose a snapshot in the saturated phase; the results are insensitive in this phase to the time chosen. There are still substantial variations in $\Sigma_{\rm d}$ and $\Sigma_{\rm g}$, but most of the local three-dimensional fluctuation in their ratio is factored out: the sightline-integrated correlation between dust and gas surface densities is quite good, in fact, a result of the dust driving the gas into overdensities in the same columns and structures of the outflow. 

We show the bivariate distribution of these two quantities for all simulations for \textit{the initial growth phases} and \textit{the saturated states} of the RDIs in Fig. 5 and Fig. 6. Firstly, we note that there is no significant differences between the simulations for all four simulations that we take into account in the analysis, at their respective early and late evolutionary stages. Moreover, we note that in the saturated regime of the RDIs the scatter of these distributions is only weakly affected, i.e. we it does not matter for which snapshot in the saturated stage we plot those quantities.  

\subsection{Extinction Variations in Space and Time}
\label{sec:extinction}

Next consider the implications for extinction and scattering, represented in the optical depth:
\begin{align}
\tau(\lambda,\,{\bf x},\,...) = \int \int \, Q(\epsilon_{\rm d},\,\lambda,\,...)\,\pi\,\epsilon_{\rm d}^{2}\,\frac{{\rm d} N_{\rm d}}{d \epsilon_{\rm d}\,d^{3}{\bf x}}\,d\ell\,d \epsilon_{\rm d}
\end{align}
where $d N_{\rm d}/ d \epsilon_{\rm d}\,d^{3}{\bf x}$ is the local number density of grains of size $\epsilon_{\rm d}$. To calculate the optical effects in detail, we would need to model scattering through the outflow, with detailed radiation-hydrodynamics simulations accounting for how the different extinction and scattering $Q$ values depend on wavelength and grain size, the incident spectrum being attenuated, and the observed wavelengths. But given the idealized nature of our simulations, such a detailed calculation would not be especially realistic. 

What we can calculate robustly without reference to this is the {\em geometric} optical depth $\tau_{\rm geo}({\bf x})$, calculated ignoring scattering and assuming $Q=1$ (i.e.\ the geometric limit), which can be immediately computed from our simulations along any sightline, integrating $d \ell$ through the box length $L_{\rm box}$ on the grid described above. Note that since our initial conditions are homogeneous, we can immediately calculate the initial homogeneous value $\langle \tau_{\rm geo}(t=0) \rangle = (15/2)\,(\mu/\alpha)$ for our assumed dust grain size distribution and definition of $\mu$ and $\alpha$. This means our {\bf HD} simulations have initial/mean $\tau_{\rm geo} \sim 0.5-1$, while our {\bf MHD} case has $\tau_{\rm geo} \sim 10$, making the latter more appropriate perhaps for comparison to RCB systems which exhibit much larger extinction in dusty ``events'' \citep{Jeffery2019}. 

However, the sightline-integrated $\tau_{\rm geo}$ is appropriate only for point-sources; if the star is the background source, its size is not infinitesimal compared to the obscuring envelope/outflow. We therefore account for finite source-size effects by computing $\tau_{\rm geo}({\bf x})$ along sightlines on a densely-sampled 2D grid, then assuming the source is a disc with uniform surface brightness and computing its effective optical depth $\tilde{\tau}_{\rm geo} = - \ln{[ N^{-1}\,\sum_{i}\,e^{-\tau_{\rm geo}({\bf x}_{i})}] }$ (where $N$ is the number of sightlines uniformly sampling the disc), which gives (by definition) the correct extincted luminosity $L_{\ast}\,e^{-\tilde{\tau}_{\rm geo}}$. 

Fig.~\ref{fig:tau_hd} shows the resulting $\tilde{\tau}_{\rm geo}$ for a {\em specific} line of sight (through the box center) as a function of time (in the saturated state), as a function of source size, for the runs {\bf HD-Q} and {\bf HD}. Of course, as the source size approaches the box scale, more of the variation is averaged-out. Note that we focus on the ``radial'' direction (line-of-sight parallel the acceleration $\hat{\bf a}$ or outflow direction); if we instead select ``perpendicular'' directions (integrating perpendicular to the outflow direction, the behavior is qualitatively quite similar, but fluctuations are somewhat smaller for the same fractional source size (although physically, in this case, background sources would be essentially point-sources, so the maximum variation of the sources sizes we consider would be most applicable). We show this for reference for the simulation {\bf HD} in Fig.~\ref{fig:tau_perp} and note that this is essentially identical for the model {\bf HD-Q}

Fig.~\ref{fig:tau_mhd} shows the resulting $\tilde{\tau}_{\rm geo}$ for a {\em specific} line of sight (through the box center) as a function of time (in the saturated state), as a function of source size for the simulation {\bf MHD-Q}. Similar to the case of {\bf HD} and {\bf HD-Q} which we present in Fig.~\ref{fig:tau_hd} we can see that as the cylinders approach larger and larger radii, the optical depth approaches the limit of the box. Qualitatively, we observe similar behavior for the simulation {\bf MHD-Q} compared to its non-MHD counterparts {\bf HD} and {\bf HD-Q} and report fluctuations on a similar scale of around $\sim 20\%$. However, we note that in specific sightline projections we find single peaks the reach up to a fluctuation scale of around $\sim 50\%$ (e.g. the peak at L$_\mathrm{Box}/c_\mathrm{s} \sim 0.65$ int the middle panel of Fig.~\ref{fig:tau_mhd}). 

Fig.~\ref{fig:std_hd} shows the $1\sigma$ variation in $\tilde{\tau}_{\rm geo}$ across {\em spatial} positions (sampling all possible non-overlapping cylinders in the box), at a given time, and Fig.~\ref{fig:pdf_hd} shows the full PDF of $\tilde{\tau}_{\rm geo}$ at several times in the saturated phase. We clearly see the instabilities grow and saturate in time; we also see that the magnitude of fluctuations decreases for very large source sizes which represent a large fraction of the box, but is weakly-dependent on source size for source radii $R_{\ast} \lesssim 0.2\,L_{\rm box}$. This indicates that the dominant modes driving the fluctuations are actually quite large-scale, coherent structures (as suggested visually in Figs.~\ref{fig:RDI-const}-\ref{fig:RDI-MHD}), and for almost any realistic envelope/outflow size (more than a couple times the stellar diameter) the fluctuations will be significant. We also see that, in saturation, the $1\,\sigma$ fluctuations in $\tilde{\tau}_{\rm geo}$ are generally $\approx 10\%$ of the box-averaged mean $\langle \tilde{\tau}_{\rm geo} \rangle$, for all three of our simulations. In the full PDFs, however, we see that the distribution is significantly non-Gaussian, with a power-law slope to higher $\tilde{\tau}_{\rm geo}$, so high-extinction events are significantly more common than Gaussian. Finally we obtain similar space and time PDFs in saturation, suggesting (within our idealized simulations) the statistics are somewhat ergodic.

In saturation we can see (viewing a time-series of different snapshots as Fig.~\ref{fig:RDI-const}) coherent dust structures move and re-shape on of order the dust crossing time (given by the drift speed $|{\bf w}_{s}|$). This suggests a minimum characteristic timescale for large variations in dust structures of characteristic size $\lambda$ in the saturated state,\\ $\Delta t_{\rm min} \sim \lambda / |{\bf w}_{s}| \sim 0.2\,{\rm yr}\,(\lambda/500\,R_{\odot})\,(\dot{M}/10^{-8}\,{\rm M_{\odot}\,yr^{-1}})^{1/2}\,\\(L_{\ast}/10^{3}\,L_{\odot})^{-1/2}\,(Q_{0}/0.2)^{-1/2}\,(v_{\rm wind}/10\,{\rm km\,s^{-1}})^{-1/2}$. Since we have shown relatively large-scale modes dominate the obscuring structures, $\lambda$ of order a few hundred $R_{\odot}$ is reasonable. In Fig.~\ref{fig:tau_hd} we can see this provides a fairly reasonable estimate, with most of the large variations occurring on a timescale a few times this $\Delta t_{\rm min}$ ($\sim$\,yr).

\subsection{Grain Size Distribution \&\ Extinction Curve Variations}
\label{sec:hist}

Since grain stopping times and either accelerations or drift velocities depend on size, large and small dust grains do not move exactly together; there can also be variations in the grain size distribution (GSD), which is directly directly affect observables, such as the shape of the observed extinction curve, along different sightlines. Fig.~\ref{fig:grainsize.dist} illustrates this by showing the variance in dust particle number as a function of grain size, for narrow cylinders (source size $R_{\ast}=0.05\,L_{\rm box}$) along $\hat{\bf a}$. This depends strongly on size of the grains, implying there will be sightlines which are significantly with shallower or more ``grey'' extinction curves (relatively more large vs.\ small grains) and also with steeper or ``redder'' extinction (more small vs.\ large). 

Interestingly, in {\bf HD-Q}, where accelerations are grain grain size independent, the largest fluctuations are in the large grains, i.e.\ they are the most-clumped. These are drifting more rapidly, producing faster RDI growth rates, and form the core structures of the dust ``filaments'' in Fig.~\ref{fig:RDI-const}, with the smallest grains better-coupled to gas so filling more space in-between. This means ``diffuse'' (lower-column) sightlines will have steeper extinction curves, compared to high-column sightlines. In {\bf HD}, where accelerations are larger for smaller grains, the drift velocity being constant means RDI growth rates are largest for smallest grains, which form the denser structures. This means ``diffuse'' (low-column) sightlines will have more-grey extinction, compared to high-column sightlines (the opposite of {\bf HD}). In either case, the magnitude of the effect is modest, but measureable: from Fig.~\ref{fig:grainsize.dist} even the most extreme sightline variations will produce factor $<2$ variation in the ratio of largest-to-smallest grains over a factor $\sim 100$ in grain size (the equivalent of changing the extinction curve $A_{\lambda} \propto \lambda^{\psi}$ by $\Delta \psi \sim 0.1$).

Similar behaviour is confirmed by the MHD-simulations in {\bf MHD-Q} and {\bf MHD} with the difference that while the mean of dust particle in the instabilities stays roughly constant, the largest fluctuations are increasing by a factor $2-3$ which is tracing the more violent and rapid build-up of the instabilities that we can observe in the presence of magnetic fields, consistent with the previous studies of \citet{seligman:2018.mhd.rdi.sims} and \citet{hopkins:2019.mhd.rdi.periodic.box.sims}. Thus, we can achieve a factor of $<4$ in the ratio of the largest to the smallest grains leading to a change of the extinction curve by $A_{\lambda} \propto \lambda^{\psi}$ by $\Delta \psi \sim 0.2$)

Of course, in real systems, large grain size variations might arise from dust physics not modelled here (condensation, growth, coagulation, shattering) -- and these processes will be strongly modified by the RDIs themselves. Moreover, in the real physical system of a cool-stellar atmosphere the magnetic field structure might by much more complex than we can model with our current setup and since the magnetic fields seem to be a crucial part in the evolution of the RDIs this could change the picture. However, the general trend is consistent between the simulations {\bf HD, HD-Q, MHD} and {\bf MHD-Q} but the time-scale on which the fluctuations change appears to be faster in the MHD-runs. Thus, this should simply be taken as a proof of concept that extinction curve variations can arise also from purely dust-dynamical processes in the outflow. 

\section{Conclusions}
\label{sec:conclusions}

We consider for the first time the optical properties of dust-gas mixtures subject to resonant drag instabilities (RDIs), focusing on parameters of interest in dusty outflows around cool stars, e.g.\ AGB outflows or RCB-like dust ejection events. Specifically, we consider idealized simulations of an initially uniform dust-gas mixture with an MRN-type size spectrum of dust grains, subject to a uniform external radiative acceleration, with explicit size-dependent dust drag forces and their back-reaction coupling the dust and gas, and consider both pure-hydrodynamic with uncharged grains and ideal-MHD with charged grains limits. We then calculate how the geometric optical depth and grain size distribution vary across sightlines in both time and space, owing purely to the dust dynamics induced by the RDIs.

We show that in these environments, the RDIs robustly drive dust clumping, producing rms variations in the geometric optical depth of $\sim 10-20\%$ (corresponding to $\sim 0.1\,$mag under AGB-like conditions, or $\sim 1\,$mag under RCB event-like conditions), with a characteristic timescale of months to a couple years. We show these fluctuations are dominated by relatively large-scale modes, so are robust to finite-source-size effects so long as the dusty envelope is a factor of a few times larger than the source. The PDF of fluctuations is broad and non-Gaussian, with some rare sightlines having little or no dust, and others more than double the mean column. We also show that the RDIs can produce variations in the grain size distribution along the line of sight, with up to factor $\sim 2$ change in the ratio of the number of largest to smallest (100x smaller) grains, corresponding to changes of $\sim 0.1$ in the logarithmic extinction curve slope. We show that all of these qualitative effects are robust to the optical properties of the grains (e.g.\ how the absorption efficiency or grain acceleration depends on grain size) or strength of magnetic fields. 

We therefore conclude that the RDIs could play a central role driving much of the observed temporal (and spatial) variation in extinction in dusty outflows from cool stars. The characteristic timescales, magnitude of the fluctuations (in both total extinction and extinction curve shape), and visual morphologies of some of the structures (compared to resolved cases) are all similar to those observed. And, as we have shown in previous work, the linear growth of the RDIs is not suppressed or eliminated by other physics we have not considered here (e.g.\ non-ideal MHD, explicit radiation transport, gas cooling), and they are unstable on essentially all scales and times for any parameters remotely similar to those considered here, so we expect them to be ubiquitous in such environments. 

Furthermore, we would like to highlight the importance of the results that we obtained based on the shape of  Fig.~ \ref{fig:grainsize.dist}. Generally, we find that in the case of a grain size independent acceleration ({\bf HD-Q, MHD-Q}) the scatter in the instabilities across different sightlines is dominated by the largest grains in the MRN-spectrum. Physically, this arises because the largest grains are drifting the fastest causing a stronger contribution to the RDIs \citep{squire2021:hd.periodic.grain.spectrum.sims}. Vice versa in the cases with a grain size dependent acceleration we find that the scatter in the smallest grains is larger and the instabilities are dominated by the smallest grains in the MRN-spectrum. This is a \textit{clear theoretical prediction} as this specific feature is a direct consequence of the presence of the RDIs. With observations, one could try to target this specific feature in atmospheres of cool stars and it would not only be possible to evaluate to which degree the RDIs contribute to the dust dynamics in cool star atmospheres but one could also directly determine the general structure of the underlying grain size spectrum given the behaviour of the large and small grains in the atmosphere of the cool stars. Specifically it would be possible to determine how strongly the grain size spectrum in cool star atmospheres resembles the classic MRN-distribution. On top of this, we find it important to point out that this result is ``stable", i.e. it is consistent between the ``pure-hydro'' and ``ideal MHD". However, MHD effects play a crucial role when it comes to the amplitude of the scatter across different sightlines indicating that the effect is more dramatic in the presence of the magnetic field. In turn one could directly infer the magnetic field strength based on the scatter in the grain size distribution in the largest and smallest grain respectively.

However, we stress that we intend this paper to be just a ``proof of concept.'' In order to identify if the RDIs could play a central role, we have considered intentionally simplified simulations where the RDIs are essentially the only physics driving variability. These are not intended to model detailed observables, as we neglect many effects, including e.g.\ more realistic 3D, non-periodic geometries, non-ideal MHD and explicit multi-wavelength radiation transport, dynamical formation and evolution of dust grains, etc. Some of these, e.g.\ dust coagulation and growth, could be themselves strongly modified by the RDIs: the RDIs not only drive strong dust clumping (promoting coagulation), but also as shown here can provide ``shadowing'' (shielding some regions of the outflow from the stellar source) which would potentially allow for more efficient grain condensation, raising the local dust-to-gas ratio, which in turn leads to more rapid growth of the RDIs. In future work we will explore some of these physics as they pertain to cool-star systems, in order to develop more detailed observational predictions and tests. 


\section*{Acknowledgments}
UPS is supported by the Simons Foundation through a Flatiron Research Fellowship (FRF) at the Center for Computational Astrophysics. The Flatiron Institute is supported by the Simons Foundation. UPS acknowledges computing time granted by the Leibniz Rechenzentrum (LRZ) in Garching under the project number pn72bu and computing time granted by the c2pap-cluster in Garching under the project number pr27mi.\\
This work was initiated as part of the Kavli Summer Program in Astrophysics, hosted at the Center for Computational Astrophysics (CCA) at the Flatiron Institute in New York. We thank the Kavli Foundation and the Simons Foundation, for their support. The Flatiron Institute is supported by the Simons Foundation.
UPS thanks Kristina Monsch, Benjamin Moster, Joseph O'Leary, Eve Ostriker, Tyler Parsotan, Marius Ramsoy and Darryl Seligman for their useful comments. UPS thanks Bruce Draine for sharing his insights on interstellar dust. 
Support for PFH was provided by an Alfred P. Sloan Research Fellowship, NSF Collaborative Research Grant 1715847 and CAREER grant 1455342, and NASA grants NNX15AT06G, JPL 1589742, 17-ATP17-0214. JS acknowledges the support of Rutherford Discovery Fellowship RDF-U001804 and Marsden Fund grant UOO1727, which are managed through the Royal Society Te Ap\=arangi. 
The authors acknowledge the computing time provided by the Caltech computing facilities on the cluster ``Wheeler" as well as allocations from XSEDE TG-AST130039 and PRAC NSF.1713353 supported by the NSF, and NASA HEC SMD-16-7592.

\datastatement{The data supporting this article are available on reasonable request to the corresponding author.} 

\bibliography{lib}

\begin{thebibliography}{}
\makeatletter
\relax
\def\mn@urlcharsother{\let\do\@makeother \do\$\do\&\do\#\do\^\do\_\do\%\do\~}
\def\mn@doi{\begingroup\mn@urlcharsother \@ifnextchar [ {\mn@doi@}
  {\mn@doi@[]}}
\def\mn@doi@[#1]#2{\def\@tempa{#1}\ifx\@tempa\@empty \href
  {http://dx.doi.org/#2} {doi:#2}\else \href {http://dx.doi.org/#2} {#1}\fi
  \endgroup}
\def\mn@eprint#1#2{\mn@eprint@#1:#2::\@nil}
\def\mn@eprint@arXiv#1{\href {http://arxiv.org/abs/#1} {{\tt arXiv:#1}}}
\def\mn@eprint@dblp#1{\href {http://dblp.uni-trier.de/rec/bibtex/#1.xml}
  {dblp:#1}}
\def\mn@eprint@#1:#2:#3:#4\@nil{\def\@tempa {#1}\def\@tempb {#2}\def\@tempc
  {#3}\ifx \@tempc \@empty \let \@tempc \@tempb \let \@tempb \@tempa \fi \ifx
  \@tempb \@empty \def\@tempb {arXiv}\fi \@ifundefined
  {mn@eprint@\@tempb}{\@tempb:\@tempc}{\expandafter \expandafter \csname
  mn@eprint@\@tempb\endcsname \expandafter{\@tempc}}}

\bibitem[\protect\citeauthoryear{{Ag{\'u}ndez}, {Cernicharo}  \&
  {Gu{\'e}lin}}{{Ag{\'u}ndez} et~al.}{2010}]{2010ApJ...724L.133A}
{Ag{\'u}ndez} M.,  {Cernicharo} J.,   {Gu{\'e}lin} M.,  2010, \mn@doi [\apjl]
  {10.1088/2041-8205/724/2/L133}, \href
  {http://adsabs.harvard.edu/abs/2010ApJ...724L.133A} {724, L133}

\bibitem[\protect\citeauthoryear{{Bai} \& {Stone}}{{Bai} \&
  {Stone}}{2010a}]{Bai2010}
{Bai} X.-N.,  {Stone} J.~M.,  2010a, \mn@doi [\apjs]
  {10.1088/0067-0049/190/2/297}, \href
  {https://ui.adsabs.harvard.edu/abs/2010ApJS..190..297B} {190, 297}

\bibitem[\protect\citeauthoryear{{Bai} \& {Stone}}{{Bai} \&
  {Stone}}{2010b}]{Bai2010DiskII}
{Bai} X.-N.,  {Stone} J.~M.,  2010b, \mn@doi [\apj]
  {10.1088/0004-637X/722/2/1437}, \href
  {https://ui.adsabs.harvard.edu/abs/2010ApJ...722.1437B} {722, 1437}

\bibitem[\protect\citeauthoryear{{Bai} \& {Stone}}{{Bai} \&
  {Stone}}{2010c}]{Bai2010DiskI}
{Bai} X.-N.,  {Stone} J.~M.,  2010c, \mn@doi [\apjl]
  {10.1088/2041-8205/722/2/L220}, \href
  {https://ui.adsabs.harvard.edu/abs/2010ApJ...722L.220B} {722, L220}

\bibitem[\protect\citeauthoryear{{Balick} \& {Frank}}{{Balick} \&
  {Frank}}{2002}]{balick:2002.pne.shape.structure.review}
{Balick} B.,  {Frank} A.,  2002, \mn@doi [\araa]
  {10.1146/annurev.astro.40.060401.093849}, \href
  {http://adsabs.harvard.edu/abs/2002ARA%26A..40..439B} {40, 439}

\bibitem[\protect\citeauthoryear{{Balick}, {Alexander}, {Hajian}, {Terzian},
  {Perinotto}  \& {Patriarchi}}{{Balick}
  et~al.}{1998}]{balick:1998.knots.fliers.cometary.tails.pne}
{Balick} B.,  {Alexander} J.,  {Hajian} A.~R.,  {Terzian} Y.,  {Perinotto} M.,
   {Patriarchi} P.,  1998, \mn@doi [\aj] {10.1086/300429}, \href
  {http://adsabs.harvard.edu/abs/1998AJ....116..360B} {116, 360}

\bibitem[\protect\citeauthoryear{{Baud} \& {Habing}}{{Baud} \&
  {Habing}}{1983}]{Baud1983}
{Baud} B.,  {Habing} H.~J.,  1983, \aap, \href
  {https://ui.adsabs.harvard.edu/abs/1983A&A...127...73B} {127, 73}

\bibitem[\protect\citeauthoryear{{Carballido}, {Stone}  \&
  {Turner}}{{Carballido} et~al.}{2008}]{Carballido2008}
{Carballido} A.,  {Stone} J.~M.,   {Turner} N.~J.,  2008, \mn@doi [\mnras]
  {10.1111/j.1365-2966.2008.13014.x}, \href
  {https://ui.adsabs.harvard.edu/abs/2008MNRAS.386..145C} {386, 145}

\bibitem[\protect\citeauthoryear{{Cox} et~al.,}{{Cox}
  et~al.}{2012}]{2012A&A...537A..35C}
{Cox} N.~L.~J.,  et~al., 2012, \mn@doi [\aap] {10.1051/0004-6361/201117910},
  \href {http://adsabs.harvard.edu/abs/2012A%26A...537A..35C} {537, A35}

\bibitem[\protect\citeauthoryear{{Deguchi}}{{Deguchi}}{1997}]{deguchi:1997.dust.envelope.pne.spherical.drag.instability.quasi.resonant}
{Deguchi} S.,  1997, in {Habing} H.~J.,  {Lamers} H.~J.~G.~L.~M.,  eds,  IAU
  Symposium Vol. 180, Planetary Nebulae, Publisher: Dordrecht: Kluwer Academic
  Publishers. p.~151

\bibitem[\protect\citeauthoryear{{Deng}, {Mayer}, {Latter}, {Hopkins}  \&
  {Bai}}{{Deng} et~al.}{2019}]{deng:2019.mri.turb.sims.gizmo.methods}
{Deng} H.,  {Mayer} L.,  {Latter} H.,  {Hopkins} P.~F.,   {Bai} X.-N.,  2019,
  \mn@doi [\apjs] {10.3847/1538-4365/ab0957}, \href
  {https://ui.adsabs.harvard.edu/abs/2019ApJS..241...26D} {241, 26}

\bibitem[\protect\citeauthoryear{{Draine}}{{Draine}}{2003}]{Draine2003}
{Draine} B.~T.,  2003, \mn@doi [\araa]
  {10.1146/annurev.astro.41.011802.094840}, \href
  {https://ui.adsabs.harvard.edu/abs/2003ARA&A..41..241D} {41, 241}

\bibitem[\protect\citeauthoryear{{Draine}}{{Draine}}{2011}]{draine:ism.book}
{Draine} B.~T.,  2011, {Physics of the Interstellar and Intergalactic Medium}.
Princeton University Press, Princeton, NJ, USA

\bibitem[\protect\citeauthoryear{{Draine} \& {Salpeter}}{{Draine} \&
  {Salpeter}}{1979}]{Draine1979}
{Draine} B.~T.,  {Salpeter} E.~E.,  1979, \mn@doi [\apj] {10.1086/157165},
  \href {https://ui.adsabs.harvard.edu/abs/1979ApJ...231...77D} {231, 77}

\bibitem[\protect\citeauthoryear{{Draine} \& {Sutin}}{{Draine} \&
  {Sutin}}{1987}]{draine:1987.grain.charging}
{Draine} B.~T.,  {Sutin} B.,  1987, \mn@doi [\apj] {10.1086/165596}, \href
  {http://adsabs.harvard.edu/abs/1987ApJ...320..803D} {320, 803}

\bibitem[\protect\citeauthoryear{{Engels}, {Kreysa}, {Schultz}  \&
  {Sherwood}}{{Engels} et~al.}{1983}]{Engels1983}
{Engels} D.,  {Kreysa} E.,  {Schultz} G.~V.,   {Sherwood} W.~A.,  1983, \aap,
  \href {https://ui.adsabs.harvard.edu/abs/1983A&A...124..123E} {124, 123}

\bibitem[\protect\citeauthoryear{{Fadeyev} \& {Fokin}}{{Fadeyev} \&
  {Fokin}}{1986}]{1986NInfo..61...48F}
{Fadeyev} Y.~A.,  {Fokin} A.~B.,  1986, Nauchnye Informatsii, \href
  {https://ui.adsabs.harvard.edu/abs/1986NInfo..61...48F} {61, 48}

\bibitem[\protect\citeauthoryear{{Fong}, {Meixner}  \& {Shah}}{{Fong}
  et~al.}{2003}]{2003ApJ...582L..39F}
{Fong} D.,  {Meixner} M.,   {Shah} R.~Y.,  2003, \mn@doi [\apjl]
  {10.1086/346034}, \href {http://adsabs.harvard.edu/abs/2003ApJ...582L..39F}
  {582, L39}

\bibitem[\protect\citeauthoryear{{Gaburov} \& {Nitadori}}{{Gaburov} \&
  {Nitadori}}{2011}]{Gaburov2011}
{Gaburov} E.,  {Nitadori} K.,  2011, \mn@doi [\mnras]
  {10.1111/j.1365-2966.2011.18313.x}, \href
  {https://ui.adsabs.harvard.edu/abs/2011MNRAS.414..129G} {414, 129}

\bibitem[\protect\citeauthoryear{{Glass}, {Schultheis}, {Blommaert}, {Sahai},
  {Stute}  \& {Uttenthaler}}{{Glass} et~al.}{2009}]{Glass2009}
{Glass} I.~S.,  {Schultheis} M.,  {Blommaert} J.~A.~D.~L.,  {Sahai} R.,
  {Stute} M.,   {Uttenthaler} S.,  2009, \mn@doi [\mnras]
  {10.1111/j.1745-3933.2009.00628.x}, \href
  {https://ui.adsabs.harvard.edu/abs/2009MNRAS.395L..11G} {395, L11}

\bibitem[\protect\citeauthoryear{{Hartquist} \& {Havnes}}{{Hartquist} \&
  {Havnes}}{1994}]{hartquist:bfield.dust.coupling.cool.star.winds}
{Hartquist} T.~W.,  {Havnes} O.,  1994, \mn@doi [Astrophysics and Space
  Science] {10.1007/BF00658063}, \href
  {http://adsabs.harvard.edu/abs/1994Ap%26SS.218...23H} {218, 23}

\bibitem[\protect\citeauthoryear{{Hill} \& {Willson}}{{Hill} \&
  {Willson}}{1979}]{Hill1979}
{Hill} S.~J.,  {Willson} L.~A.,  1979, \mn@doi [\apj] {10.1086/157038}, \href
  {https://ui.adsabs.harvard.edu/abs/1979ApJ...229.1029H} {229, 1029}

\bibitem[\protect\citeauthoryear{{H{\"o}fner} \& {Olofsson}}{{H{\"o}fner} \&
  {Olofsson}}{2018}]{2018A&ARv..26....1H}
{H{\"o}fner} S.,  {Olofsson} H.,  2018, \mn@doi [\aapr]
  {10.1007/s00159-017-0106-5}, \href
  {https://ui.adsabs.harvard.edu/abs/2018A&ARv..26....1H} {26, 1}

\bibitem[\protect\citeauthoryear{{Hopkins}}{{Hopkins}}{2015}]{Hopkins2015}
{Hopkins} P.~F.,  2015, \mn@doi [\mnras] {10.1093/mnras/stv195}, \href
  {https://ui.adsabs.harvard.edu/abs/2015MNRAS.450...53H} {450, 53}

\bibitem[\protect\citeauthoryear{{Hopkins}}{{Hopkins}}{2016}]{hopkins:cg.mhd.gizmo}
{Hopkins} P.~F.,  2016, \mn@doi [\mnras] {10.1093/mnras/stw1578}, \href
  {http://adsabs.harvard.edu/abs/2016MNRAS.462..576H} {462, 576}

\bibitem[\protect\citeauthoryear{{Hopkins}}{{Hopkins}}{2017}]{hopkins:gizmo.public.release}
{Hopkins} P.~F.,  2017, ArXiv e-prints, arXiv:1712.01294, \href
  {http://adsabs.harvard.edu/abs/2017arXiv171201294H} {}

\bibitem[\protect\citeauthoryear{{Hopkins} \& {Lee}}{{Hopkins} \&
  {Lee}}{2016}]{hopkins.2016:dust.gas.molecular.cloud.dynamics.sims}
{Hopkins} P.~F.,  {Lee} H.,  2016, \mn@doi [\mnras] {10.1093/mnras/stv2745},
  \href {http://adsabs.harvard.edu/abs/2016MNRAS.456.4174H} {456, 4174}

\bibitem[\protect\citeauthoryear{{Hopkins} \& {Raives}}{{Hopkins} \&
  {Raives}}{2016}]{hopkins:mhd.gizmo}
{Hopkins} P.~F.,  {Raives} M.~J.,  2016, \mn@doi [\mnras]
  {10.1093/mnras/stv2180}, \href
  {http://adsabs.harvard.edu/abs/2016MNRAS.455...51H} {455, 51}

\bibitem[\protect\citeauthoryear{{Hopkins} \& {Squire}}{{Hopkins} \&
  {Squire}}{2018a}]{hopkins:2018.mhd.rdi}
{Hopkins} P.~F.,  {Squire} J.,  2018a, \mn@doi [\mnras]
  {10.1093/mnras/sty1604}, \href
  {http://adsabs.harvard.edu/abs/2018MNRAS.479.4681H} {479, 4681}

\bibitem[\protect\citeauthoryear{{Hopkins} \& {Squire}}{{Hopkins} \&
  {Squire}}{2018b}]{hopkins:2017.acoustic.RDI}
{Hopkins} P.~F.,  {Squire} J.,  2018b, \mn@doi [\mnras]
  {10.1093/mnras/sty1982}, \href
  {http://adsabs.harvard.edu/abs/2018MNRAS.480.2813H} {480, 2813}

\bibitem[\protect\citeauthoryear{{Hopkins}, {Squire}  \& {Seligman}}{{Hopkins}
  et~al.}{2020}]{hopkins:2019.mhd.rdi.periodic.box.sims}
{Hopkins} P.~F.,  {Squire} J.,   {Seligman} D.,  2020, \mn@doi [\mnras]
  {10.1093/mnras/staa1046}, \href
  {https://ui.adsabs.harvard.edu/abs/2020MNRAS.496.2123H} {496, 2123}

\bibitem[\protect\citeauthoryear{{Hopkins}, {Rosen}, {Squire}, {Panopoulou},
  {Soliman}, {Seligman}  \& {Steinwandel}}{{Hopkins}
  et~al.}{2021}]{hopkins2021:mhd.rdi.stratified.box.sims}
{Hopkins} P.~F.,  {Rosen} A.~L.,  {Squire} J.,  {Panopoulou} G.~V.,  {Soliman}
  N.~H.,  {Seligman} D.,   {Steinwandel} U.~P.,  2021, arXiv e-prints, \href
  {https://ui.adsabs.harvard.edu/abs/2021arXiv210704608H} {p. arXiv:2107.04608}

\bibitem[\protect\citeauthoryear{{Jeffery} \& {Hambsch}}{{Jeffery} \&
  {Hambsch}}{2019}]{Jeffery2019}
{Jeffery} C.~S.,  {Hambsch} F.~J.,  2019, \mn@doi [\mnras]
  {10.1093/mnras/stz1600}, \href
  {https://ui.adsabs.harvard.edu/abs/2019MNRAS.487.4128J} {487, 4128}

\bibitem[\protect\citeauthoryear{{Johansen}, {Youdin}  \& {Mac Low}}{{Johansen}
  et~al.}{2009}]{Johansen2009}
{Johansen} A.,  {Youdin} A.,   {Mac Low} M.-M.,  2009, \mn@doi [\apjl]
  {10.1088/0004-637X/704/2/L75}, \href
  {https://ui.adsabs.harvard.edu/abs/2009ApJ...704L..75J} {704, L75}

\bibitem[\protect\citeauthoryear{{Karovicova}, {Wittkowski}, {Ohnaka},
  {Boboltz}, {Fossat}  \& {Scholz}}{{Karovicova} et~al.}{2013}]{Karovicova2013}
{Karovicova} I.,  {Wittkowski} M.,  {Ohnaka} K.,  {Boboltz} D.~A.,  {Fossat}
  E.,   {Scholz} M.,  2013, \mn@doi [\aap] {10.1051/0004-6361/201322376}, \href
  {https://ui.adsabs.harvard.edu/abs/2013A&A...560A..75K} {560, A75}

\bibitem[\protect\citeauthoryear{{Khouri} et~al.,}{{Khouri}
  et~al.}{2016}]{Khouri2016}
{Khouri} T.,  et~al., 2016, \mn@doi [\aap] {10.1051/0004-6361/201628435}, \href
  {https://ui.adsabs.harvard.edu/abs/2016A&A...591A..70K} {591, A70}

\bibitem[\protect\citeauthoryear{{Kudritzki} \& {Puls}}{{Kudritzki} \&
  {Puls}}{2000}]{Kudritzki2000}
{Kudritzki} R.-P.,  {Puls} J.,  2000, \mn@doi [\araa]
  {10.1146/annurev.astro.38.1.613}, \href
  {https://ui.adsabs.harvard.edu/abs/2000ARA&A..38..613K} {38, 613}

\bibitem[\protect\citeauthoryear{{Lee}, {Hopkins}  \& {Squire}}{{Lee}
  et~al.}{2017}]{lee:dynamics.charged.dust.gmcs}
{Lee} H.,  {Hopkins} P.~F.,   {Squire} J.,  2017, \mn@doi [\mnras]
  {10.1093/mnras/stx1097}, \href
  {http://adsabs.harvard.edu/abs/2017MNRAS.469.3532L} {469, 3532}

\bibitem[\protect\citeauthoryear{{MacGregor} \& {Stencel}}{{MacGregor} \&
  {Stencel}}{1992}]{macgregor:grains.cool.star.eventually.decouple}
{MacGregor} K.~B.,  {Stencel} R.~E.,  1992, \mn@doi [\apj] {10.1086/171820},
  \href {http://adsabs.harvard.edu/abs/1992ApJ...397..644M} {397, 644}

\bibitem[\protect\citeauthoryear{{Mastrodemos}, {Morris}  \&
  {Castor}}{{Mastrodemos}
  et~al.}{1996}]{mastrodemos:dust.gas.coupling.cool.winds.spherical.symmetry}
{Mastrodemos} N.,  {Morris} M.,   {Castor} J.,  1996, \mn@doi [\apj]
  {10.1086/177741}, \href {http://adsabs.harvard.edu/abs/1996ApJ...468..851M}
  {468, 851}

\bibitem[\protect\citeauthoryear{{Mathis}, {Rumpl}  \& {Nordsieck}}{{Mathis}
  et~al.}{1977}]{Mathis1977}
{Mathis} J.~S.,  {Rumpl} W.,   {Nordsieck} K.~H.,  1977, \mn@doi [\apj]
  {10.1086/155591}, \href
  {https://ui.adsabs.harvard.edu/abs/1977ApJ...217..425M} {217, 425}

\bibitem[\protect\citeauthoryear{{Matsuura} et~al.,}{{Matsuura}
  et~al.}{2009}]{matsuura:2009.pne.knots}
{Matsuura} M.,  et~al., 2009, \mn@doi [\apj] {10.1088/0004-637X/700/2/1067},
  \href {http://adsabs.harvard.edu/abs/2009ApJ...700.1067M} {700, 1067}

\bibitem[\protect\citeauthoryear{{McDonald} \& {Zijlstra}}{{McDonald} \&
  {Zijlstra}}{2016}]{McDonald2016}
{McDonald} I.,  {Zijlstra} A.~A.,  2016, \mn@doi [\apjl]
  {10.3847/2041-8205/823/2/L38}, \href
  {https://ui.adsabs.harvard.edu/abs/2016ApJ...823L..38M} {823, L38}

\bibitem[\protect\citeauthoryear{{Morris}}{{Morris}}{1993}]{morris:1993.cool.wind.dust.drag.instability.slow.saturated.mode}
{Morris} M.,  1993, in {Schwarz} H.~E.,  ed.,  European Southern Observatory
  Conference and Workshop Proceedings Vol. 46, European Southern Observatory
  Conference and Workshop Proceedings, Garching. p.~60

\bibitem[\protect\citeauthoryear{{Moseley}, {Squire}  \& {Hopkins}}{{Moseley}
  et~al.}{2019}]{moseley:2018.acoustic.rdi.sims}
{Moseley} E.~R.,  {Squire} J.,   {Hopkins} P.~F.,  2019, \mn@doi [\mnras]
  {10.1093/mnras/stz2128}, \href
  {https://ui.adsabs.harvard.edu/abs/2019MNRAS.489..325M} {489, 325}

\bibitem[\protect\citeauthoryear{{O'Dell}, {Balick}, {Hajian}, {Henney}  \&
  {Burkert}}{{O'Dell} et~al.}{2002}]{odell:2002.pne.knots.review}
{O'Dell} C.~R.,  {Balick} B.,  {Hajian} A.~R.,  {Henney} W.~J.,   {Burkert} A.,
   2002, \mn@doi [\aj] {10.1086/340726}, \href
  {http://adsabs.harvard.edu/abs/2002AJ....123.3329O} {123, 3329}

\bibitem[\protect\citeauthoryear{{O'dell} \& {Handron}}{{O'dell} \&
  {Handron}}{1996}]{odell:1996.pne.cometary.knots.from.instabilities}
{O'dell} C.~R.,  {Handron} K.~D.,  1996, \mn@doi [\aj] {10.1086/117902}, \href
  {http://adsabs.harvard.edu/abs/1996AJ....111.1630O} {111, 1630}

\bibitem[\protect\citeauthoryear{{Ohnaka}, {Weigelt}  \& {Hofmann}}{{Ohnaka}
  et~al.}{2017a}]{Ohnaka2017Natur}
{Ohnaka} K.,  {Weigelt} G.,   {Hofmann} K.~H.,  2017a, \mn@doi [\nat]
  {10.1038/nature23445}, \href
  {https://ui.adsabs.harvard.edu/abs/2017Natur.548..310O} {548, 310}

\bibitem[\protect\citeauthoryear{{Ohnaka}, {Weigelt}  \& {Hofmann}}{{Ohnaka}
  et~al.}{2017b}]{Ohnaka2017A&A}
{Ohnaka} K.,  {Weigelt} G.,   {Hofmann} K.~H.,  2017b, \mn@doi [\aap]
  {10.1051/0004-6361/201629761}, \href
  {https://ui.adsabs.harvard.edu/abs/2017A&A...597A..20O} {597, A20}

\bibitem[\protect\citeauthoryear{{Owocki}}{{Owocki}}{2010}]{Owocki2010}
{Owocki} S.,  2010, in {Leitherer} C.,  {Bennett} P.~D.,  {Morris} P.~W.,
  {Van Loon} J.~T.,  eds,  Astronomical Society of the Pacific Conference
  Series Vol. 425, Hot and Cool: Bridging Gaps in Massive Star Evolution.
  p.~199

\bibitem[\protect\citeauthoryear{{Owocki}}{{Owocki}}{2014}]{Owocki2014}
{Owocki} S.,  2014, arXiv e-prints, \href
  {https://ui.adsabs.harvard.edu/abs/2014arXiv1409.2084O} {p. arXiv:1409.2084}

\bibitem[\protect\citeauthoryear{{Pan}, {Padoan}, {Scalo}, {Kritsuk}  \&
  {Norman}}{{Pan} et~al.}{2011}]{Pan2011}
{Pan} L.,  {Padoan} P.,  {Scalo} J.,  {Kritsuk} A.~G.,   {Norman} M.~L.,  2011,
  \mn@doi [\apj] {10.1088/0004-637X/740/1/6}, \href
  {https://ui.adsabs.harvard.edu/abs/2011ApJ...740....6P} {740, 6}

\bibitem[\protect\citeauthoryear{{Puls}, {Vink}  \& {Najarro}}{{Puls}
  et~al.}{2008}]{Puls2008}
{Puls} J.,  {Vink} J.~S.,   {Najarro} F.,  2008, \mn@doi [\aapr]
  {10.1007/s00159-008-0015-8}, \href
  {https://ui.adsabs.harvard.edu/abs/2008A&ARv..16..209P} {16, 209}

\bibitem[\protect\citeauthoryear{{Puls}, {Sundqvist}  \& {Markova}}{{Puls}
  et~al.}{2015}]{Puls2015}
{Puls} J.,  {Sundqvist} J.~O.,   {Markova} N.,  2015, in {Meynet} G.,  {Georgy}
  C.,  {Groh} J.,   {Stee} P.,  eds,  Proceedings of the International
  Astronomical Union, IAU Symposium Vol. 307, New Windows on Massive Stars. pp
  25--36 (\mn@eprint {arXiv} {1409.3582}), \mn@doi{10.1017/S174392131400622X}

\bibitem[\protect\citeauthoryear{{Rao}}{{Rao}}{2008}]{Rao2008}
{Rao} N.~K.,  2008, in {Werner} A.,  {Rauch} T.,  eds,  Astronomical Society of
  the Pacific Conference Series Vol. 391, Hydrogen-Deficient Stars. p.~25

\bibitem[\protect\citeauthoryear{{Reimers}}{{Reimers}}{1975}]{Reimers1975}
{Reimers} D.,  1975, {Circumstellar envelopes and mass loss of red giant
  stars.}.
pp 229--256

\bibitem[\protect\citeauthoryear{{Sabin}, {Wade}  \& {L{\`e}bre}}{{Sabin}
  et~al.}{2015}]{Sabin2015}
{Sabin} L.,  {Wade} G.~A.,   {L{\`e}bre} A.,  2015, \mn@doi [\mnras]
  {10.1093/mnras/stu2227}, \href
  {https://ui.adsabs.harvard.edu/abs/2015MNRAS.446.1988S} {446, 1988}

\bibitem[\protect\citeauthoryear{{Sandin} \& {H{\"o}fner}}{{Sandin} \&
  {H{\"o}fner}}{2003}]{sandin:agb.wind.sims}
{Sandin} C.,  {H{\"o}fner} S.,  2003, \mn@doi [\aap]
  {10.1051/0004-6361:20030515}, \href
  {http://adsabs.harvard.edu/abs/2003A%26A...404..789S} {404, 789}

\bibitem[\protect\citeauthoryear{{Seligman}, {Hopkins}  \& {Squire}}{{Seligman}
  et~al.}{2019}]{seligman:2018.mhd.rdi.sims}
{Seligman} D.,  {Hopkins} P.~F.,   {Squire} J.,  2019, \mn@doi [\mnras]
  {10.1093/mnras/stz666}, \href
  {https://ui.adsabs.harvard.edu/abs/2019MNRAS.485.3991S} {485, 3991}

\bibitem[\protect\citeauthoryear{{Simis}, {Icke}  \& {Dominik}}{{Simis}
  et~al.}{2001}]{simis:2001.shells.around.dust.agb.winds}
{Simis} Y.~J.~W.,  {Icke} V.,   {Dominik} C.,  2001, \mn@doi [\aap]
  {10.1051/0004-6361:20010341}, \href
  {http://adsabs.harvard.edu/abs/2001A%26A...371..205S} {371, 205}

\bibitem[\protect\citeauthoryear{{Smith}}{{Smith}}{2014}]{Smith2014}
{Smith} N.,  2014, \mn@doi [\araa] {10.1146/annurev-astro-081913-040025}, \href
  {https://ui.adsabs.harvard.edu/abs/2014ARA&A..52..487S} {52, 487}

\bibitem[\protect\citeauthoryear{{Soker}}{{Soker}}{2000}]{soker:agb.star.magnetic.cycle.instabilities}
{Soker} N.,  2000, \mn@doi [\apj] {10.1086/309326}, \href
  {http://adsabs.harvard.edu/abs/2000ApJ...540..436S} {540, 436}

\bibitem[\protect\citeauthoryear{{Soker}}{{Soker}}{2002}]{soker:2002.arcs.around.agb.stars.from.instability}
{Soker} N.,  2002, \mn@doi [\apj] {10.1086/339573}, \href
  {http://adsabs.harvard.edu/abs/2002ApJ...570..369S} {570, 369}

\bibitem[\protect\citeauthoryear{{Squire} \& {Hopkins}}{{Squire} \&
  {Hopkins}}{2018a}]{squire:rdi.ppd}
{Squire} J.,  {Hopkins} P.~F.,  2018a, \mn@doi [\mnras] {10.1093/mnras/sty854},
  \href {http://adsabs.harvard.edu/abs/2018MNRAS.477.5011S} {477, 5011}

\bibitem[\protect\citeauthoryear{{Squire} \& {Hopkins}}{{Squire} \&
  {Hopkins}}{2018b}]{squire.hopkins:RDI}
{Squire} J.,  {Hopkins} P.~F.,  2018b, \mn@doi [\apjl]
  {10.3847/2041-8213/aab54d}, \href
  {http://adsabs.harvard.edu/abs/2018ApJ...856L..15S} {856, L15}

\bibitem[\protect\citeauthoryear{{Squire}, {Moroianu}  \& {Hopkins}}{{Squire}
  et~al.}{2021}]{squire2021:hd.periodic.grain.spectrum.sims}
{Squire} J.,  {Moroianu} S.,   {Hopkins} P.~F.,  2021, arXiv e-prints, \href
  {https://ui.adsabs.harvard.edu/abs/2021arXiv211011422S} {p. arXiv:2110.11422}

\bibitem[\protect\citeauthoryear{{Stewart}, {Tuthill}, {Nicholson}  \&
  {Hedman}}{{Stewart} et~al.}{2016}]{Stewart2016}
{Stewart} P.~N.,  {Tuthill} P.~G.,  {Nicholson} P.~D.,   {Hedman} M.~M.,  2016,
  \mn@doi [\mnras] {10.1093/mnras/stw045}, \href
  {https://ui.adsabs.harvard.edu/abs/2016MNRAS.457.1410S} {457, 1410}

\bibitem[\protect\citeauthoryear{{Tej}, {Lan{\c{c}}on}, {Scholz}  \&
  {Wood}}{{Tej} et~al.}{2003}]{Tej2003}
{Tej} A.,  {Lan{\c{c}}on} A.,  {Scholz} M.,   {Wood} P.~R.,  2003, \mn@doi
  [\aap] {10.1051/0004-6361:20031430}, \href
  {https://ui.adsabs.harvard.edu/abs/2003A&A...412..481T} {412, 481}

\bibitem[\protect\citeauthoryear{{Weingartner} \& {Draine}}{{Weingartner} \&
  {Draine}}{2001}]{Weingartner2001}
{Weingartner} J.~C.,  {Draine} B.~T.,  2001, \mn@doi [\apj] {10.1086/318651},
  \href {https://ui.adsabs.harvard.edu/abs/2001ApJ...548..296W} {548, 296}

\bibitem[\protect\citeauthoryear{{Whitelock}, {Menzies}, {Catchpole}, {Feast},
  {Roberts}  \& {Marang}}{{Whitelock} et~al.}{1991}]{Whitelock1991}
{Whitelock} P.~A.,  {Menzies} J.~W.,  {Catchpole} R.~M.,  {Feast} M.~W.,
  {Roberts} G.,   {Marang} F.,  1991, \mn@doi [\mnras]
  {10.1093/mnras/250.3.638}, \href
  {https://ui.adsabs.harvard.edu/abs/1991MNRAS.250..638W} {250, 638}

\bibitem[\protect\citeauthoryear{{Whitelock}, {Feast}, {van Loon}  \&
  {Zijlstra}}{{Whitelock} et~al.}{2003}]{Whitelock2003}
{Whitelock} P.~A.,  {Feast} M.~W.,  {van Loon} J.~T.,   {Zijlstra} A.~A.,
  2003, \mn@doi [\mnras] {10.1046/j.1365-8711.2003.06514.x}, \href
  {https://ui.adsabs.harvard.edu/abs/2003MNRAS.342...86W} {342, 86}

\bibitem[\protect\citeauthoryear{{Willson} \& {Hill}}{{Willson} \&
  {Hill}}{1979}]{Willson1979}
{Willson} L.~A.,  {Hill} S.~J.,  1979, \mn@doi [\apj] {10.1086/156911}, \href
  {https://ui.adsabs.harvard.edu/abs/1979ApJ...228..854W} {228, 854}

\bibitem[\protect\citeauthoryear{{Winters}, {Dominik}  \& {Sedlmayr}}{{Winters}
  et~al.}{1994}]{1994A&A...288..255W}
{Winters} J.~M.,  {Dominik} C.,   {Sedlmayr} E.,  1994, \aap, \href
  {http://adsabs.harvard.edu/abs/1994A%26A...288..255W} {288, 255}

\bibitem[\protect\citeauthoryear{{Wittkowski}, {Boboltz}, {Ohnaka}, {Driebe}
  \& {Scholz}}{{Wittkowski} et~al.}{2007}]{Wittkowski2007}
{Wittkowski} M.,  {Boboltz} D.~A.,  {Ohnaka} K.,  {Driebe} T.,   {Scholz} M.,
  2007, \mn@doi [\aap] {10.1051/0004-6361:20077168}, \href
  {https://ui.adsabs.harvard.edu/abs/2007A&A...470..191W} {470, 191}

\bibitem[\protect\citeauthoryear{{Woitke}}{{Woitke}}{2006a}]{woitke:2d.agb.wind.simulations}
{Woitke} P.,  2006a, \mn@doi [\aap] {10.1051/0004-6361:20054202}, \href
  {http://adsabs.harvard.edu/abs/2006A%26A...452..537W} {452, 537}

\bibitem[\protect\citeauthoryear{{Woitke}}{{Woitke}}{2006b}]{woitke:2d.rad.pressure.dust.agb.wind.models}
{Woitke} P.,  2006b, \mn@doi [\aap] {10.1051/0004-6361:20066322}, \href
  {http://adsabs.harvard.edu/abs/2006A%26A...460L...9W} {460, L9}

\bibitem[\protect\citeauthoryear{{Wood}}{{Wood}}{1979}]{Wood1979}
{Wood} P.~R.,  1979, \mn@doi [\apj] {10.1086/156721}, \href
  {https://ui.adsabs.harvard.edu/abs/1979ApJ...227..220W} {227, 220}

\bibitem[\protect\citeauthoryear{{Young}, {Highberger}, {Arnett}  \&
  {Ziurys}}{{Young} et~al.}{2003}]{young:2003.clumpy.wind.models}
{Young} P.~A.,  {Highberger} J.~L.,  {Arnett} D.,   {Ziurys} L.~M.,  2003,
  \mn@doi [\apjl] {10.1086/379813}, \href
  {http://adsabs.harvard.edu/abs/2003ApJ...597L..53Y} {597, L53}

\bibitem[\protect\citeauthoryear{{Zhao-Geisler}, {Quirrenbach}, {K{\"o}hler}
  \& {Lopez}}{{Zhao-Geisler} et~al.}{2012}]{Zhao-Geisler2012}
{Zhao-Geisler} R.,  {Quirrenbach} A.,  {K{\"o}hler} R.,   {Lopez} B.,  2012,
  \mn@doi [\aap] {10.1051/0004-6361/201118150}, \href
  {https://ui.adsabs.harvard.edu/abs/2012A&A...545A..56Z} {545, A56}

\bibitem[\protect\citeauthoryear{{Ziurys}, {Milam}, {Apponi}  \&
  {Woolf}}{{Ziurys} et~al.}{2007}]{2007Natur.447.1094Z}
{Ziurys} L.~M.,  {Milam} S.~N.,  {Apponi} A.~J.,   {Woolf} N.~J.,  2007,
  \mn@doi [\nat] {10.1038/nature05905}, \href
  {http://adsabs.harvard.edu/abs/2007Natur.447.1094Z} {447, 1094}

\makeatother
\end{thebibliography}
\bsp




\label{lastpage}
\end{document}